\documentclass[twocolumn]{aastex61}
\usepackage{amsmath,amstext}
\usepackage[T1]{fontenc}
\usepackage{apjfonts} 
\usepackage[figure,figure*]{hypcap}
\makeatletter

\newcommand{\Rmnum}[1]{\expandafter\@slowromancap\romannumeral #1@}
\makeatother

\usepackage{color}

\usepackage{soul}   
\usepackage[colorinlistoftodos]{todonotes}

\shorttitle{The Nature of DES1, Eri\,III and Tuc\,V}
\shortauthors{Conn et al.}

\begin{document}


\title{On The Nature of ultra-faint Dwarf Galaxy Candidates I:\\ DES1, Eridanus\,III and Tucana\,V}

\author{Blair C. Conn}
\affiliation{Research School of Astronomy and Astrophysics, Australian National University, Canberra, ACT 2611, Australia}

\author{Helmut Jerjen}
\affiliation{Research School of Astronomy and Astrophysics, Australian National University, Canberra, ACT 2611, Australia}

\author{Dongwon Kim} 
\affiliation{Research School of Astronomy and Astrophysics, Australian National University, Canberra, ACT 2611, Australia}

\author{Mischa Schirmer} 
\affiliation{Gemini Observatory, Casilla 603, La Serena, Chile}

\correspondingauthor{Blair C. Conn}
\email{blair.conn@anu.edu.au}





\begin{abstract}
We use deep Gemini/GMOS-S $g,r$ photometry to study the three ultra-faint dwarf galaxy candidates DES1, Eridanus\,III (Eri\,III) and Tucana\,V (Tuc\,V).  
Their total luminosities, $M_V$(DES1)\,$ = -1.42\pm0.50$ and $M_V$(Eri\,III)\,$  = -2.07\pm0.50$, and mean metallicities, [Fe/H]\,$=-2.38^{+0.21}_{-0.19}$ and [Fe/H]\,$=-2.40^{+0.19}_{-0.12}$,
are consistent with them being ultra-faint dwarf galaxies as they fall just outside the 1-sigma confidence band of the luminosity-metallicity relation for Milky Way satellite galaxies.  
However, their positions in the size-luminosity relation suggests that they are star clusters. Interestingly, DES1 and Eri\,III are at relatively large Galactocentric distances with DES1 located at $D_{GC} = 74\pm$4 kpc and Eri\,III at $D_{GC} = 91\pm$4 kpc. In projection both objects are in the tail of gaseous filaments trailing the Magellanic Clouds and have similar 3D-separations from the Small Magellanic Cloud (SMC): $\Delta D_{SMC,DES1}$ = 31.7\,kpc and $\Delta D_{SMC,Eri\,III}$ = 41.0\,kpc, respectively. It is plausible that these stellar systems are metal-poor SMC satellites. Tuc\,V represents an interesting phenomenon in its own right. Our deep photometry at the nominal position of Tuc\,V reveals a low-level excess of stars at various locations across the GMOS field without a well-defined centre. A SMC Northern Overdensity-like isochrone would be an adequate match to the Tuc\,V colour-magnitude diagram, and the proximity to the SMC ($12\fdg 1$; $\Delta D_{SMC,Tuc\,V}=13$\,kpc) suggests that Tuc\,V is either a chance grouping of stars related to the SMC halo or a star cluster in an advanced stage of dissolution.
\end{abstract}



\section{Introduction}
In recent years, around 35 new Milky Way satellites (dwarf galaxies and star clusters) have been discovered \citep{Balbinot2013, Belokurov2014,Laevens2014, Bechtol2015, Drlica-Wagner2015, Kim2, Kim2015b, KimJerjen2015a, KimJerjen2015b, Koposov2015, Laevens2015a, Laevens2015b, Martin2015, Kim2016, Luque2016, Martin2016b, Torrealba2016a, Torrealba2016b, Koposov2017}. This is a dramatic jump in number and 
once their true nature has been established these objects will provide crucial empirical input for testing near-field cosmology predictions and  verifying formation scenarios of the Milky Way. However, since the majority of the discoveries are based on shallow SDSS\footnote{Sloan Digital Sky Survey}, Pan-STARRS1\footnote{Panoramic Survey Telescope and Rapid Response System, \citet{2016arXiv161205560C}} or DES\footnote{Dark Energy Survey, http://des.ncsa.illinois.edu/releases/sva1D} imaging surveys, most new objects themselves are still poorly constrained in terms of their stellar population, structure parameters, distance and luminosity. The only path forward to accurately determine these fundamental properties is to analyse deep photometric follow-up observations. 

In this paper, we use deep Gemini/GMOS-S $g,r$ photometry to derive more accurate constraints on the three ultra-faint dwarf galaxy candidates DES1 \citep{Luque2016}, Eridanus\,III ~\citep[DES J0222.7-5217]{Bechtol2015,Koposov2015} and Tucana\,V ~\citep[DES J2337-6316]{Drlica-Wagner2015}. DES1 was detected in first-year Dark Energy Survey data with a peak Poisson significance of 11.6 as a compact Milky Way companion at $\alpha(J2000)=0^\mathrm{h}33^\mathrm{m}59\fs7$ and $\delta(J2000)=-49^{\circ}\,02'\,20''\,$ located at approximately 80\,kpc.

Its total luminosity is estimated in the range $-$3.00 $\leqslant$ M$_V$ $\leqslant$ $-$2.21. Eridanus\,III at $\alpha(J2000)=02^\mathrm{h}22^\mathrm{m}45\fs5$, $\delta(J2000)=-52^{\circ}17'\,01''$ detected at a significance level of 10.1 resides at a heliocentric distance of $\sim$87 kpc with a total luminosity of $M_V = -2.0\pm0.3$. Tucana\,V was discovered at $\alpha(J2000)=23^\mathrm{h}37^\mathrm{m}24\fs0$, $\delta(J2000)=-63^{\circ}\,16'\,12''\,$ (J2000) with a peak significance of 8.0 at an estimated distance of 55$\pm$9 kpc and has a total luminosity of $M_V = -1.60\pm0.49$.

\begin{figure*}
\begin{center}
\includegraphics[width=1.0\hsize]{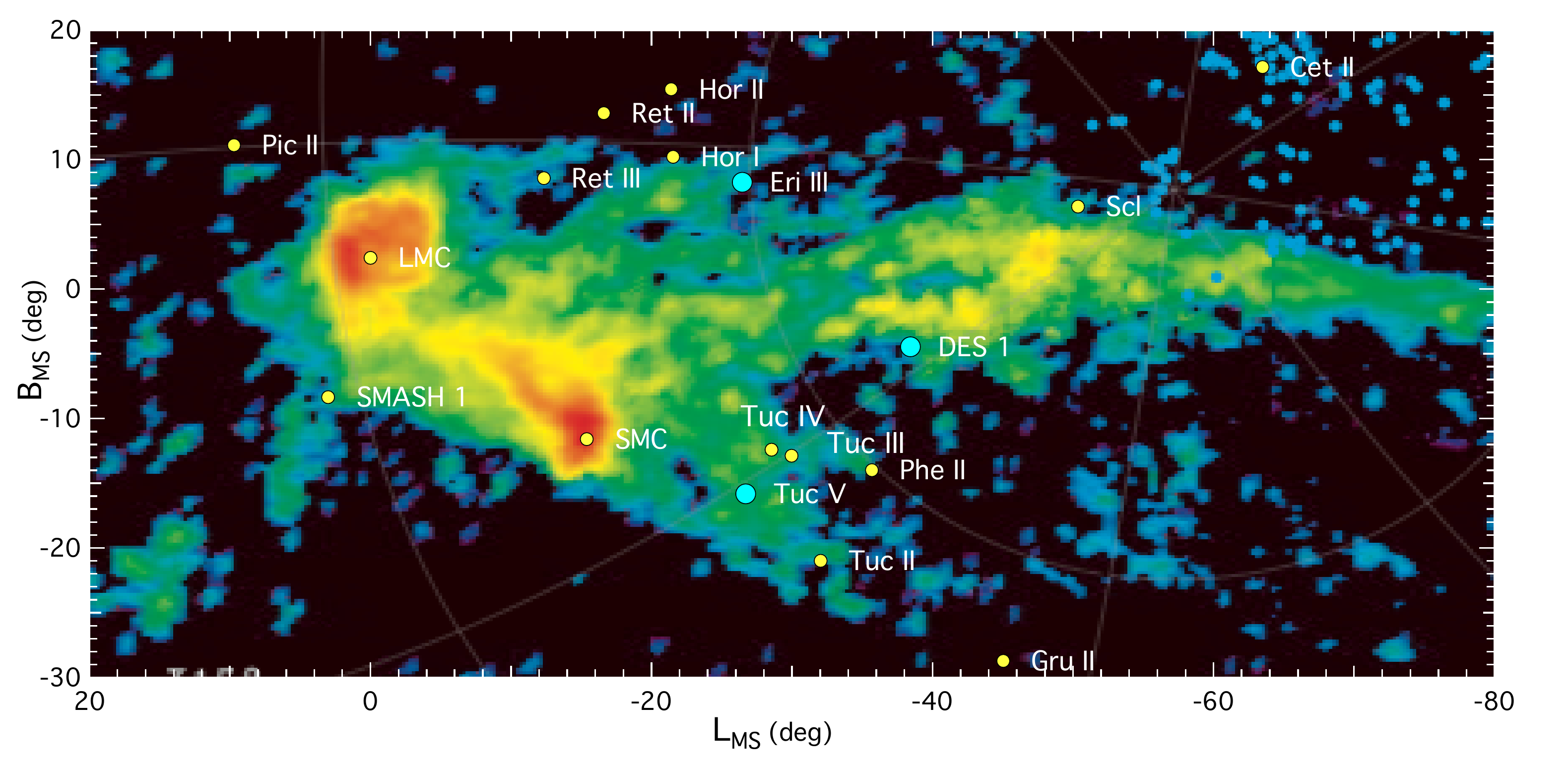}
\caption{On sky distribution of all known Milky Way satellite candidates in the distance range $30<D_{GC}<100$\,kpc with respect to the Magellanic Clouds and the neutral hydrogen gas of the Magellanic stream. The HI column density ($\log(N_{HI})$ in units of cm$^{-2}$) is shown over six orders of magnitudes, ranging from $\log(N_{HI})=16$ (black) to 22 (red). For more details we refer to \citet{2010ApJ...723.1618N}. The three candidates discussed in this study are highlighted in cyan.}
\label{fig:MCs}
\end{center}
\end{figure*}

Interestingly, all three objects have half-light radii ($r_{h,DES1}\sim10$\,pc, $r_{h,Eri\,III}= 14.0^{+12.5}_{-2.6}$\,pc, $r_{h,Tuc\,V}= 17\pm6$\,pc, ) that puts them in the transition zone between ultra-faint star clusters and dwarf galaxies. 
Figure~\ref{fig:MCs} shows the location of these three stellar overdensities among others with respect to the Magellanic Clouds and the gaseous Magellanic Stream.

Each of these objects reside at the very limit of the DES photometry ($g_{lim}\sim 23$) in which they were discovered. This introduced large uncertainties into all of their known properties including the half-light radius, which can be used to discriminate between a baryon-dominated star clusters and dark matter dominated dwarf galaxy. These data presented here will significantly improve our understanding of these faint stellar systems and allow us to refine their locations in the size-luminosity plane and also in the luminosity-metallicity parameter space where there is a known relation between these parameters for dwarf galaxies ~\citep{Kirby2013}. Additionally, by probing several magnitudes below the main sequence turn-off (MSTO) we can take advantage of the stellar mass differences of main sequence stars to probe for any evidence of mass segregation as witnessed in the stellar cluster Kim\,2 ~\citep{Kim2}. Evidence of mass segregation can confirm a system as being purely baryonic and so may provide a unique opportunity to resolve their origins with photometry. Through these relations we can test the likelihood of their true nature as star clusters or dwarf galaxies.

\section{Observations and Data Reduction}
\begin{figure}
\begin{center}
\includegraphics[width=1.0\hsize]{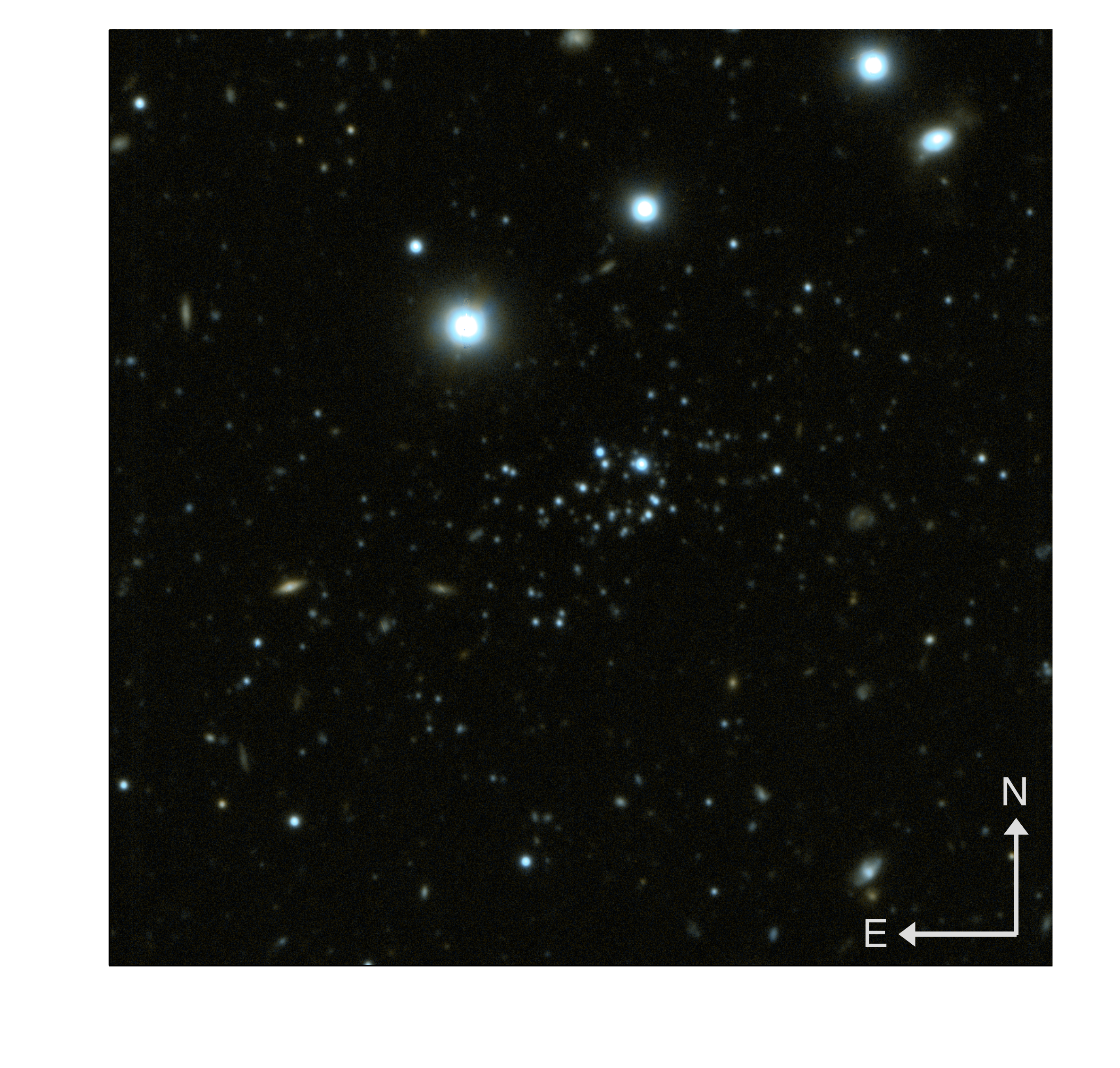}
\caption{False color RGB image of DES1, made using {\sc Aladin Sky Atlas v8.040}, with a $\sim$2$'$$\times$2$'$ field of view. The $g$-band co-added image was used for the Blue and the $r$-band co-added image for the Red. DES1 is the small overdensity of stars in the centre of this field. The arrows in the lower right corner have a length of 15 arcsec.}
\label{fig:DES1}
\end{center}
\end{figure}

\begin{figure}
\begin{center}
\includegraphics[width=1.0\hsize]{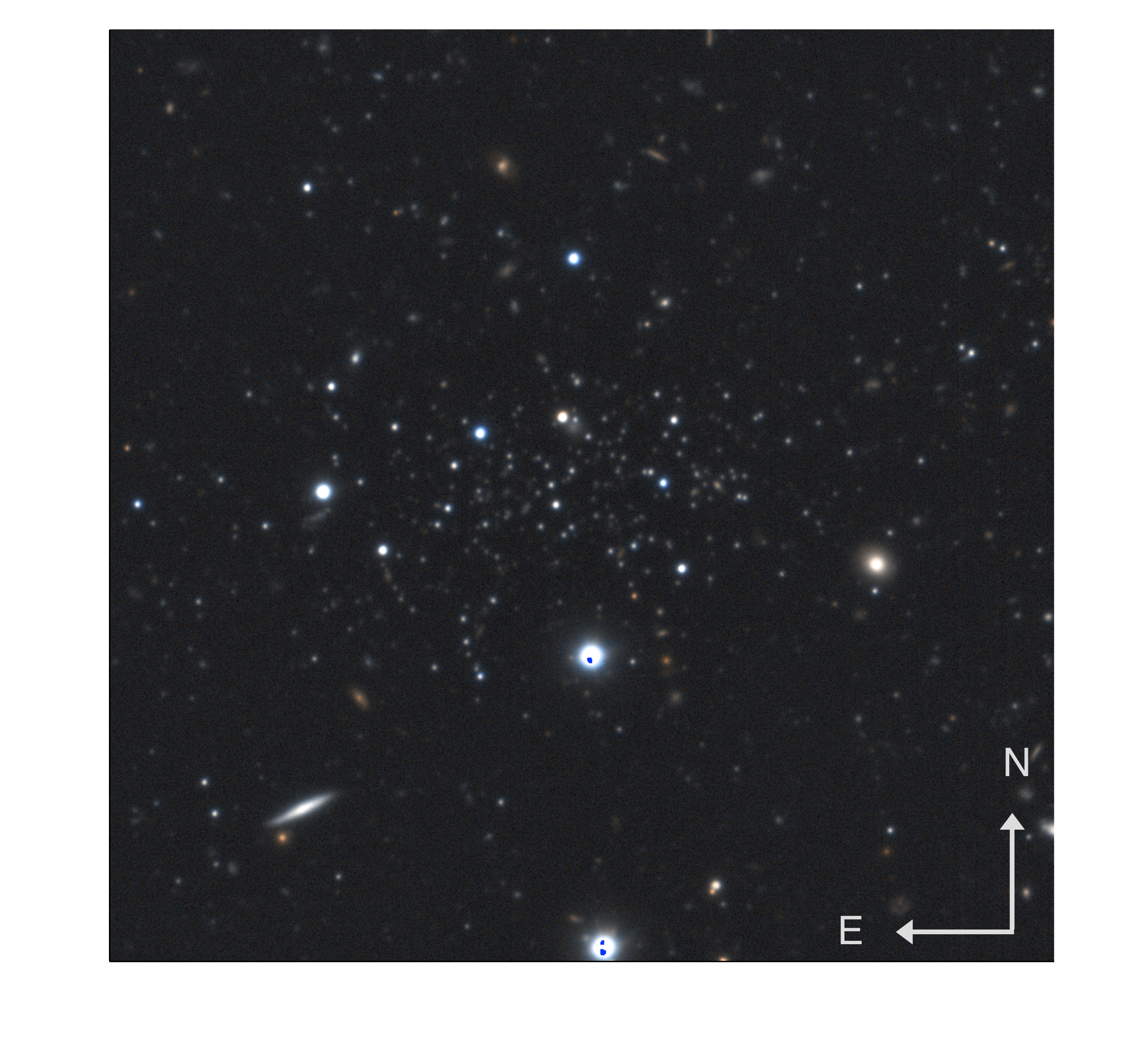}
\caption{As per Figure~\ref{fig:DES1}, false colour RGB image of Eridanus\,III with a $\sim$2$'$$\times$2$'$ field of view. The $g$-band co-added image was used for the Blue and the $r$-band co-added image for the Red. Eridanus\,III is clearly visible as an overdensity of stars in the centre of this field. The arrows in the lower right corner have a length of 15 arcsec.}
\label{fig:EriIII}
\end{center}
\end{figure}

\begin{figure}
\begin{center}
\includegraphics[width=1.0\hsize]{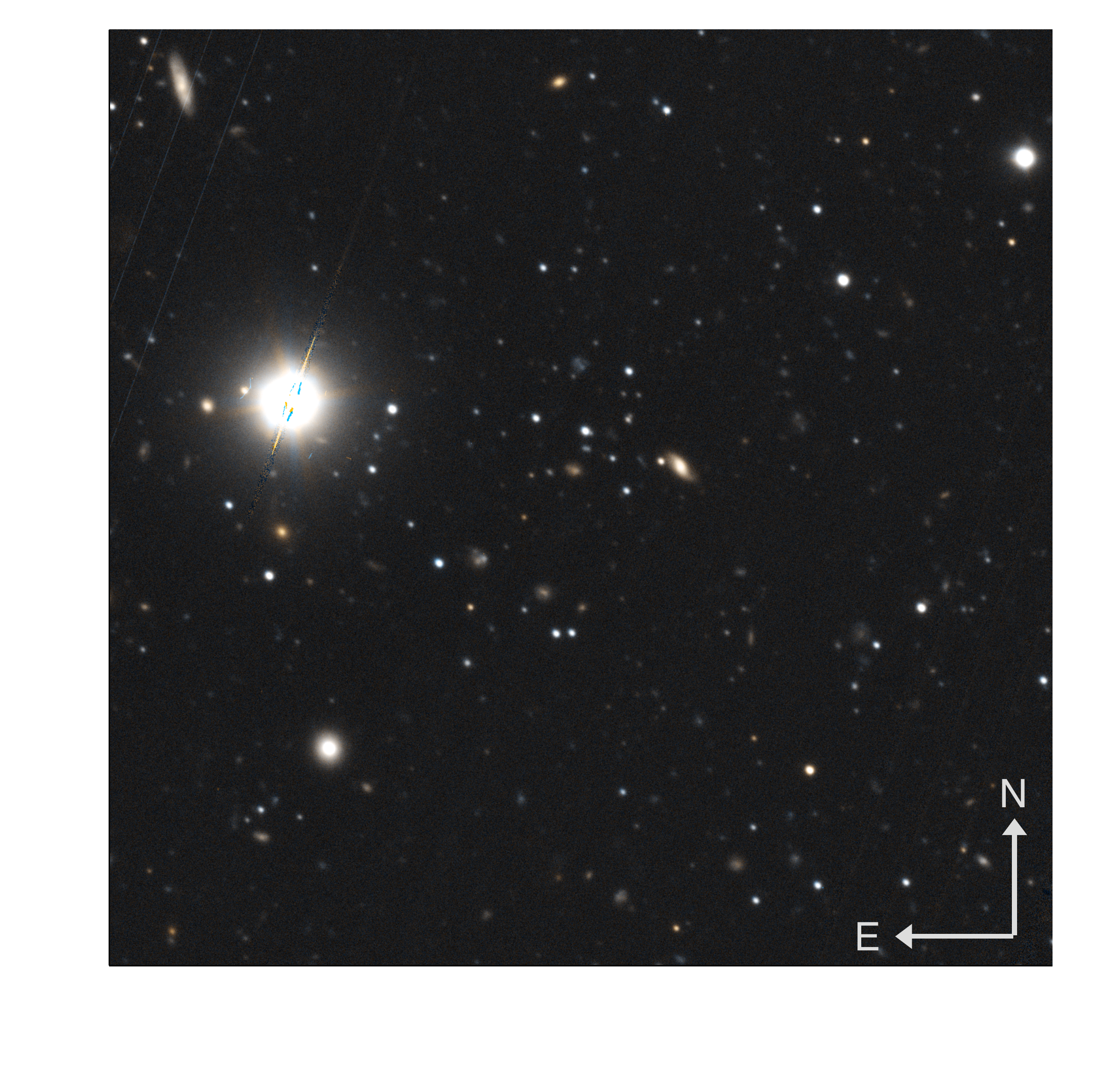}
\caption{As per Figure~\ref{fig:DES1}, false colour RGB image of Tucana V with a $\sim$2$'$$\times$2$'$ field of view. The $g$-band co-added image was used for the Blue and the $r$-band co-added image for the Red. There is a small concentration of brighter stars noticeable in the centre of the field, but there are no obvious fainter stars associated to this group. The arrows in the lower right corner have a length of 15 arcsec.}
\label{fig:TucV}
\end{center}
\end{figure}

\begin{table*}
\caption{Observing Log}
\centering
\begin{tabular}{lrccccccc}
\hline
Field & Right Ascension & Declination & Position Angle & Filter & Observation& Airmass & Exposure& Seeing\\
 &  (deg, J2000) & (deg, J2000) & (deg) & & Date & & (sec) & (\arcsec)\\
 \hline\hline
DES 1  & 8.4987   &       $-$49.0389    &    180 &   g\_G0325   &      2016-08-30  &    1.058 - 1.056 &  600   &  0.55     \\
            &     &         &    180 &   r\_G0326   &      2016-08-30  &    1.058 - 1.067 &  600   &  0.52     \\\hline

Eridanus III   & 35.6897  &       $-$52.2837    &    180 &   g\_G0325   &      2016-08-30  &    1.081 - 1.094 &  600  &   0.60    \\
(DES J0222.7-5217)                     &   &       &    180 &   r\_G0326   &      2016-08-30,-31  &    1.078 - 1.101 &  600   &  0.43     \\\hline
  Tucana V      &  354.3500   &       $-$63.2700    &    160 &   g\_G0325   &      2016-09-27  &    1.259 - 1.297 &  520   &   0.51      \\
(DES J2337-6316)      &   &        &    160 &   r\_G0326   &      2016-09-27  &    1.316 - 1.356 &  520  &  0.51    \\\hline
\hline
\end{tabular}
\end{table*}\label{table:data}

The imaging data were obtained with the Gemini Multi-Object Spectrograph South (GMOS-S) at the 8m diameter Gemini South Telescope through Program ID: GS-2016B-Q-7. The observing conditions, following the Gemini Observatory standards, were dark, clear skies (SB50\footnote{SB50 - Sky Brightness 50$^{th}$ percentile}/CC50\footnote{CC50 - Cloud Cover 50$^{th}$ percentile}) and seeing typically better than 0.6 arcsecond (IQ20\footnote{IQ20 - Image Quality 20$^{th}$ percentile}) on the nights of August 30 and 31, September 27, 2016 (see Table~\ref{table:data}). By taking advantage of the excellent seeing ($0\farcs4-0\farcs6$) we were able to utilize the $1\times1$ binning mode of GMOS-S and achieve a pixel scale of $0\farcs08$ per pixel. The field of view is $5\farcm5\times 5\farcm5$ and each object was observed in the $g$ (g\_G0325) and $r$ (r\_G0326) bands with a short 60s exposure centred on the target followed by three dithered exposures of 600s each. Figures~\ref{fig:DES1}, \ref{fig:EriIII} \& \ref{fig:TucV} present the false-colour images of the co-added frames for DES1, Eri\,III and Tuc\,V respectively.

The basic data reduction steps of generating master biases and master twilight flats, bias subtraction and flat fielding, astrometry and co-addition have been performed using the {\sc theli} pipeline \citep{2013ApJS..209...21S}. Point Spread Function (PSF) photometry has been undertaken on the co-added files using {\sc dolphot} \citep{2000PASP..112.1383D}. {\sc dolphot} parameters have been adjusted to minimise the residuals by adjusting the PSF solution. In particular, we have employed the sum of a Lorentzian and a circular Gaussian model PSF to achieve better residuals. 

\subsection{Photometric Calibration}\label{sec:photcalib}

\begin{table}
\caption{Photometric calibration results \label{table:calib}}
\centering
\begin{tabular}{lcc}
\hline
 & $g$ band & $r$ band \\ \hline\hline 
Colour term $(g -r)$ & $+0.026^{+0.045}_{-0.046}$& $-0.059^{+0.042}_{-0.041}$\\ 
DES1 Offset& $-3.213^{+0.051}_{-0.052}$ & $-2.979^{+0.048}_{-0.049}$ \\
Eridanus\,III Offset& $-3.237^{+0.034}_{-0.034}$ &  $-2.696^{+0.028}_{-0.029}$  \\
Tucana\,V Offset& $-3.162^{+0.034}_{-0.034}$& $-2.798^{+0.030}_{-0.031}$ \\
\hline
\end{tabular}
\tablecomments{Colour terms and magnitude offsets derived from comparison with APASS calibrated DES photometry. All photometry assumed an instrumental zero point of 30.00 for both filters prior to the offsets being applied. The offset values listed are a combination of the true zeropoint correction and the atmospheric extinction correction.}
\end{table}

The photometry generated by {\sc dolphot} was crossmatched with APASS\footnote{The AAVSO Photometric All-Sky Survey}~\citep{2015AAS...22533616H} calibrated DECam photometry\footnote{DECam photometry generated using the procedures  outlined in \citet{KimJerjen2015b}.} using the built-in routines of {\sc topcat} \citep{2005ASPC..347...29T} and then quality cuts were applied to the matched stars. These cuts first removed objects with extremely large photometric errors followed by cuts on the colour error ($\sigma_{(g-r)} < 0.3$), the range in colour ($0.0<g-r<+1.5$), object sharpness in both filters (sharpness$^{2} < 0.1$), object type (Objtype = 1) and photometry quality flag (class = 0). The resulting subset was fit with a linear function to determine the colour term, zero point offset and atmospheric extinction correction for calibration. Since the colour term is related to the physical differences between the GMOS-S and APASS filter sets, all objects will therefore require the same colour term in the calibration. A nominal instrumental zero point of 30.00 was chosen for all fields and filters then a correction is applied that calibrates the data with APASS. This offset encapsulates both the true photometric zero point and the atmospheric extinction correction, since they cannot be separated in this dataset. Due to the relatively small number of APASS stars available in each field to determine the photometric calibration of the data, we utilized the python package {\sc emcee} \citep{2013PASP..125..306F} to perform a Markov Chain Monte Carlo analysis of all the calibration fits simultaneously. This leveraged all of the data to establish the best colour term while allowing the remaining zeropoint and atmospheric extinction correction to stay unique to each field. Table~\ref{table:calib} lists the colour terms and offsets, with their corresponding errors, used to calibrate the data.

\subsection{Catalogue Generation}\label{sec:catalogue}
The raw instrumental magnitudes from {\sc dolphot} were corrected using the results from $\S$\ref{sec:photcalib} creating a catalogue of photometrically calibrated objects. These form the basis for the analysis presented in this paper. The criteria for selecting stellar objects from this catalogue is less stringent than that used in the calibration step and consisted of finding objects where:
\begin{itemize} 
	\item in either filter, sharpness$^{2} \leq 0.1$
	\item in both filters, signal-to-noise ratio $\geq3.5$
    \item and the object type corresponds to "good stars" (Objtype = 1). 
\end{itemize}
Spurious or saturated objects were again identified and removed based on either their extremely large magnitude errors or zero magnitude error respectively.

\subsection{Artificial Star Experiments}
\begin{figure}
	\centering
	\includegraphics[width=1.0\hsize]{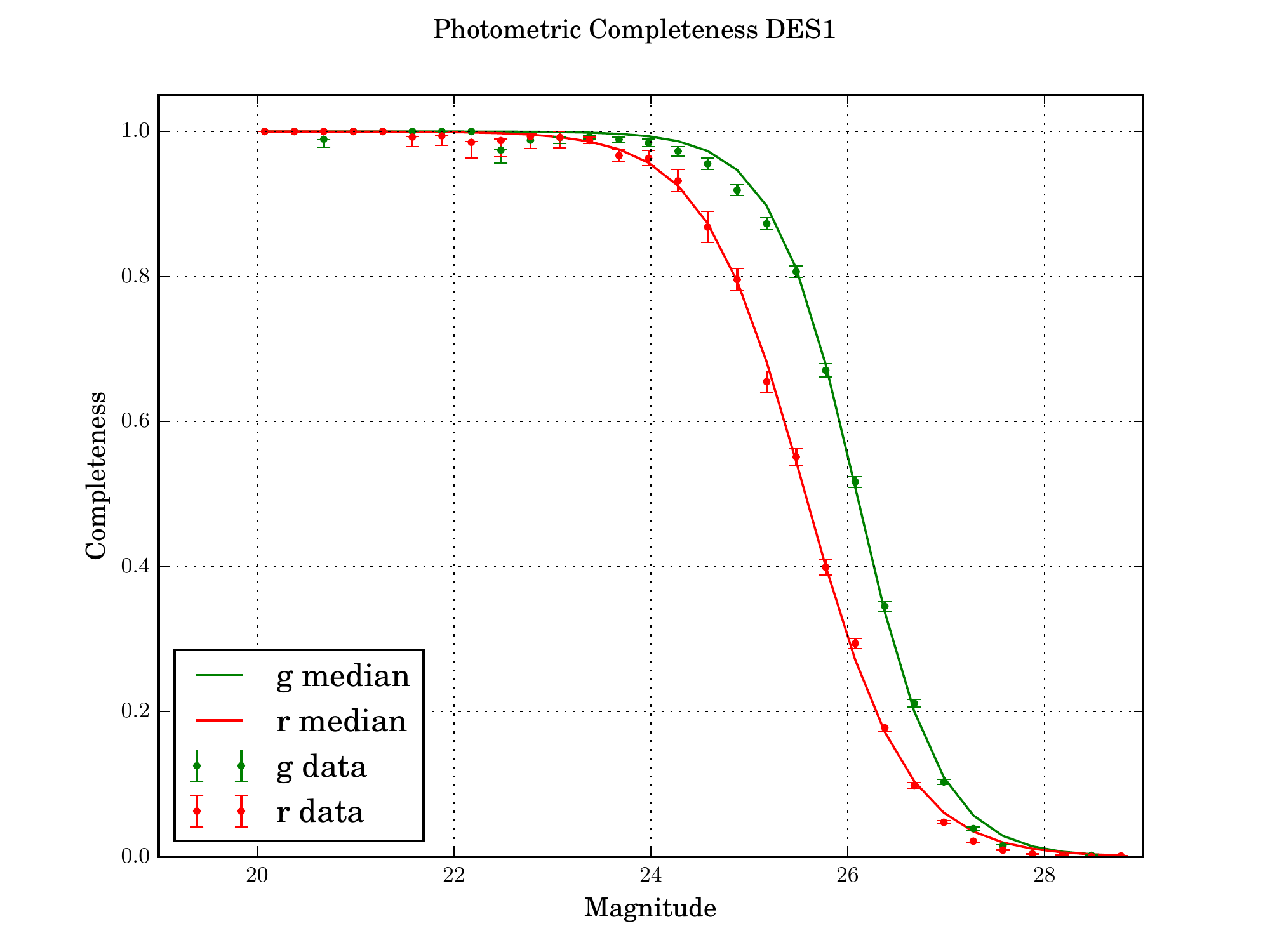}
	\includegraphics[width=1.0\hsize]{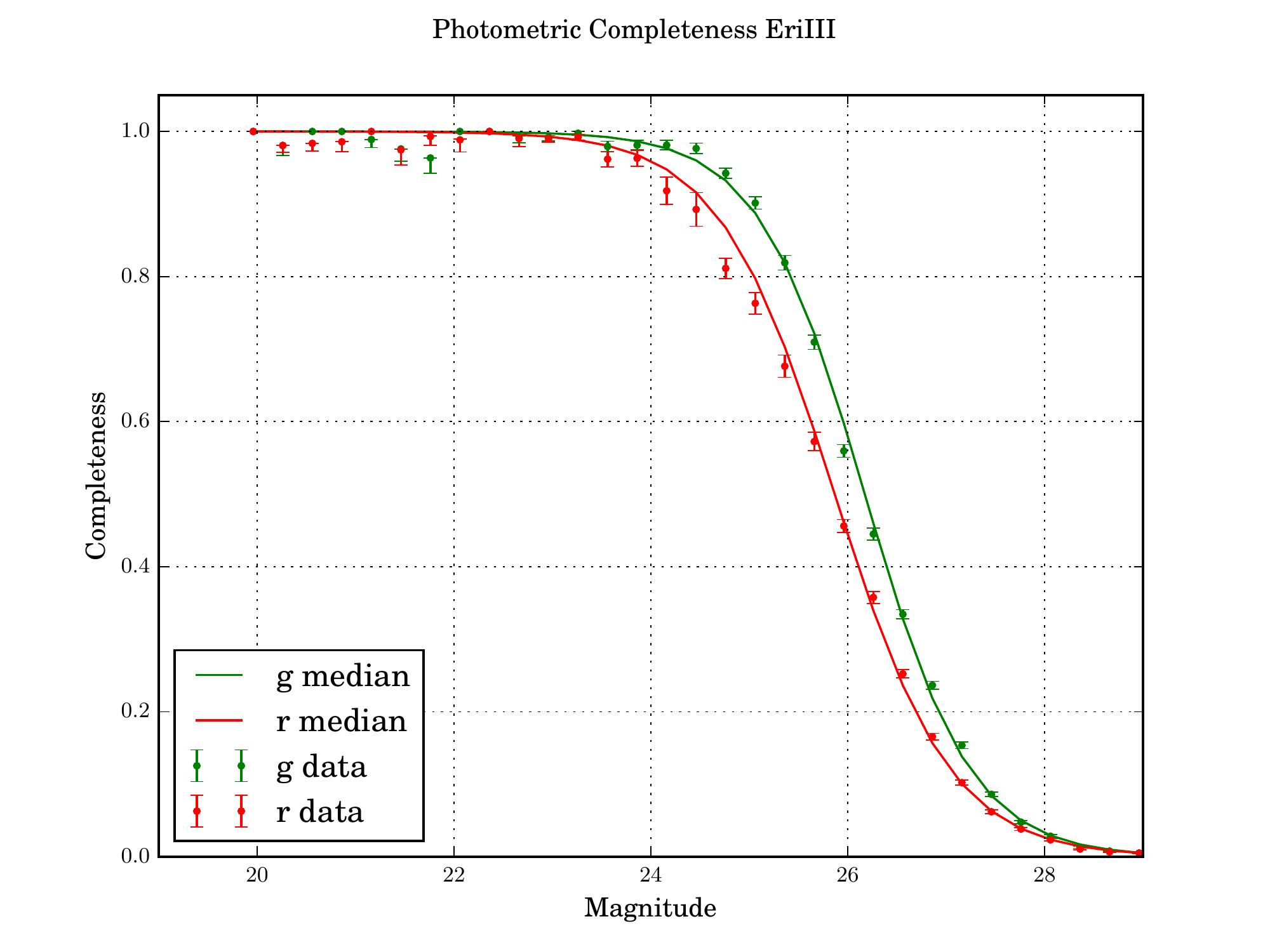}
    \includegraphics[width=1.0\hsize]{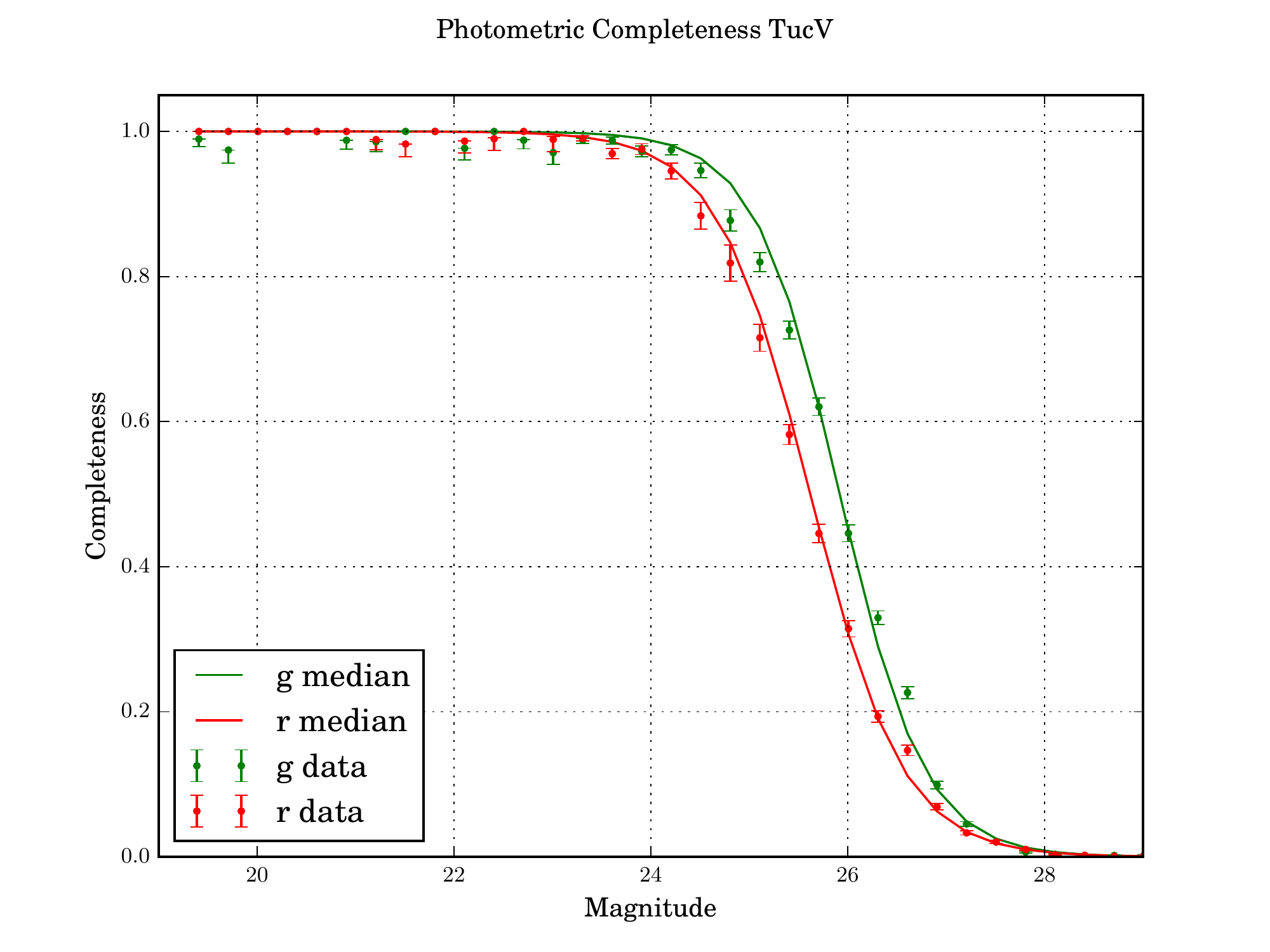}
	\caption{\label{fig:completeness} Recovery rate for artificial stars in the field of DES1 (top panel), Eri\,III (middle panel) and Tuc\,V (bottom panel). 
	The red and green points show the photometric completeness per 0.3 magnitude bin, for the $r-$ and $g-$bands respectively, while the solid lines show best chi-square fit of the Logistic function (Eqn.~\ref{eqn:logistic}).}
\end{figure}

To determine the completeness of our photometry, we have performed an artificial star experiment in each field using {\sc dolphot}'s built-in routine. The input catalogue of artificial stars was generated by taking the cumulative histogram of the magnitudes, from the data, at approximately 0.3 magnitude intervals. A base level of around 70 stars per magnitude bin was added to ensure that the brighter magnitudes were sufficiently populated and the colour of each star was randomly selected between $-1.3<(g-r)<+2.0$. This approach allows the artificial star distribution to mimic the actual stellar distribution in the data and forces each subsequent magnitude bin to have more artificial stars than the previous bin. Therefore, as the intrinsic photometric completeness of the data drops with fainter magnitudes, the number of artificial stars injected increases to compensate for the expected decrease in recovered stars. This helps ensure that at the faint end of the photometry we have confidence that the ratio of recovered stars to injected stars is robust. For DES1 and Eri\,III, a single star at the bright end of the photometry contributes about $\pm$1.4\% to the resultant completeness, while at the faint end a single star contributes only $\pm$0.015\%. A single star at the faint end of the Tuc\,V photometry contributes $\pm$0.026\% due to the fewer number of artificial stars used when compared to the other two fields. This list is then supplied to {\sc dolphot} along with the pixel position of each artificial star and these are added individually into the frame to avoid potential crowding issues. {\sc dolphot} then determines whether it can recover the star or not and Figure~\ref{fig:completeness} shows the number ratio of recovered stars to input stars. The artificial stars are subjected to the same selection criteria used when selecting real stars from the data as outlined in $\S$\ref{sec:catalogue}. To calculate the 50\% completeness level, the data in Figure~\ref{fig:completeness} has been fit with a Logistic function\footnote{The Logistic function was developed in 1838 by Pierre-Fran\c{c}ois Verhulst of Ghent, Belgium, see \citet{Bacaer2011} for a short history on the topic.}: 

\begin{eqnarray}\label{eqn:logistic}
Completeness = \frac{1}{1 + e^{(m - mc)/\lambda}}\\
Error = \sqrt{N C(1 - C)}/N
\end{eqnarray}

\begin{table}
\caption{50\,percent Photometric Completeness Estimates}\label{table:completeness}
\centering
\begin{tabular}{lccc}
\hline
Object & No.~of Artificial Stars & $mc_{g}$ & $mc_{r}$ \\\hline\hline
DES1 & 75,307 & $26.092^{+0.015}_{-0.014}$ & $25.569^{+0.017}_{-0.017}$ \\
Eridanus III & 75,392 &  26.156$^{+0.018}_{-0.018}$ & 25.839$^{+0.020}_{-0.020}$ \\
Tucana\,V & 42,462 & $25.923^{+0.023}_{-0.024}$& $25.592^{+0.025}_{-0.026}$\\
\hline
\end{tabular}
\end{table}

\noindent where $m$ is the magnitude, $mc$ is the 50\% completeness value and $\lambda$ is roughly the width of the rollover. For the error: $N$ is the number of artificial stars per bin and $C$ is the completeness in that bin. Table~\ref{table:completeness} lists the 50\% photometric completeness estimates for each field and the number of artificial stars used in the experiment and median solution from the Markov Chain Monte Carlo (MCMC) fitting routine {\sc emcee} \citep{2013PASP..125..306F}. 

We note that the recovery rate for each magnitude bin is derived from a large number of artificial stars distributed over the entire GMOS-S field. As such the level of completeness reflects any variation of the recovery rate across the field. For instance, a bright foreground star will inhibit the recovery of artificial stars in its vicinity. The error bars shown represent the statistical uncertainties in the recovery rate due to the chosen sample size in the artificial star experiment.

Given the known presence of a stellar overdensity in each field, we used the results of the artificial star test to generate a rough radial photometric completeness profile. Each field was sampled using 4 concentric annuli around an inner circle with a radius of 44 arcseconds. Each annuli has the same area as the inner circle and we found that there was no radial dependence in the photometric completeness of the data out to a radius of 98$"$ arcseconds. The variation in the 50\% completeness level between the annuli was of the order $\sim$0.06 magnitudes for DES1 and Eri\,III and $\sim$0.2 magnitudes for Tuc\,V, in both filters. The errors on the fit typically doubled, although Tuc\,V with fewer artificial stars, had a larger variation as expected. These results confirmed that crowding is not an issue in any of these fields and for this reason, the photometric completeness results quoted in Table~\ref{table:completeness} and used throughout this paper, were derived from the entire field as this ensured the smallest errors on the completeness estimates.

\subsection{Colour-Magnitude Diagrams}
\begin{figure*}
\begin{center} 
\includegraphics[width=0.32\hsize]{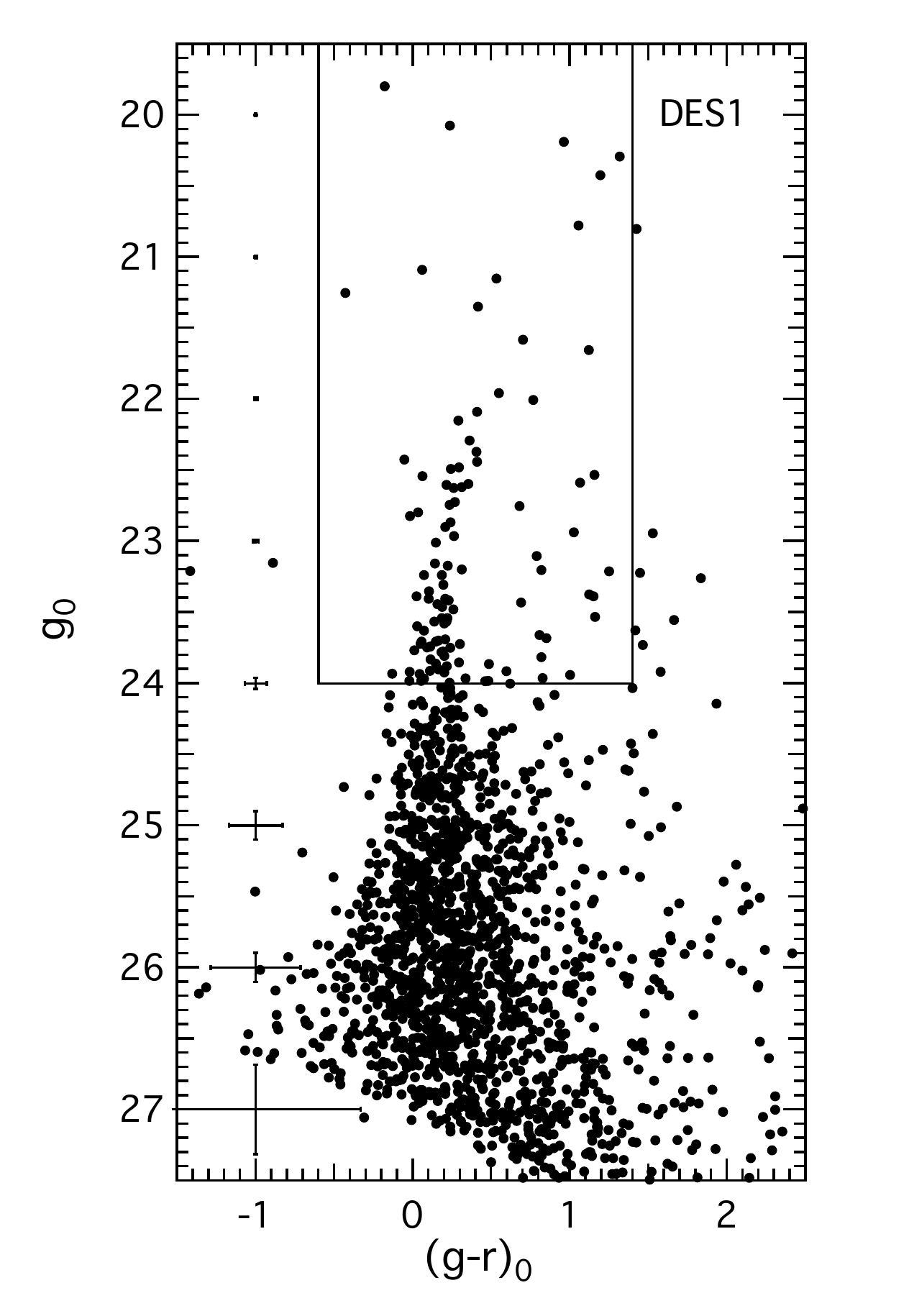}\hspace{-0.5cm}
\includegraphics[width=0.32\hsize]{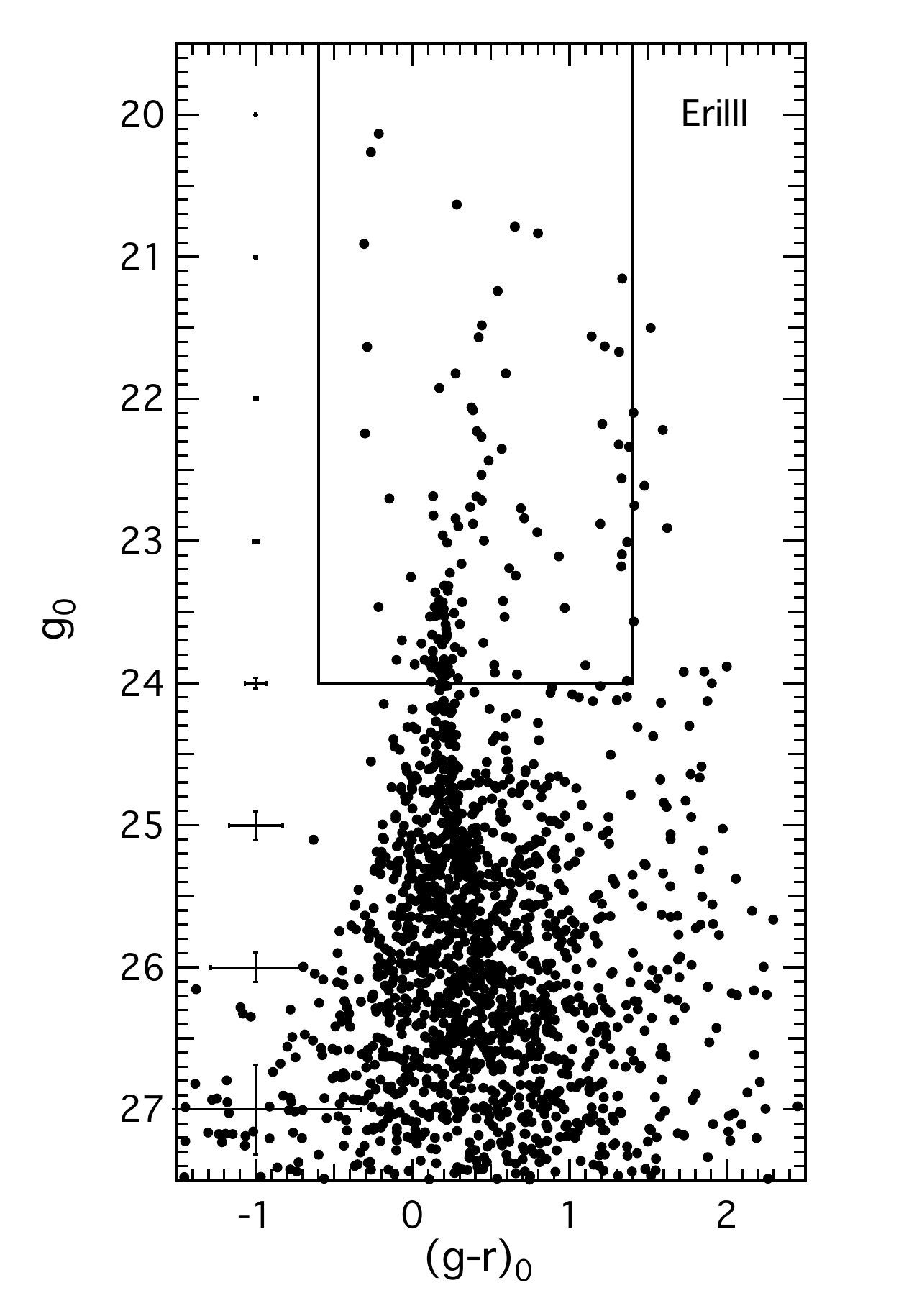}\hspace{-0.5cm}
\includegraphics[width=0.32\hsize]{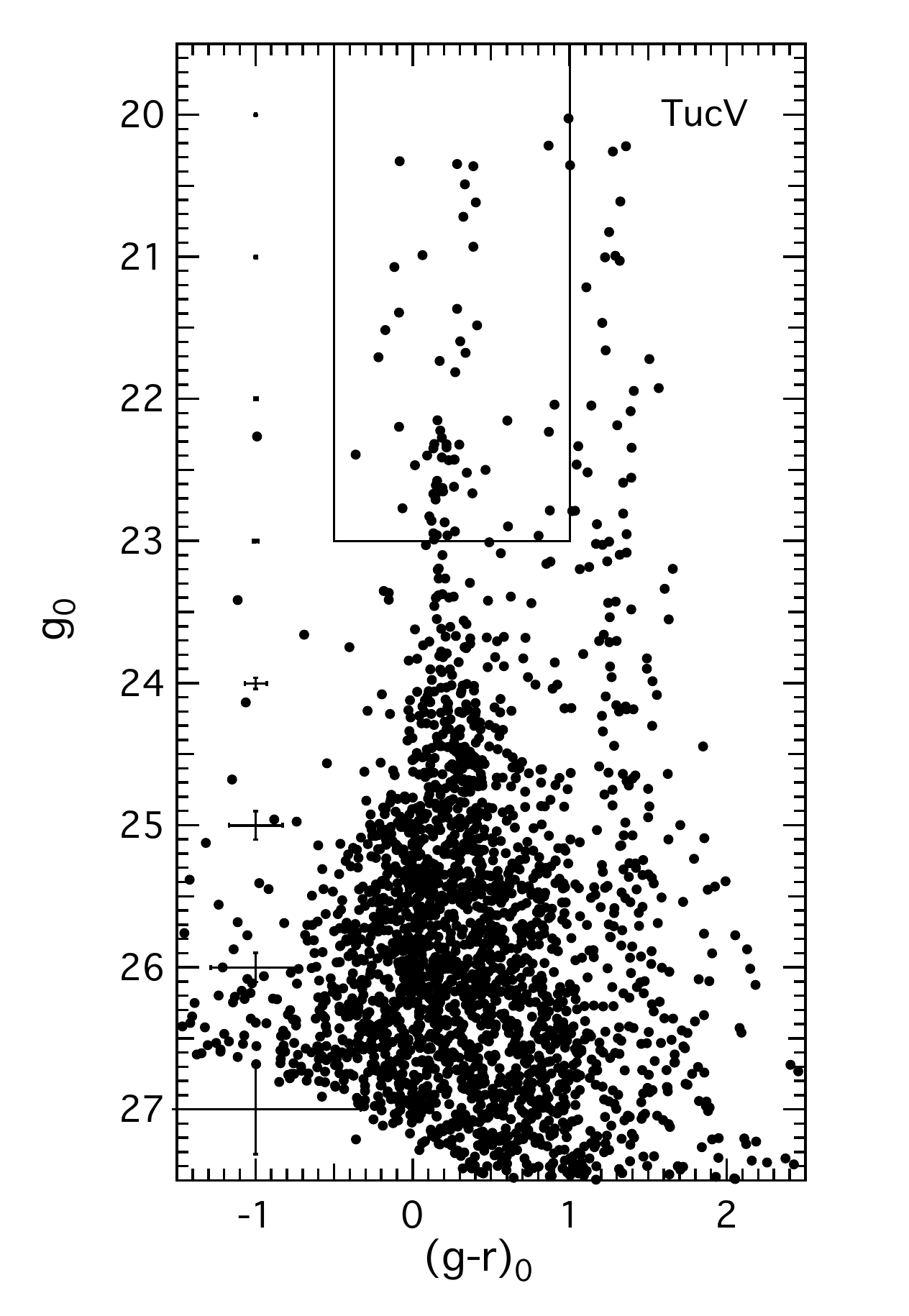}\hspace{-0.5cm}
\caption{The $g_\circ$ vs. $(g-r)_\circ$ colour-magnitude diagrams of stars in the $5\farcm5\times 5\farcm5$ GMOS-S field centred on the ultra-faint stellar systems DES1, Eridanus\,III and Tucana\,V. The rectangular boxes correspond to the colour-magnitude windows presented in the discovery papers \citep{Luque2016,Koposov2015, Drlica-Wagner2015}.  The error bars running vertically along the colour $(g-r)_\circ=-1$ in 1 mag intervals represent the typical photometric uncertainties.  \label{fig:cmd_field}}
\end{center}
\end{figure*}

The panels in Figure~\ref{fig:cmd_field} show the extinction-corrected $(g-r)_\circ$ vs. $g_\circ$ colour-magnitude diagrams (CMDs) of the entire GMOS-S fields using all objects classified as stars from our photometric analysis (see $\S$\ref{sec:catalogue}) that were found in the vicinity of each ultra-faint stellar system. The calibrated photometry was corrected for Galactic extinction based on the reddening map by \citet{SFD1998} and the correction coefficients from \citet{Schlafly2011}. The CMDs reveal stars $\sim4$ magnitudes fainter than the main-sequence turn-off (MSTO) and down to the 50\% completeness level $g_{lim}\sim 26$. The rectangular boxes correspond approximately to the colour-magnitude windows presented in the discovery papers: DES1 \citep*{Luque2016}, Eri\,III \citep{Koposov2015, Bechtol2015}, Tuc\,V \citep{Drlica-Wagner2015}.

\section{Parameter Analysis}\label{sec:param_analysis}
\begin{figure*}
\begin{center} 
\includegraphics[width=0.35\hsize]{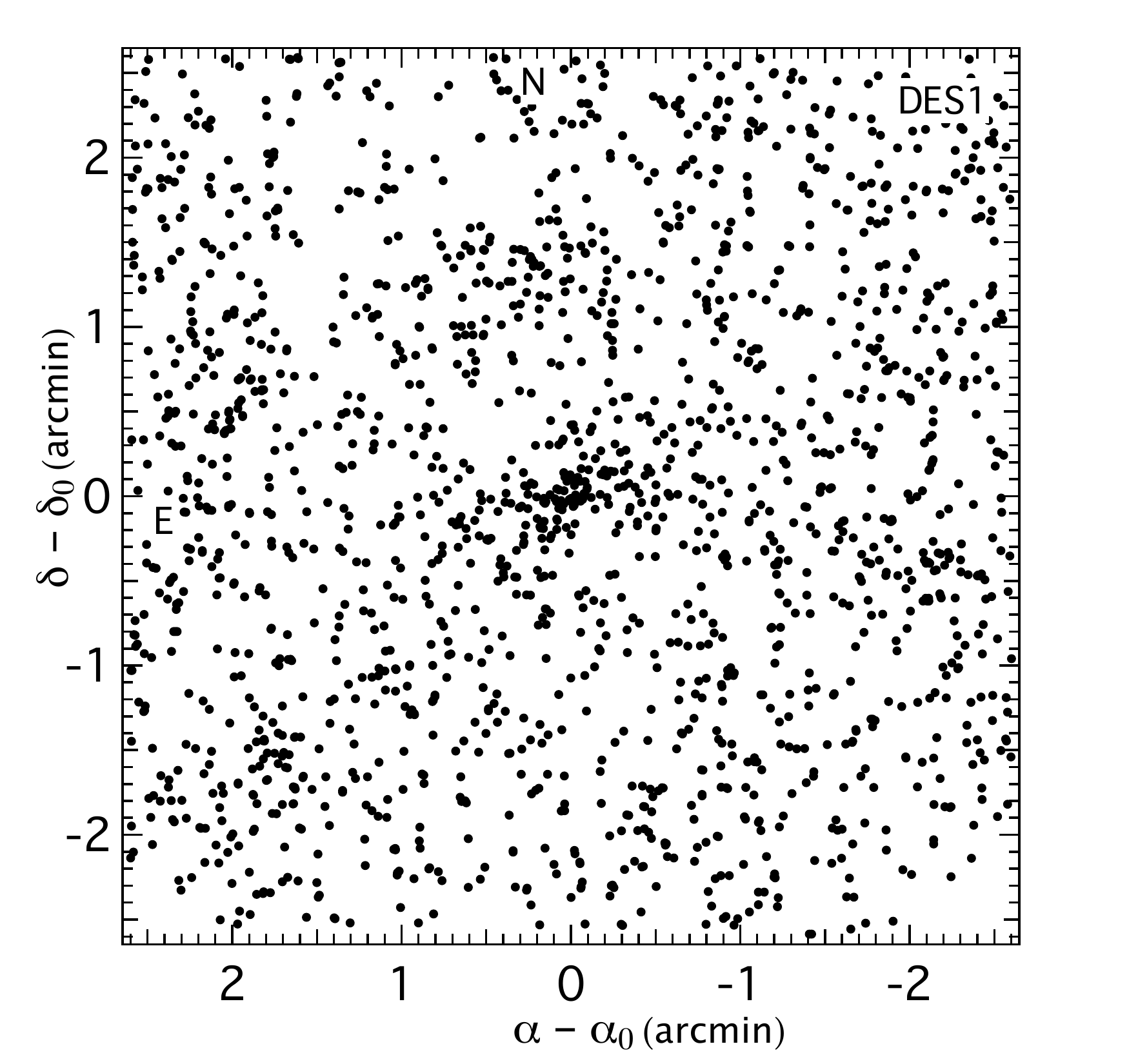}\hspace{-0.55cm}
\includegraphics[width=0.35\hsize]{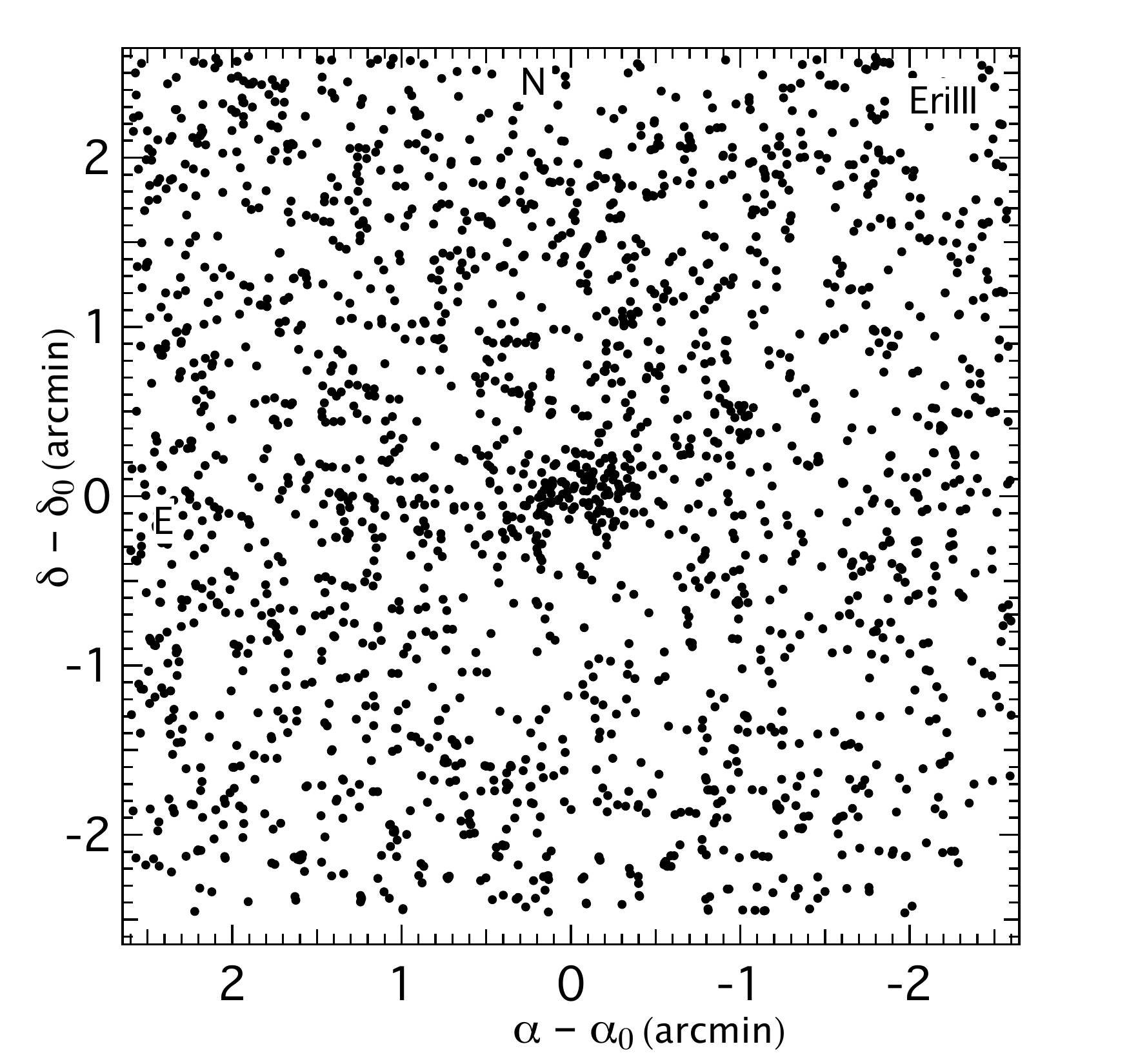}\hspace{-0.55cm}
\includegraphics[width=0.35\hsize]{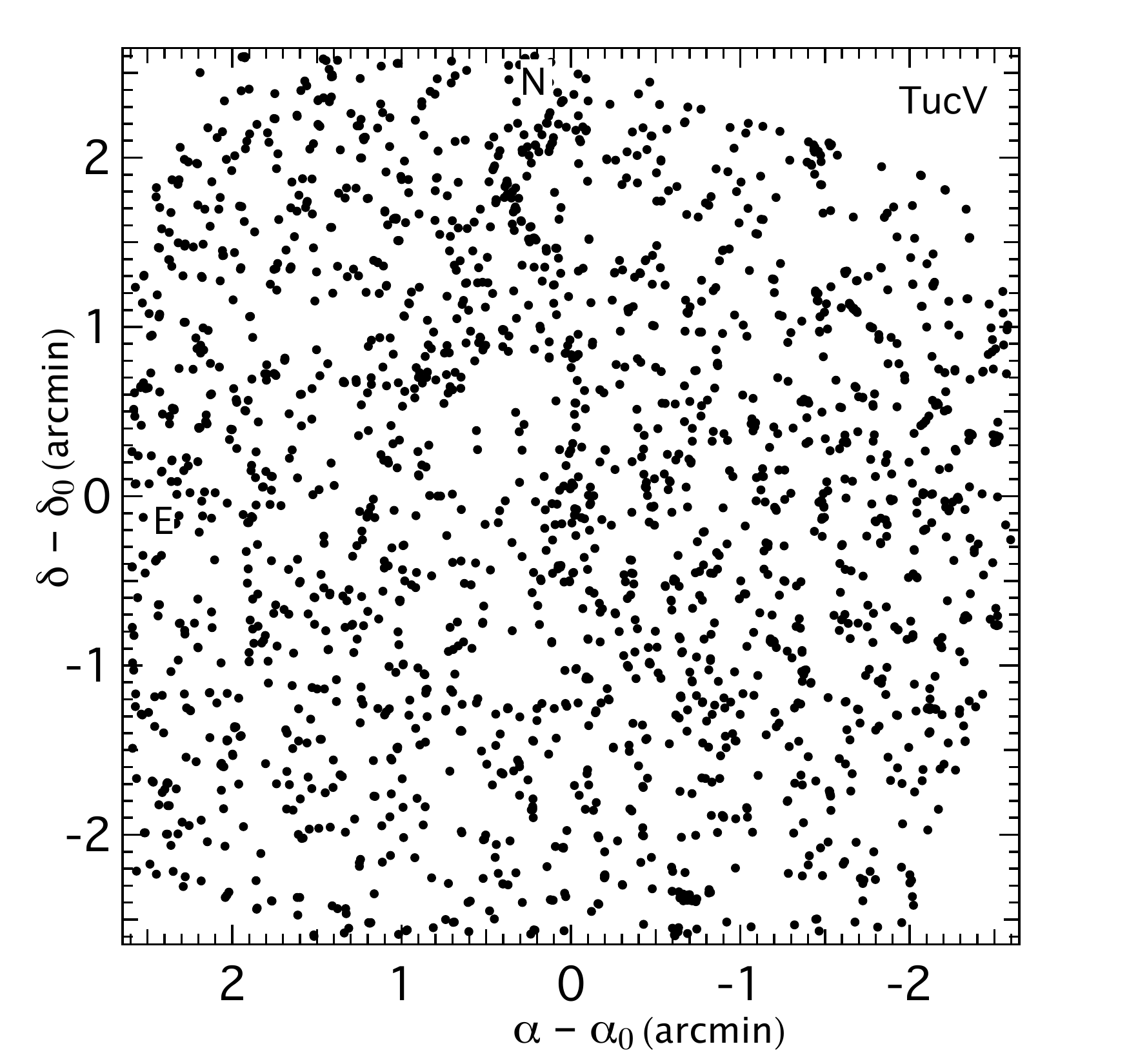}
\caption{Distribution of all objects classified as stars in the GMOS-S field centred on the ultra-faint stellar system DES1 (left), Eri\,III (centre) and Tuc\,V (right). In the cases of DES1 and Eri\,III, an elongated grouping of stars is clearly visible, while the Tuc\,V field shows no apparent stellar overdensity. \label{fig:stellar_distribution}}
\end{center}
\end{figure*}

For determining the fundamental properties of each ultra-faint stellar system: mean age, mean metallicity $\langle$[Fe/H]$\rangle$, the [$\alpha$/Fe]$_{avg}$ ratio, heliocentric distance ($D_\odot$), central coordinates ($\alpha_0,\delta_0$), position angle from north to east ($\theta$), ellipticity ($\epsilon=1-\frac{b}{a}$) and half-light radius ($r_h$) we employed an iterative process. The CMDs of the entire field for each object can be seen in Figure~\ref{fig:cmd_field}, with their on-sky distribution shown in Figure~\ref{fig:stellar_distribution}. First, we established the Dartmouth model isochrone \citep{Dartmouth} that best fits the CMD of the entire GMOS-S field (Figure~\ref{fig:cmd_field}) using the maximum likelihood (ML) method introduced in \cite{Frayn2002}. This method was used in our previous studies \citep{KimJerjen2015a, Kim2, Kim2016}. We calculated the maximum-likelihood values $\mathcal{L}_i$ over a grid of Dartmouth isochrones as defined by Equations\,1 and 2 in \citet{Fadely2011}. The grid points in the multi-dimensional parameter space cover ages from 7.0--13.5\,Gyr, a broad range of chemical composition $-2.5\leq$ [Fe/H] $\leq-0.5$\,dex, $-0.2\leq$ [$\alpha$/Fe] $\leq +0.6$\,dex, and a distance interval $(m-M)\pm 0.5$, where $(m-M)$ is the initial guess for the distance modulus for the object from the discovery papers.  Grid steps were 0.5\,Gyr, 0.1\,dex, 0.2\,dex, and 0.05\,mag, respectively. For each object, we present the matrix of likelihood values after interpolation and smoothing over two grid points.

The best fitting model isochrone was then used to identify stars that are sufficiently close to the stellar population of the object in colour-magnitude space. These stars were defined to have a $g_*$-band magnitude in the interval $19.5<g_\circ<27.0$ and a colour ($g_*- r_*$) that fulfils the requirement:
\begin{eqnarray}\label{eqn:iso}
\frac{1}{\sqrt{2\pi \sigma^2_{tot}}} \exp(-((g_*-r_*)-(g-r)_{iso})^2/2\sigma_{tot}^2) >0.5,
\end{eqnarray}
where $(g-r)_{iso}$ is the colour of the model isochrone at $g_*$ and $\sigma^2_{tot}=\sigma_{int}^2+\sigma^2_{g_*}+\sigma^2_{r_*}$. The quantity $\sigma_{int}=0.1$\,mag was chosen as the intrinsic colour width of the isochrone mask and $\sigma^2_{g_*} $, $\sigma^2_{r_*}$ are the photometric uncertainties of a star. 
Restricting the measurement of the parameters $\alpha_0,\delta_0, \theta, \epsilon, r_h$ on this sub-sample reduces the level of contamination in the R.A.-DEC distribution and thus significantly increases the number ratio between member stars of the stellar system and foreground. 

To determine the centre coordinates ($\alpha_0$, $\delta_0$) and structural parameters of the stellar system we employed the ML routine from \cite{Martin2008}, which was previously used by us e.g.~in \citet{Kim2016} based on the likely member stars, i.e.~stars that are within the isochrone mask. We used a 2-dimensional elliptical exponential profile plus foreground:
\begin{eqnarray}\label{eqn:elliptexp}
E(r,r_h,\Sigma_\circ,\Sigma_f)=\Sigma_\circ\exp(-1.68r/r_h)+\Sigma_f
\end{eqnarray}
to model the member star distribution on the sky where 
\begin{eqnarray}
 r = \left\{\left[\frac{1}{1-\epsilon}(x\cos\theta-y\sin\theta)\right]^2 + (x\sin\theta+y\cos\theta)^2\right\}^{1/2} \nonumber
\end{eqnarray}

\noindent is the elliptical (semi-major axis) radius and $(x,y)$ the spatial position of a star, $\epsilon$ the ellipticity of the distribution, $\theta$ the positional angle of the major axis, $r_h$ is the half-light radius, $\Sigma_\circ$ the central star density, and $\Sigma_f$ the foreground star density.
Based on the first estimates for these quantities we constructed a new CMD from stars that are within 
an ellipse with a semi-major axis length $a=3.9r_h$, semi-minor axis length $b=a(1-\epsilon)$ and position angle $\theta$ of the nominal centre of the stellar overdensity. Assuming an underlying exponential profile, this area contains 90\,percent of the total number of member stars and we refer to it as the 90\% ellipse in the following sections. We then re-calculated refined values for age, $\langle$[Fe/H]$\rangle$, [$\alpha$/Fe] \& $D_\odot$, and generated the associated isochrone mask to re-calculate $\alpha_0$, $\delta_0$, $\theta$, $\epsilon$ and $r_h$. This process of measuring the two sets of parameters typically converged to the final values after 2-3 iterations. 
Given the relative small number of bright stars in DES1 and Eri\,III around the MSTO we investigated the effect that individual stars have on the age, $\langle$[Fe/H]$\rangle$, [$\alpha$/Fe], and distance by running a Jackknife experiment: each star within 0.5\,mag above and below the MSTO was dropped once from the sample and the ML analysis repeated. The observed variations were well within the quoted uncertainties for each parameter confirming internal consistency of the results.
We finally calculated the number of stars $N_*$ that belong to the overdensity with Equation\,5 from~\cite{Martin2008} and
fitted a King profile \citep{1966AJ.....71...64K}:
\begin{eqnarray}\label{eqn:king}
E(r,r_c,r_t,\Sigma_\circ)=\Sigma_\circ\left(\frac{1}{\sqrt{1+r^2/r_c^2}} - \frac{1}{\sqrt{1+r_t^2/r_c^2}}\right)^2
\end{eqnarray}
at the stellar distribution using the final values for the centre coordinates, position angle and ellipticity. The $r_t$ and $r_c$ parameters are the tidal and core radii. All parameters derived in this section are summarized in Tables~\ref{tab:DES1parameters} and \ref{tab:EriIII_parameters}. We will discuss the results for DES1 in $\S$\ref{sec:DES1prop} and Eridanus\,III in $\S$\ref{sec:EriIIIprop}. The special case Tuc\,V is discussed in $\S$\ref{sec:TucVProp}.

\section{Properties of DES1}\label{sec:DES1prop}

\begin{figure}
\begin{center} 
\includegraphics[width=1\hsize]{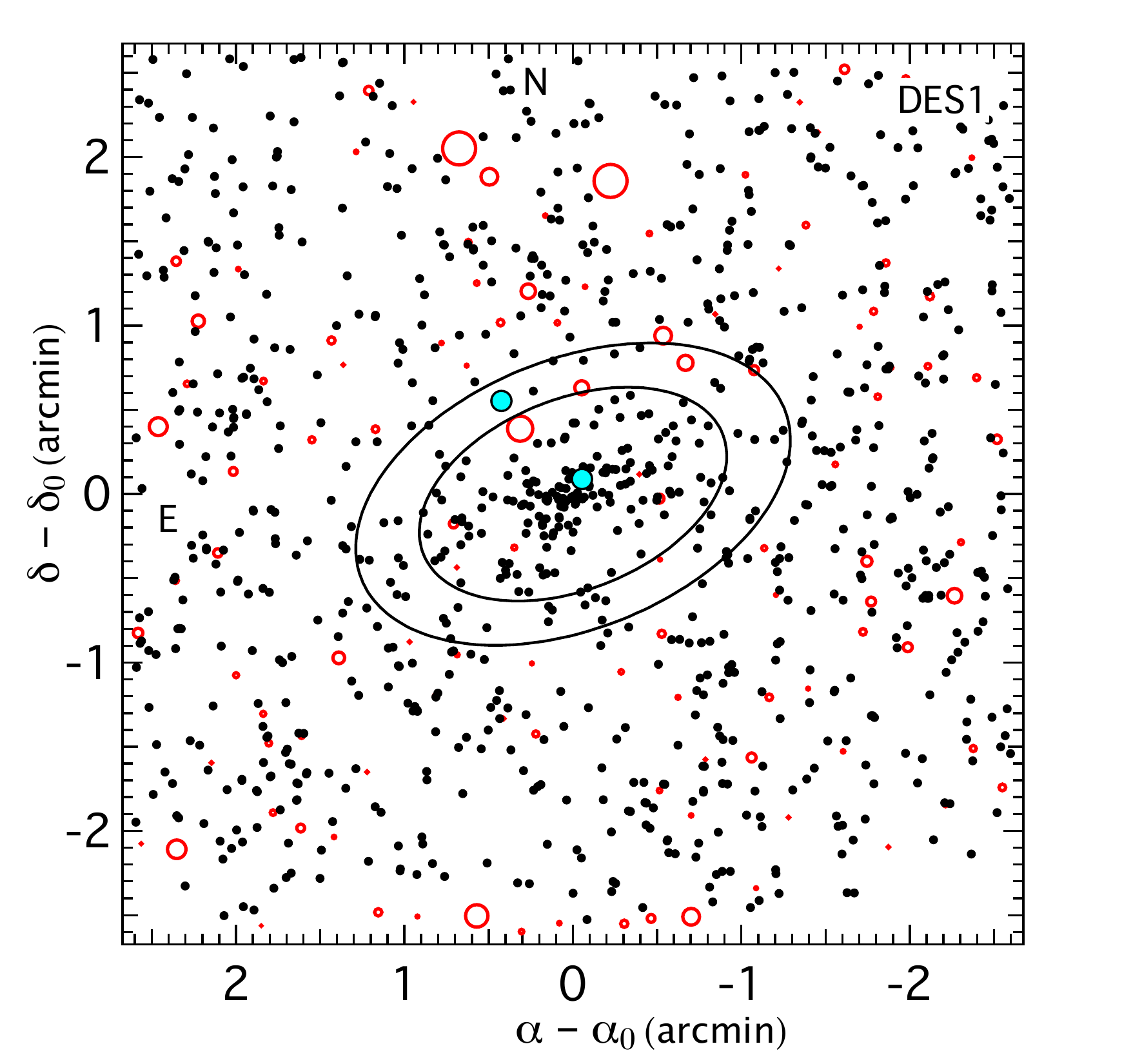}\\ 
\includegraphics[width=1\hsize]{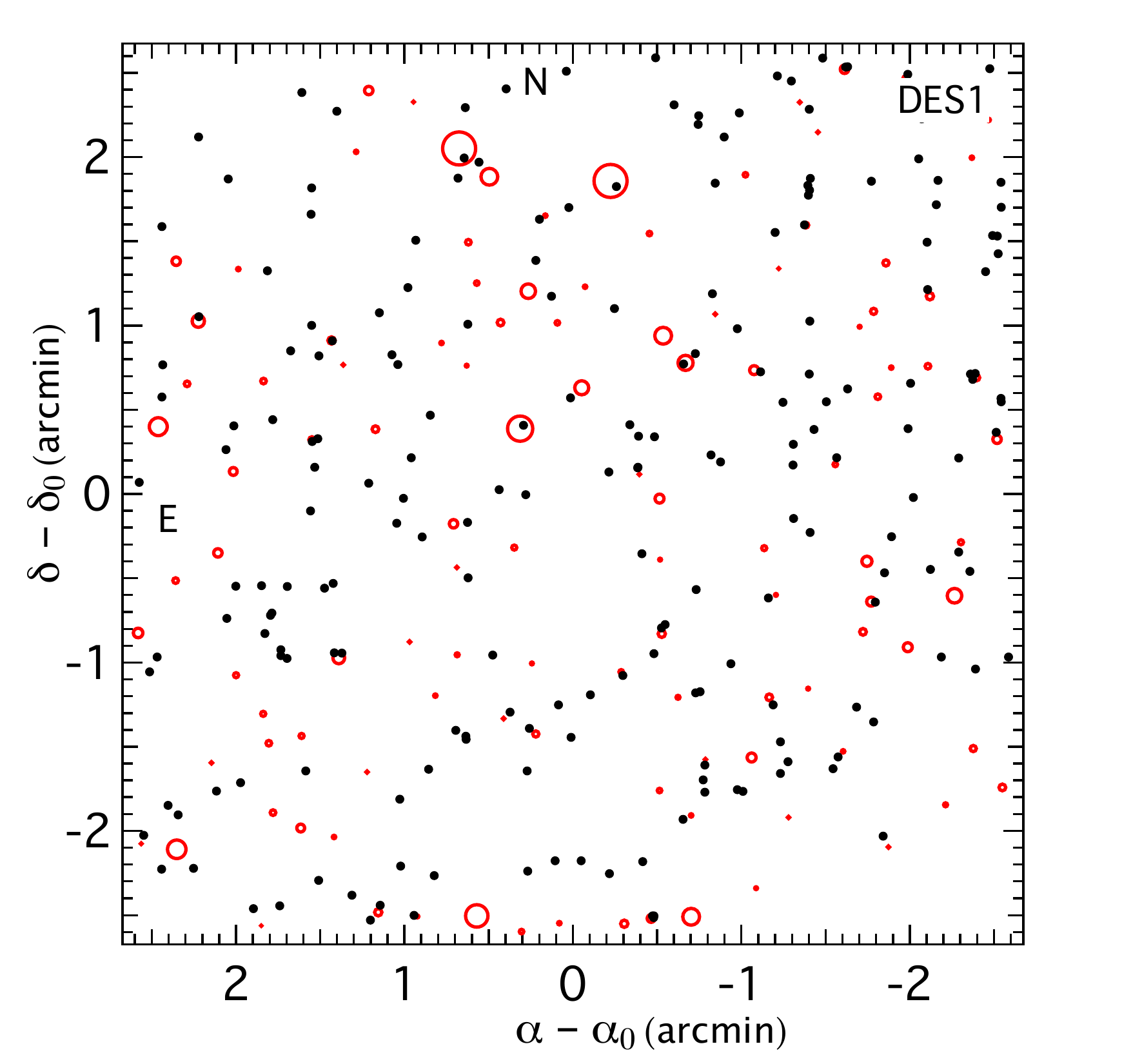}
\caption{{\it Top}: Selecting only stars that are sufficiently close to the best-fitting isochrone (for more details see $\S$\ref{sec:param_analysis}) significantly increases the contrast between stars associated to the overdensity and Galactic foreground. The elongated concentration of DES1 stars becomes more prominent and structural parameters can be determined with higher accuracy. The two ellipses have a position angle of $112^\circ$ and a semi-major axis length of $3.9r_h$ and $5.5 r_h$, respectively. The inner ellipse borders the region that contains 90 percent of the DES1 stellar population, assuming an exponential radial profile. The outer ellipse has twice the area of the inner ellipse. Two horizontal branch star candidates are highlighted in cyan. The open red circles are objects from the {\sc AllWISE} catalogue \citep{Wright2010}, scaled to reflect their magnitudes. These objects highlight the position of the bright objects in the field both foreground stars and background galaxies. {\it Bottom}: Distribution of non-stellar objects selected in the same manner as those in the top panel. {\sc AllWISE} objects are again plotted as open red circles.}
\label{fig:DES1stellar_distribution}
\end{center}
\end{figure}

\begin{figure}
\begin{center} 
\includegraphics[width=0.95\hsize]{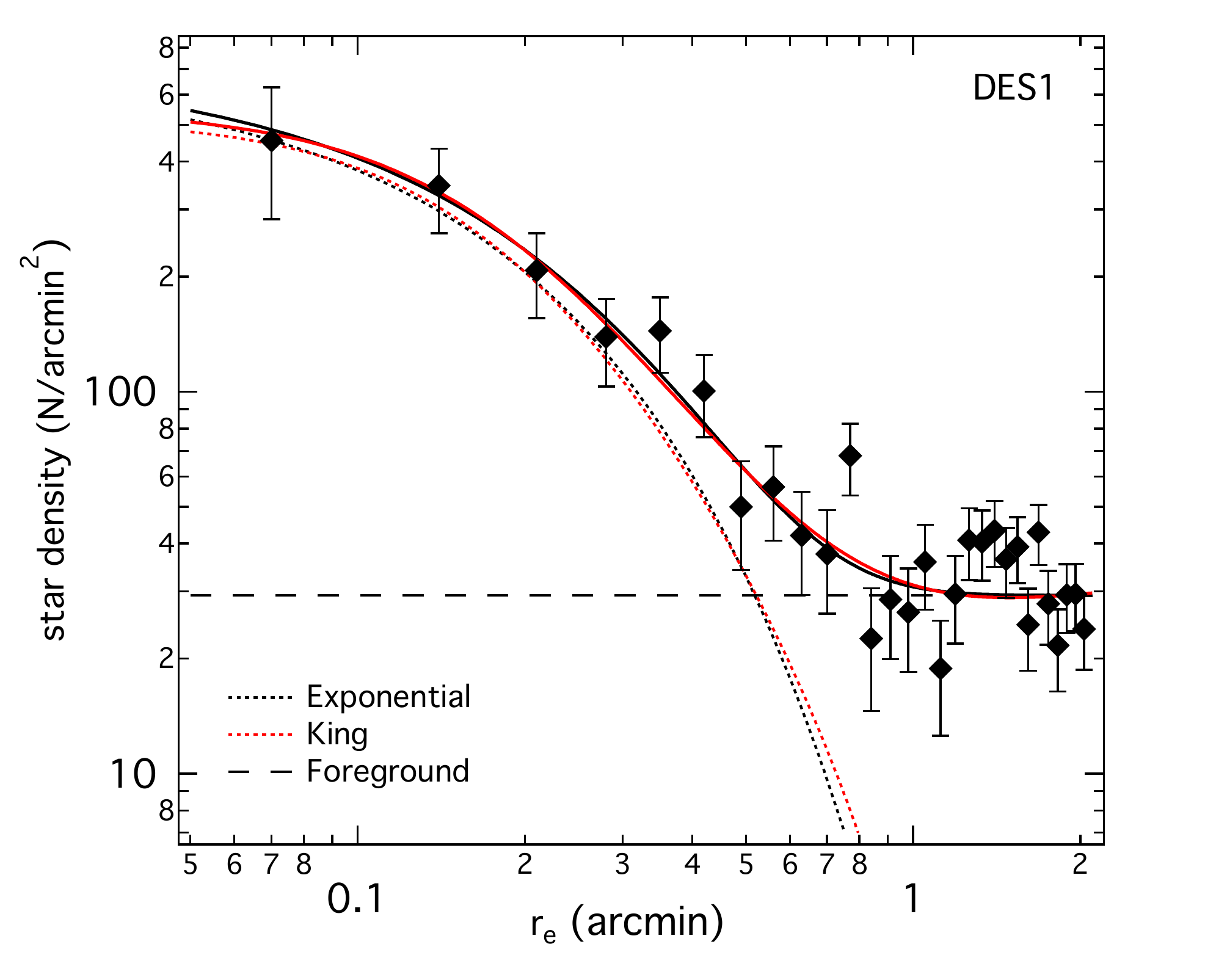}
\caption{ {\it Top Panel}: Radial density profile of DES1. The best-fitting Exponential (black dotted) and King (red dotted) profiles are superimposed on the data points. The horizontal dashed line is the density of the foreground stars. The solid black and red lines represent the profiles + foreground. The error bars were derived from Poisson statistics.}\label{fig:DES1rad_profile}
\end{center}
\end{figure}

\begin{figure*}
\begin{center} 
\includegraphics[width=0.3\hsize]{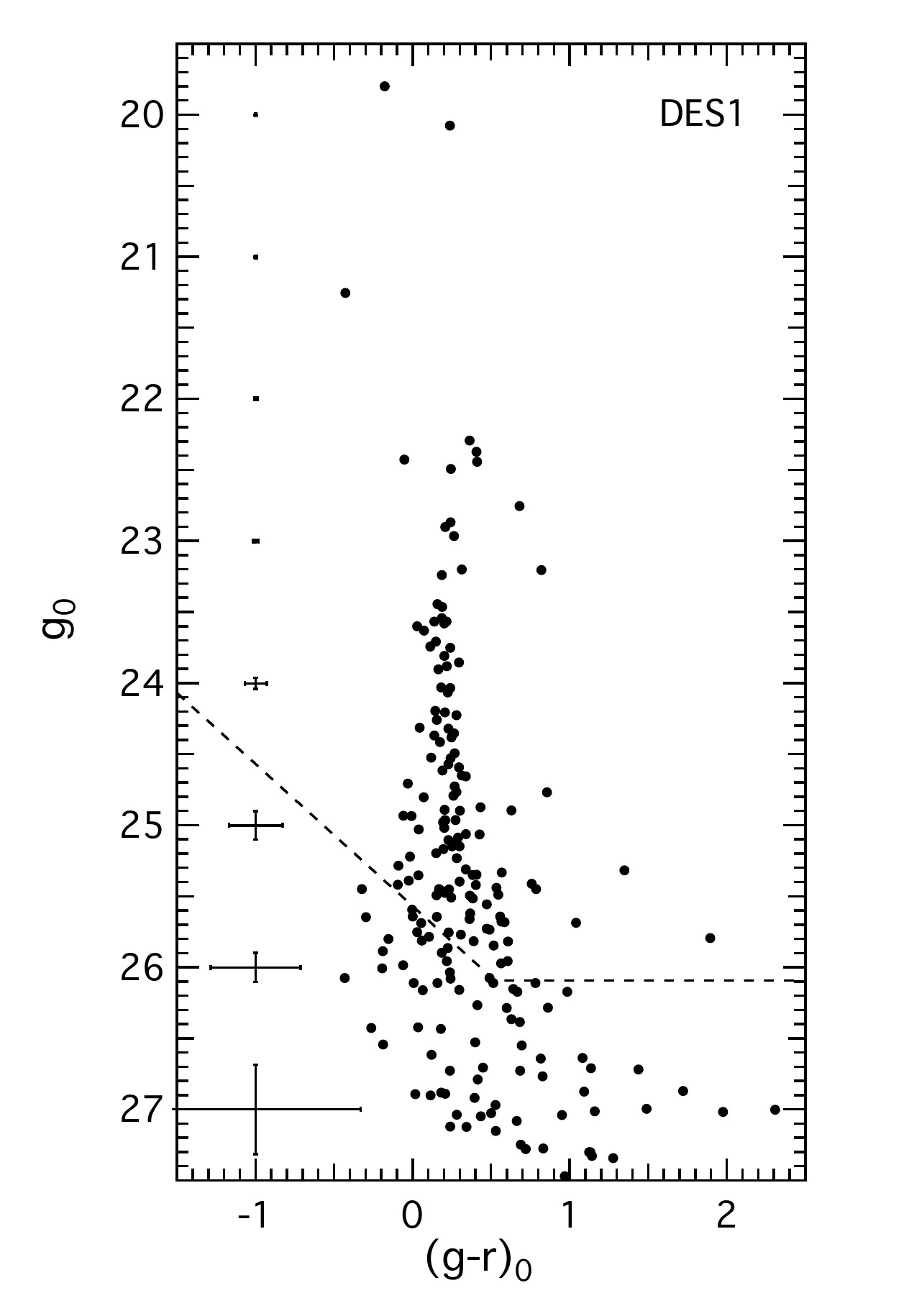}\hspace{-0.2cm}
\includegraphics[width=0.3\hsize] {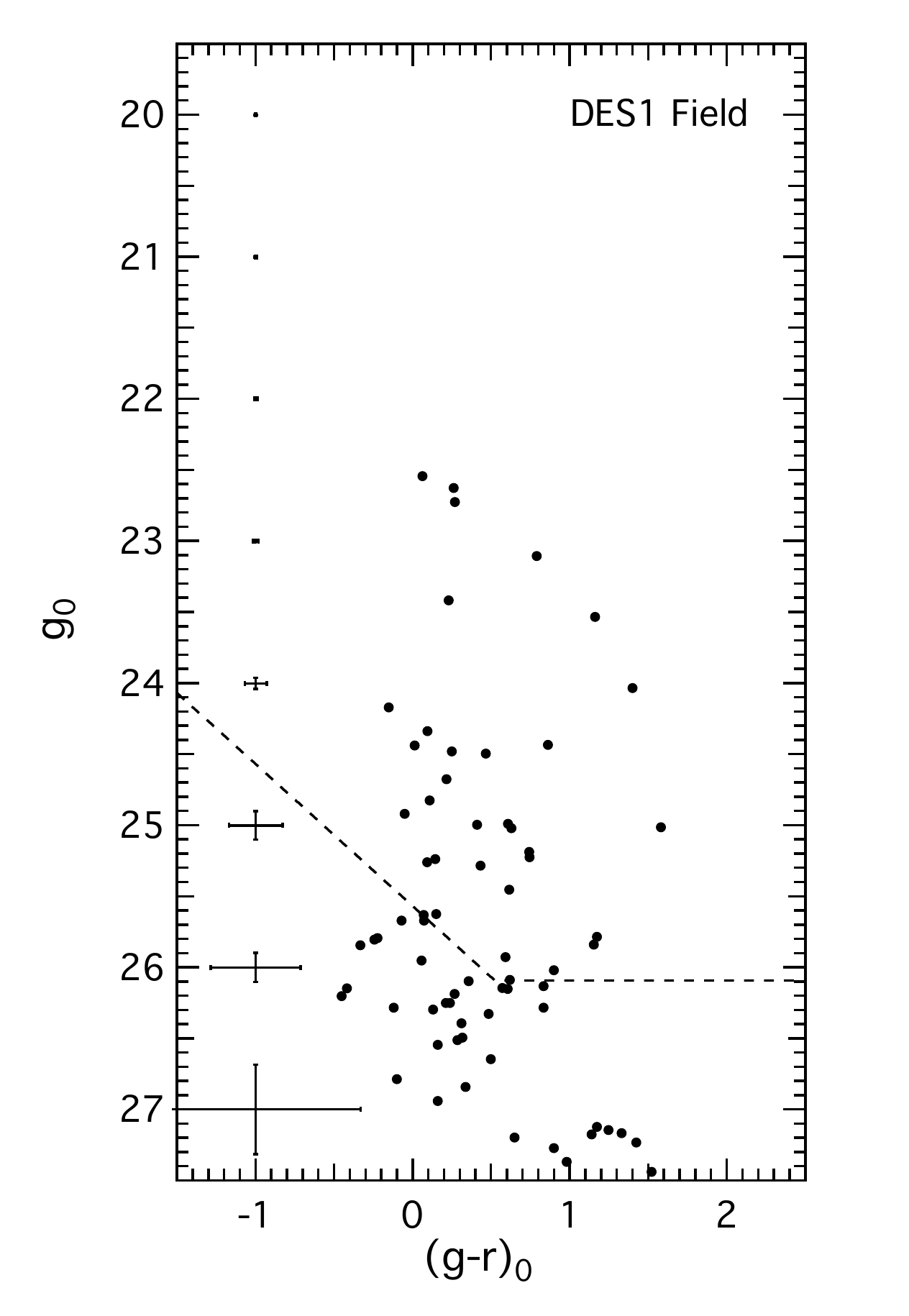}\hspace{-0.2cm}
\includegraphics[width=0.298\hsize]{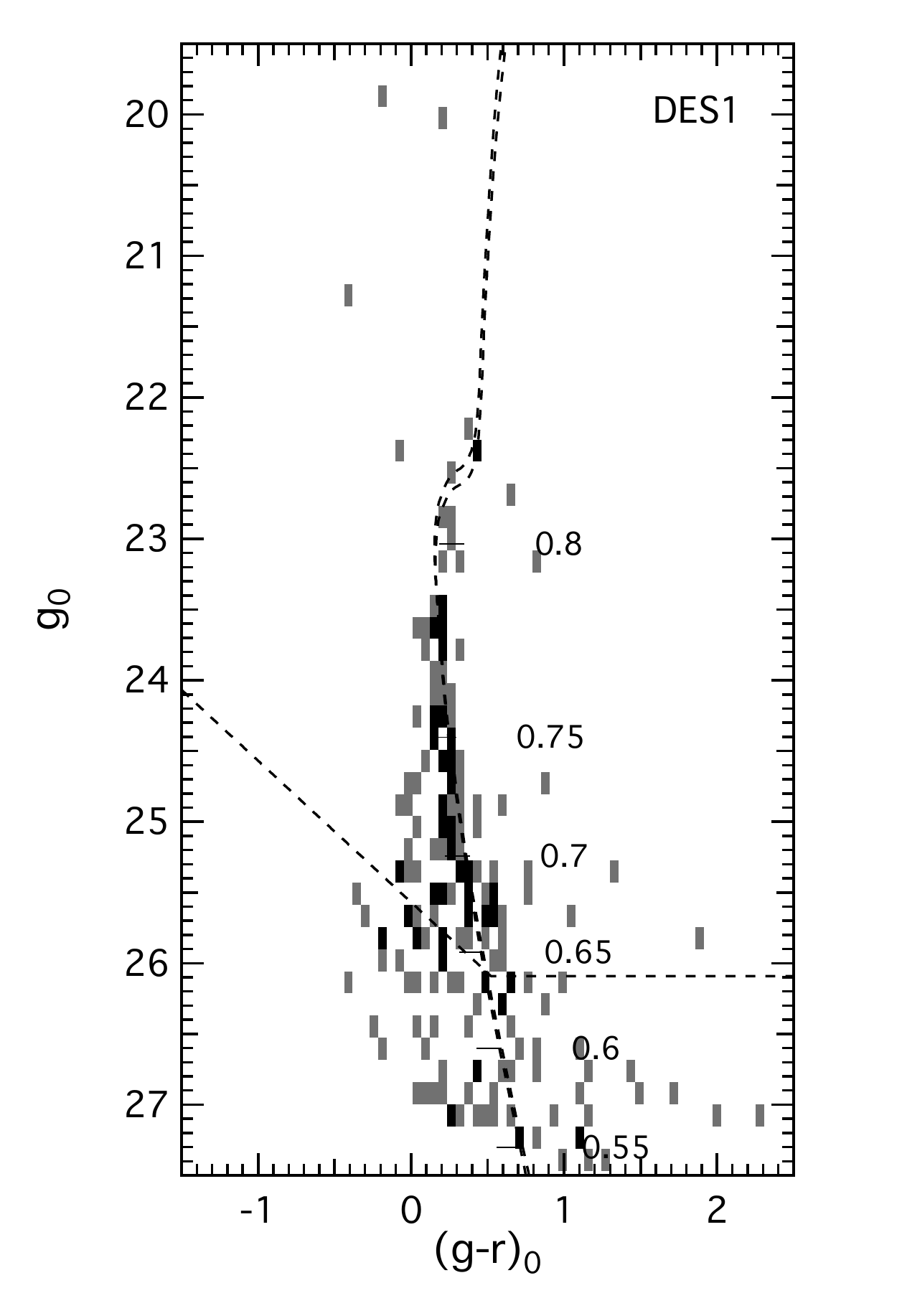}
\caption{{\it Left panel:} The colour-magnitude diagram of all stars within the ellipse centred on the nominal celestial coordinates of DES1 shown in Figure\,\ref{fig:DES1stellar_distribution}. The major and minor axes are $0\farcm84$ and $0\farcm49$ with a position angle of 112 degrees, respectively. 
{\it Middle panel:} Comparison CMD of the stars between the inner and the outer circles, showing the distribution of the foreground stars in colour-magnitude space.
{\it Right panel:} Hess diagram of the foreground-subtracted CMD superimposed with the best-fitting Dartmouth isochrones as dashed lines bracketing the 1-$\sigma$ confidence level of the metallicity estimate. Both isochrones are 11.2\,Gyr, [$\alpha$/Fe]$=+0.2$, $m-M=19.40$\,mag with the left isochrone having an [Fe/H]$=-2.50$, while the right has an [Fe/H]$=-2.17$. Note: [Fe/H]=$-2.50$ is the lowest metallicity available for the Dartmouth model isochrones.} The masses of main-sequence stars in solar mass units are marked to show the covered mass range. In all three panels the dashed line represents the 50-percent completeness limit as determined with artificial star test and the MCMC method.
\label{fig:cmdiso_DES1}
\end{center}
\end{figure*}

\begin{figure}
\begin{center} 
\includegraphics[width=0.95\hsize]{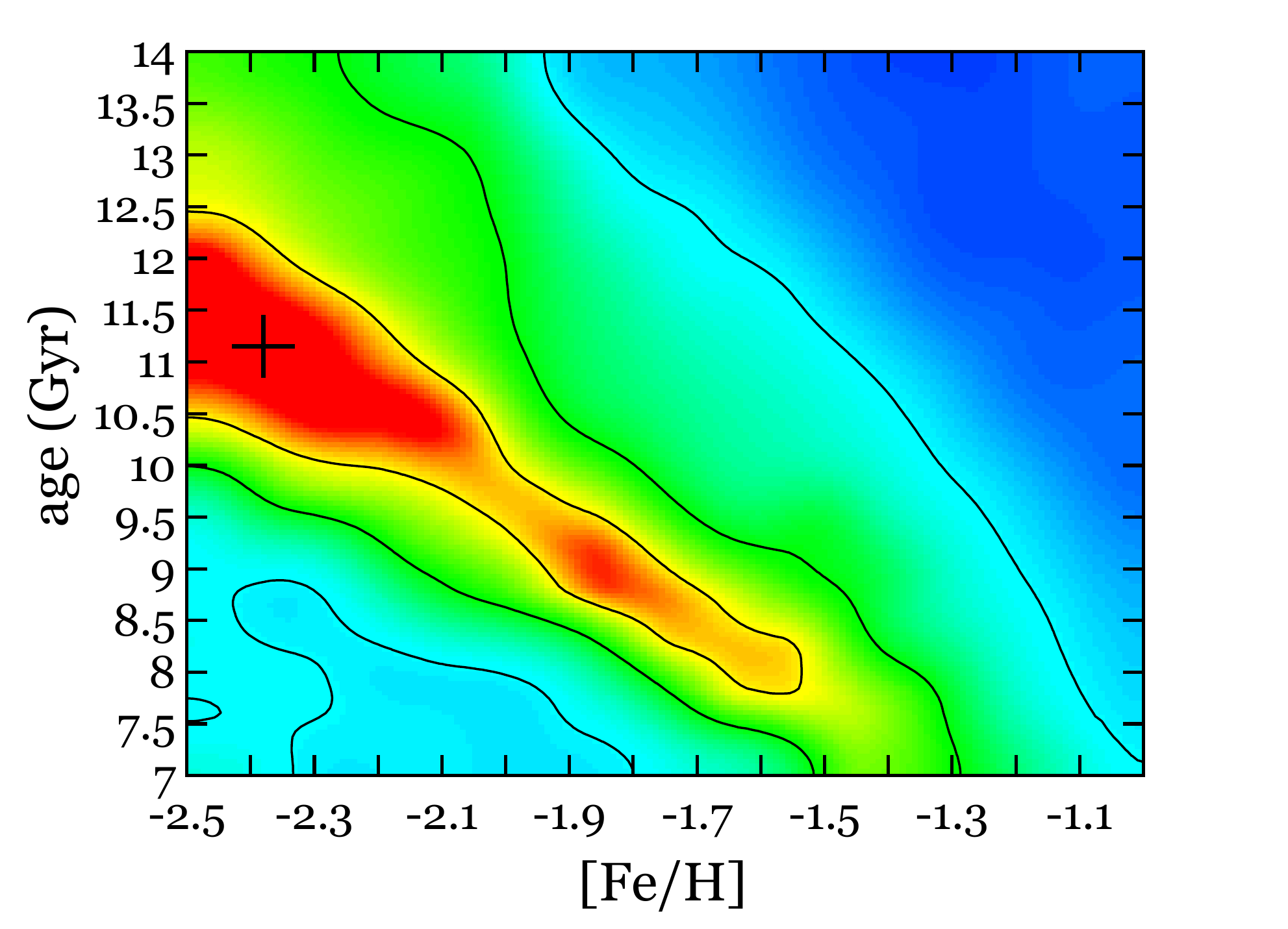}
\caption{Smoothed maximum likelihood density map in age-metallicity space for all stars within the 90\% ellipse around DES1. Contour lines show the 68\%, 95\%, and 99\% confidence levels. The diagonal flow of the contour lines reflects the age-metallicity degeneracy inherent to such an isochrone fitting procedure. The 1D marginalized parameters around the best fit (cross) with uncertainties are listed in Table~\ref{tab:DES1parameters}.}\label{fig:DES1_age_metal}
\end{center}
\end{figure}

The analysis outlined in $\S$\ref{sec:param_analysis} is highly iterative and produces many intermediate results, here we present the outcomes of that process. The on-sky distribution of the DES1 stars (Figure~\ref{fig:DES1stellar_distribution}) and their corresponding radial profile (Figure~\ref{fig:DES1rad_profile}). The stellar population of those stars inside the 90\% ellipse (Figure~\ref{fig:cmdiso_DES1}) and their most likely age and metallicity properties (Figure~\ref{fig:DES1_age_metal}). Finally, we generate the luminosity function of the system, Figure~\ref{fig:LF_DES1} and estimate its absolute magnitude. The parameters for DES1 are listed in Table~\ref{tab:DES1parameters}.

\subsection{Structural Parameters}\label{sec:DES1struct}
Figure \ref{fig:DES1stellar_distribution} highlights the on-sky distribution of the DES1 ultra-faint dwarf galaxy candidate with stars selected based on their proximity to the best-fitting isochrone. The inner ellipse with a semi-major axis length of 3.9$r_h$ encompasses 90\% of the DES1 stellar population and the outer ellipse has a semi-major axis length of 5.5$r_h$. The size of the outer ellipse is chosen such that the area difference between the two ellipses is equivalent to the area of the inner ellipse. Stars located between the inner and outer ellipse are then used to populate the comparison field CMD. The values for the central coordinates $(\alpha_0,\delta_0)$, position angle ($\theta$), and ellipticity ($\epsilon$) that best describe the stellar distribution of DES1 are listed in Table~\ref{tab:DES1parameters}. DES1 is a rather elongated system with an apparent axis ratio of 0.59, which translates to an ellipticity of $\epsilon=0.41^{+0.03}_{-0.06}$. The position angle is $\theta=112^\circ\pm3^\circ$. 

Overplotted in Figure~\ref{fig:DES1stellar_distribution} are the positions of two putative horizontal branch stars as cyan circles and as red open cirlces are objects from the {\sc AllWISE}\protect{\footnote{All Wide-field Infrared Survey Explorer mission, http://wise2.ipac.caltech.edu/docs/release/allwise/}} survey \citep{Wright2010}. The {\sc AllWISE} objects are scaled in size to reflect their magnitude and highlight the position of the bright objects in the field. These include both bright foreground stars and bright background galaxies.

Figure~\ref{fig:DES1rad_profile} shows the star number density in elliptical annuli around DES1, where $r_e$ is the elliptical radius. Overplotted are the best-fitting Exponential (black dotted) and King (blue dotted) profiles using the modal values from the ML analysis. The error bars were derived from Poisson statistics. We measure a half-light radius of $r_h=5.5^{+0.8}_{-0.7}$\,pc, which is similar in size to e.g.~Mu\~{n}oz\,1 \citep[$r_h=7.1$\,pc, $M_V=-0.4\pm0.9$,][]{Munoz2012} and SMASH1 \citep[$r_h=7.1^{+3.5}_{-2.4}$\,pc, $M_V=-1.0\pm0.9$,][]{Martin2016b}. The latter object is considered to be a star cluster tidally disrupted by the Large Magellanic Cloud (LMC).

\subsection{Stellar population}

The colour-magnitude diagram of all stars within the 90\% ellipse of DES1 is shown in the left panel of Figure~\ref{fig:cmdiso_DES1}. The red giant branch (RGB) and subgiant branch are completely absent. However, we notice two horizontal branch (HB) star candidates at $19.7<g_\circ<20.0$, $-0.3<(g-r)_\circ <+0.3$. The main sequence (MS) of DES1 is well defined down to $g\approx26.0$\,mag, below which star numbers are getting scarce. We note that our photometry is around 50\% complete at this magnitude and it is clearly beginning to influence our ability to identify probable DES1 members. The middle panel shows the comparison CMD of field stars distributed over an equal-sized area, between the 90\% ellipse and the concentric ellipse with the same ellipticity, position angle and a semi-major axis length of $5.50r_h$. 
The field CMD was then used to statistically decontaminate the DES1 CMD: for every field star we removed the nearest DES1 star in color-magnitude space if it is within the error ellipse defined by the $1\sigma$ photometric uncertainties in $g_\circ$ and $(g-r)_\circ$.
The right panel shows the foreground-subtracted Hess diagram with the best-fitting Dartmouth isochrone superimposed. 

Figure~\ref{fig:DES1_age_metal} shows the smoothed maximum likelihood density map of the age-metallicity space and the location of the best-fit is highlighted with a cross. The stars used to generate this map were selected from inside the 90\% ellipse as seen in Figure~\ref{fig:DES1stellar_distribution}. The isochrone
which best represents the DES1 features has an age of 11.2\,Gyr, a metallicity of [Fe/H]$=-2.38$ and [$\alpha$/Fe]$=+0.2$, shifted to a distance modulus of $m-M=19.40$\,mag ($D_\odot=76$\,kpc). The isochrones shown in the right panel of Figure~\ref{fig:cmdiso_DES1} bracket the 1-$\sigma$ estimate of the [Fe/H] value.

\subsection{Luminosity Function and Total Luminosity}\label{sec:DES1Lum}
\begin{figure}
\begin{center}
\includegraphics[width=0.98\hsize]{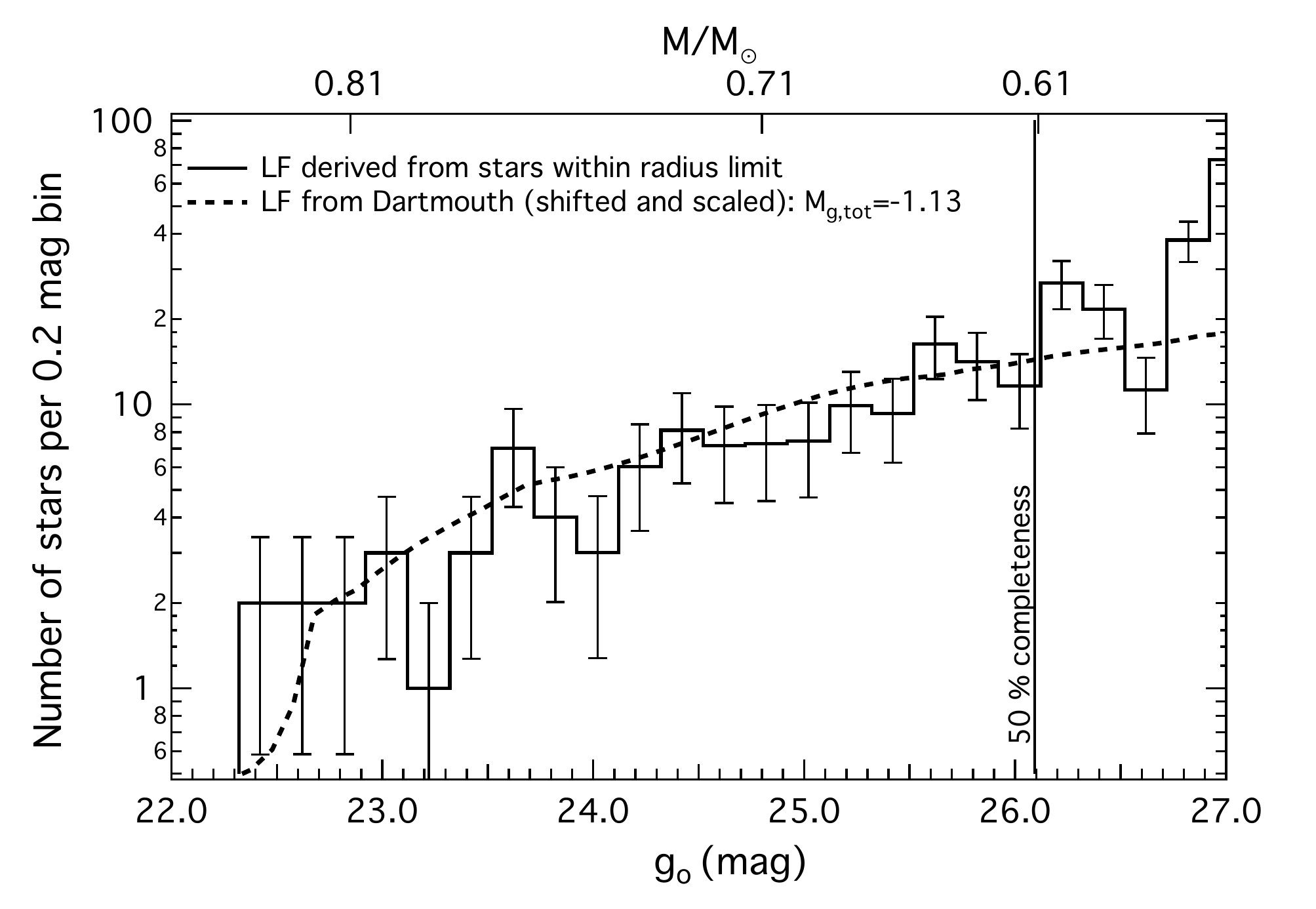}\hspace{-0.2cm}
\caption{Completeness-corrected DES1 luminosity function of all stars that are within the isochrone mask and within the 90\% ellipse. The best-fitting Dartmouth model luminosity function shifted by the distance modulus $19.40$\,mag and scaled to a total luminosity of $M_g=-1.13$\,mag is overplotted.}
 \label{fig:LF_DES1} 
\end{center}
\end{figure}

The total $V$-band luminosity of DES1 is estimated from all stars that are within the isochrone mask and within the 90\% ellipse. For that purpose, the observed $g$-band luminosity function is corrected for photometric incompleteness using the Logistic function (Eqn.~\ref{eqn:logistic}) with the parameters as determined in Section\,2.2 for this system. We then scaled the normalised theoretical luminosity function (LF) associated to the best-fitting Dartmouth isochrone (11.2\,Gyr, [Fe/H]$=-2.38$, [$\alpha$/Fe]$=+0.2$ shifted to a distance of 76\,kpc) to the observed level in the magnitude interval, $22.0<g_o<26.2$ (see Figure\,\ref{fig:LF_DES1}). The theoretical model LF is based on a power law initial mass function with a Salpeter slope of $-2.35$. The model LF follows closely the observed LF over the entire magnitude range. We calculated the integrated flux of DES1 to be $M_{g}=-1.13\pm0.2$\,mag.  A comparison with the total flux of the corresponding Dartmouth LF in the $V$-band yields a colour of $g-V=0.29$,which then converts the $M_{g}$ luminosity into $M_{V}=-1.42$.  Since the method of fitting the LF relies on the overall shape of the DES1 LF instead of individual stellar flux, the result is statistically resistant to the inclusion of some Galactic foreground stars. However, the exclusion of bright member stars of the system can still carry uncertainties of up to $\sim$25 percent ($\sim0.5$\,mag). Hence, a realistic estimate of the total luminosity of DES1 with error is $M_{V}=-1.42\pm0.50$. For example, adding the fluxes of the two HB candidates increases the total absolute $V$ magnitude of DES1 to $M_V=-1.73$\,mag.  All derived parameters presented in this section are summarized in Table~\ref{tab:DES1parameters}.

\begin{table}
\caption{Derived properties and structural parameters of DES1, see $\S$\ref{sec:param_analysis} for details on the listed parameters.
\label{tab:DES1parameters} }
{
\begin{center}
\begin{tabular}{l|ccccc}
\hline
 & \bf{DES1} \\ \hline\hline
$\alpha_0$\,(J2000) & 
$00^\mathrm{h}33^\mathrm{m}59\fs8\pm0\fs4$ & 
 \\
$\delta_0$\,(J2000) & 
$-49^\circ 02'\, 19'' \pm5''$ & 
 \\
$\theta$ (deg) & $112^{\circ}\pm3^{\circ}$ \\
$\epsilon$ & $0.41^{+0.03}_{-0.06}$  \\
$r_c$ (arcmin) & $0.116^{+0.040}_{-0.037}$ \\
$r_h$ (arcmin) & $0.245^{+0.036}_{-0.027}$ \\
$r_t$ (arcmin) & $2.114^{+0.729}_{-0.707}$ \\
$N_*$  & $54\pm 7$ \\
$E(B-V)$ (mag) & 0.0103 \\       
$A_g$ & 0.039\\
$A_r$ & 0.027\\
$(m-M)$ & $19.40\pm 0.12$  \\
$D_{\odot}$ (kpc) & $76\pm 4$ \\
$r_h$ (pc) & $5.5^{+0.8}_{-0.7}$   \\
age (Gyr) &   $11.2^{+1.0}_{-0.9}$  \\
$\langle [$Fe/H$] \rangle$ (dex) &  $-2.38^{+0.21}_{-0.19}$  \\
$[\alpha$/Fe$]_{\rm avg}$ (dex)&  $+0.2^{+0.1}_{-0.1}$ \\
$M_V$ (mag) & $-1.42\pm0.50$   \\
\hline\hline
\end{tabular}
\end{center}
}
\end{table}

\section{Properties of Eridanus\,III}\label{sec:EriIIIprop}
\begin{figure}
\begin{center} 
\includegraphics[width=1\hsize]{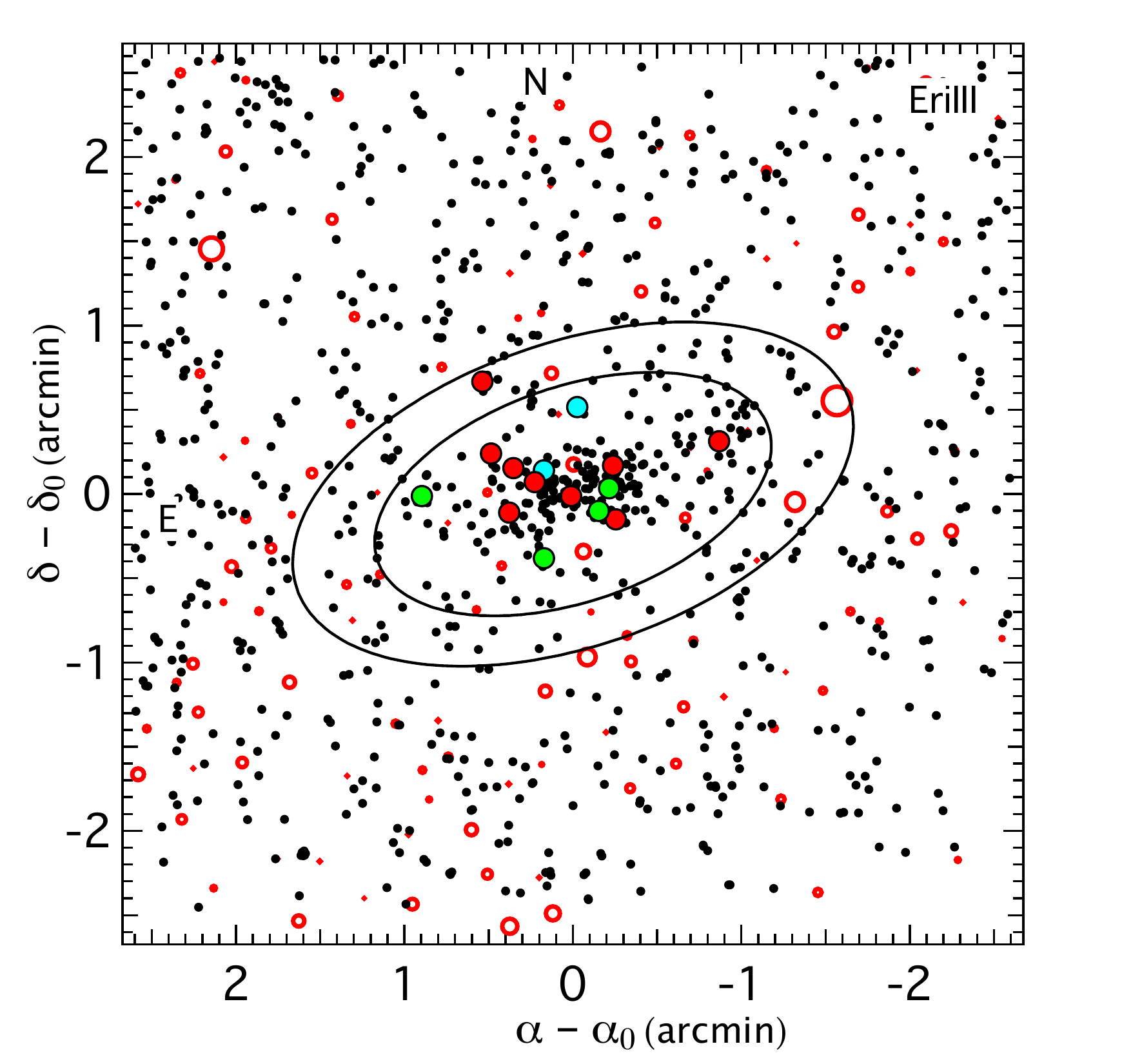}\\
\includegraphics[width=1\hsize]{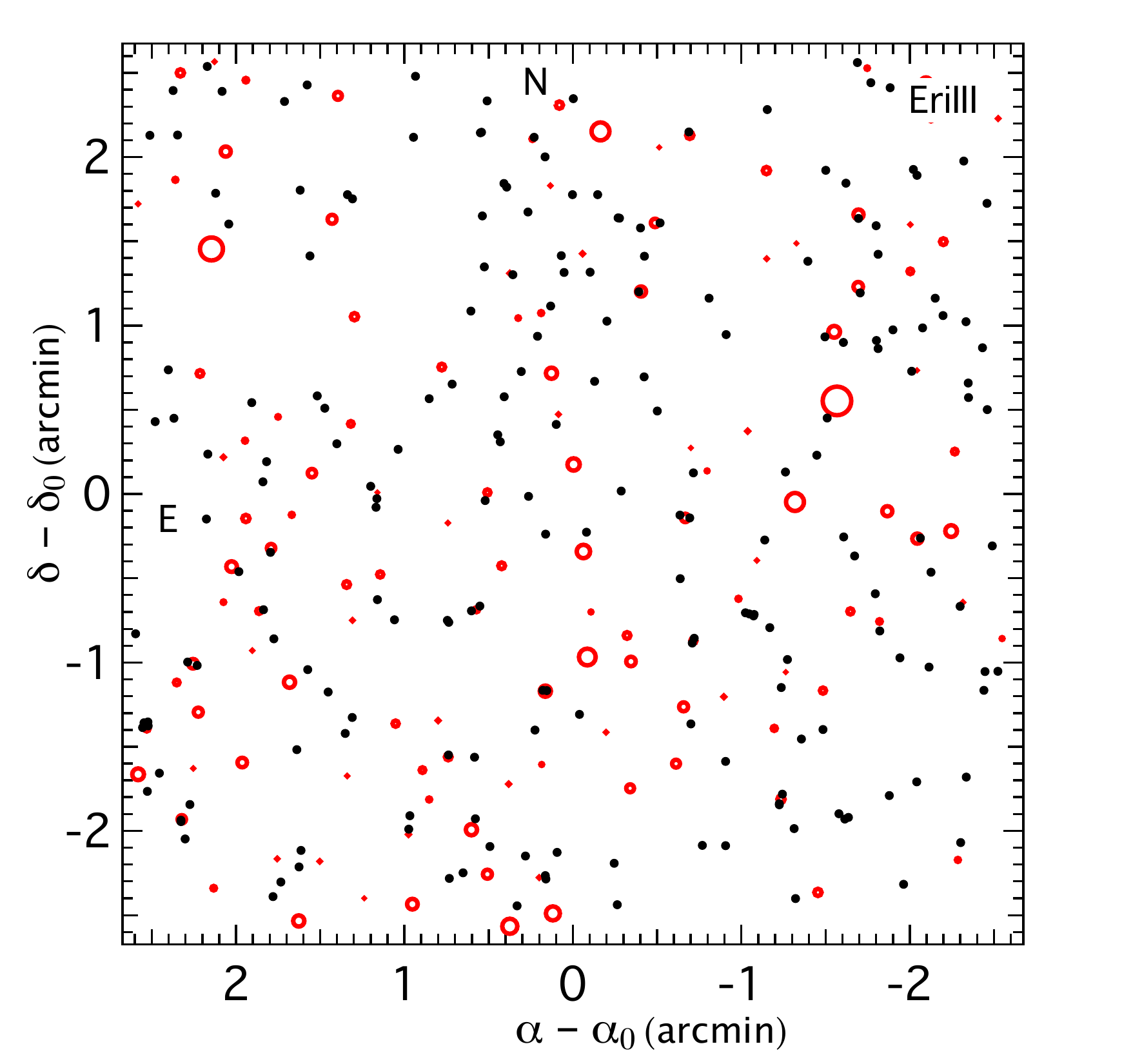}
\caption{
{\it Top}: Selecting only stars that are sufficiently close to the best-fitting isochrone (for more details see $\S$\ref{sec:param_analysis}) significantly increases the contrast between stars associated to the overdensity and Galactic foreground. The elongated concentration of Eri\,III stars becomes more prominent and structural parameters can be determined with higher accuracy. The two ellipses have a position angle of $109^\circ$ and a semi-major axis length of $3.9r_h$ and $5.5 r_h$, respectively. The inner ellipse borders the region that contains 90 percent of the DES1 stellar population, assuming an exponential radial profile. The outer ellipse has twice the area of the inner ellipse. The locations of the two HB, nine RGB and four BS candidates are highlighted in cyan, red and green respectively. The open red circles are objects from the {\sc AllWISE} catalogue \citep{Wright2010}, scaled to reflect their magnitudes. These objects highlight the position of the bright objects in the field both foreground stars and background galaxies. {\it Bottom}: Distribution of non-stellar objects selected in the same manner as those in the top panel. {\sc AllWISE} objects are again plotted as open red circles.
\label{fig:EriIIIstellar_distribution}}
\end{center}
\end{figure}

\begin{figure}
\begin{center} 
\includegraphics[width=0.99\hsize]{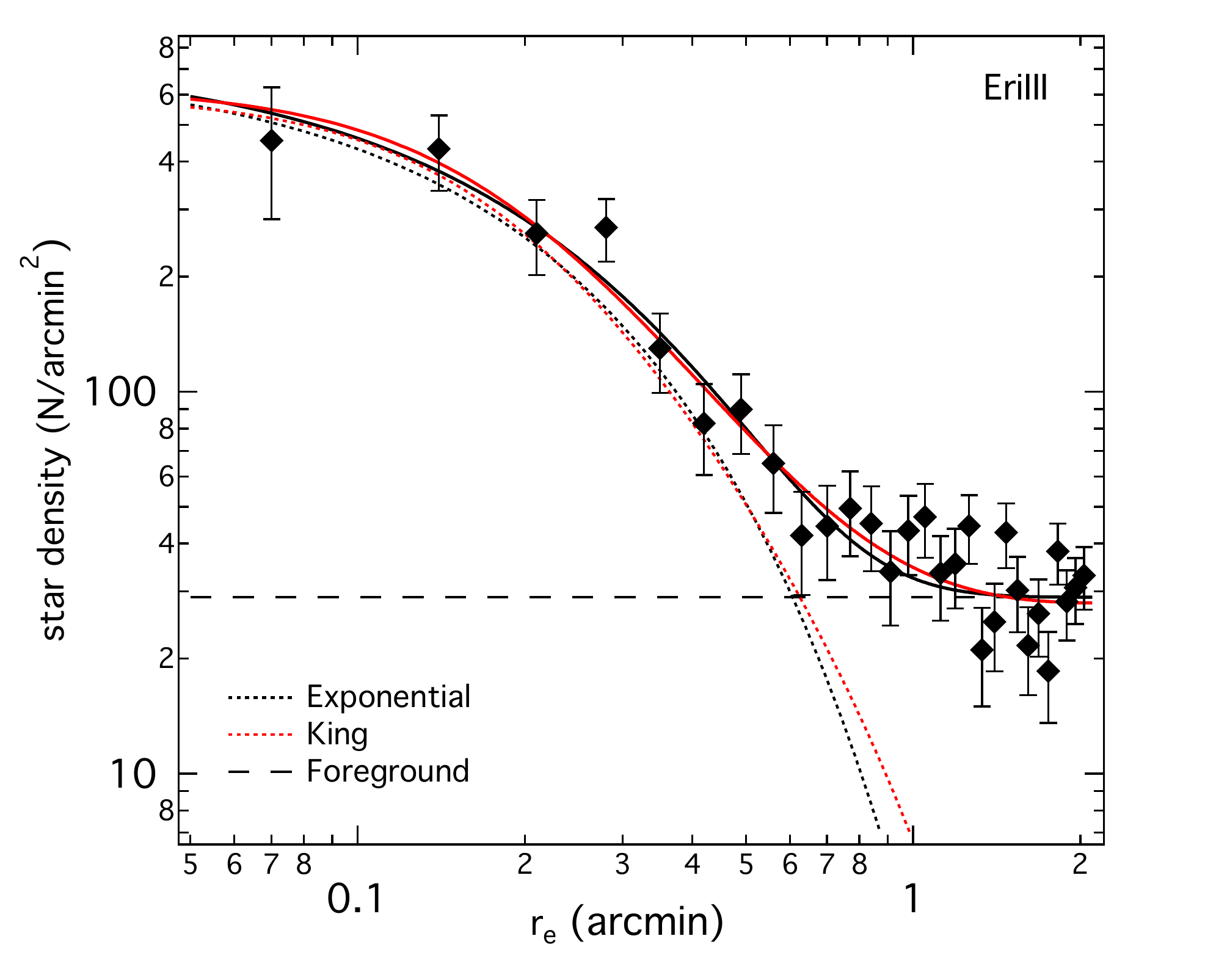}
\caption{Radial density profile of Eri\,III. The best-fitting Exponential (black dotted) and 
King (red dotted) profiles are superimposed on the data points. The horizontal dashed line is the 
density of the foreground stars. The solid black and red lines represent the profiles + foreground. 
The error bars were derived from Poisson statistics.\label{fig:EriIIIrad_profile}}
\end{center}
\end{figure}

\begin{figure*}
\begin{center} 
\includegraphics[width=0.3\hsize]{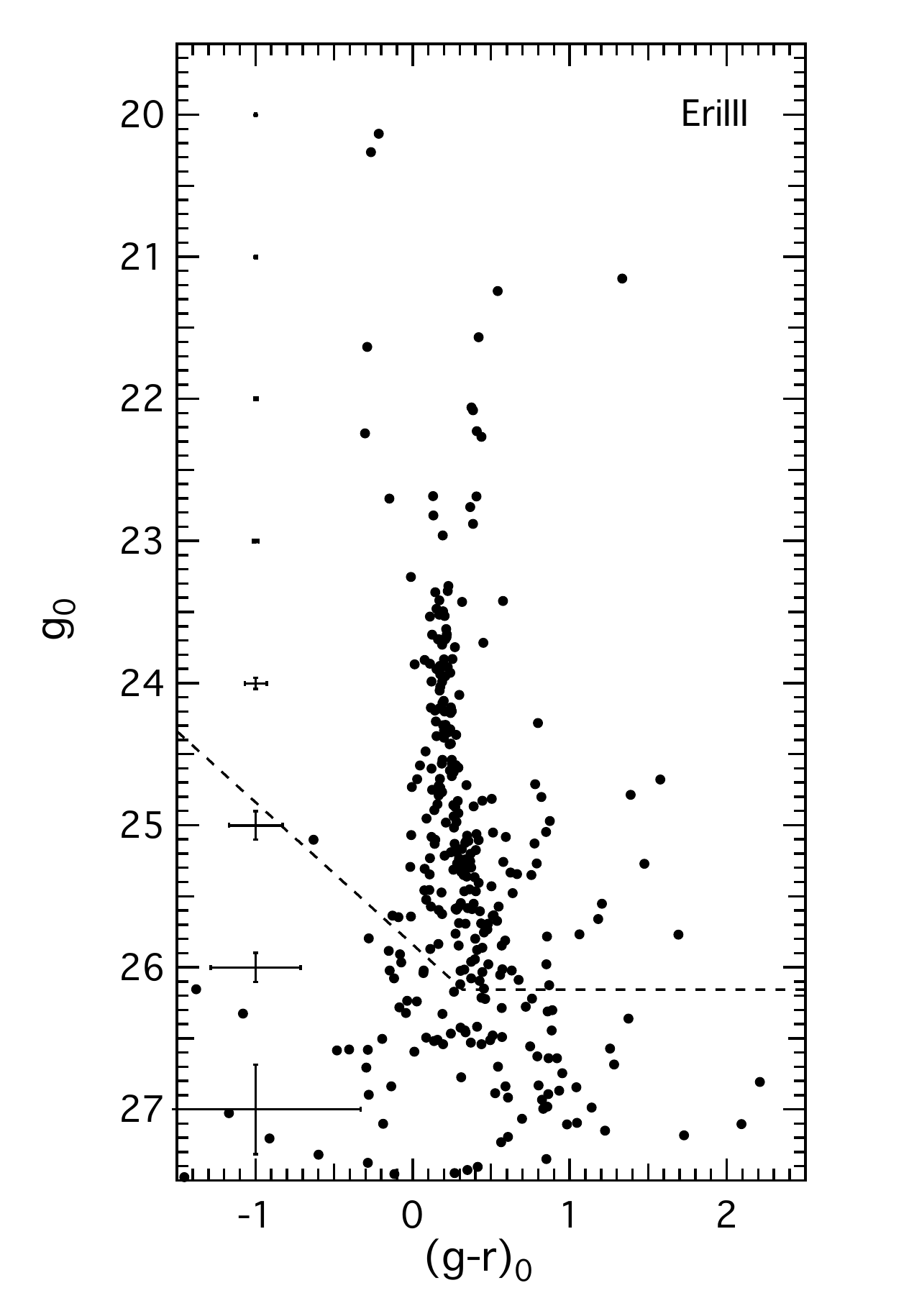}\hspace{-0.2cm}
\includegraphics[width=0.3\hsize]{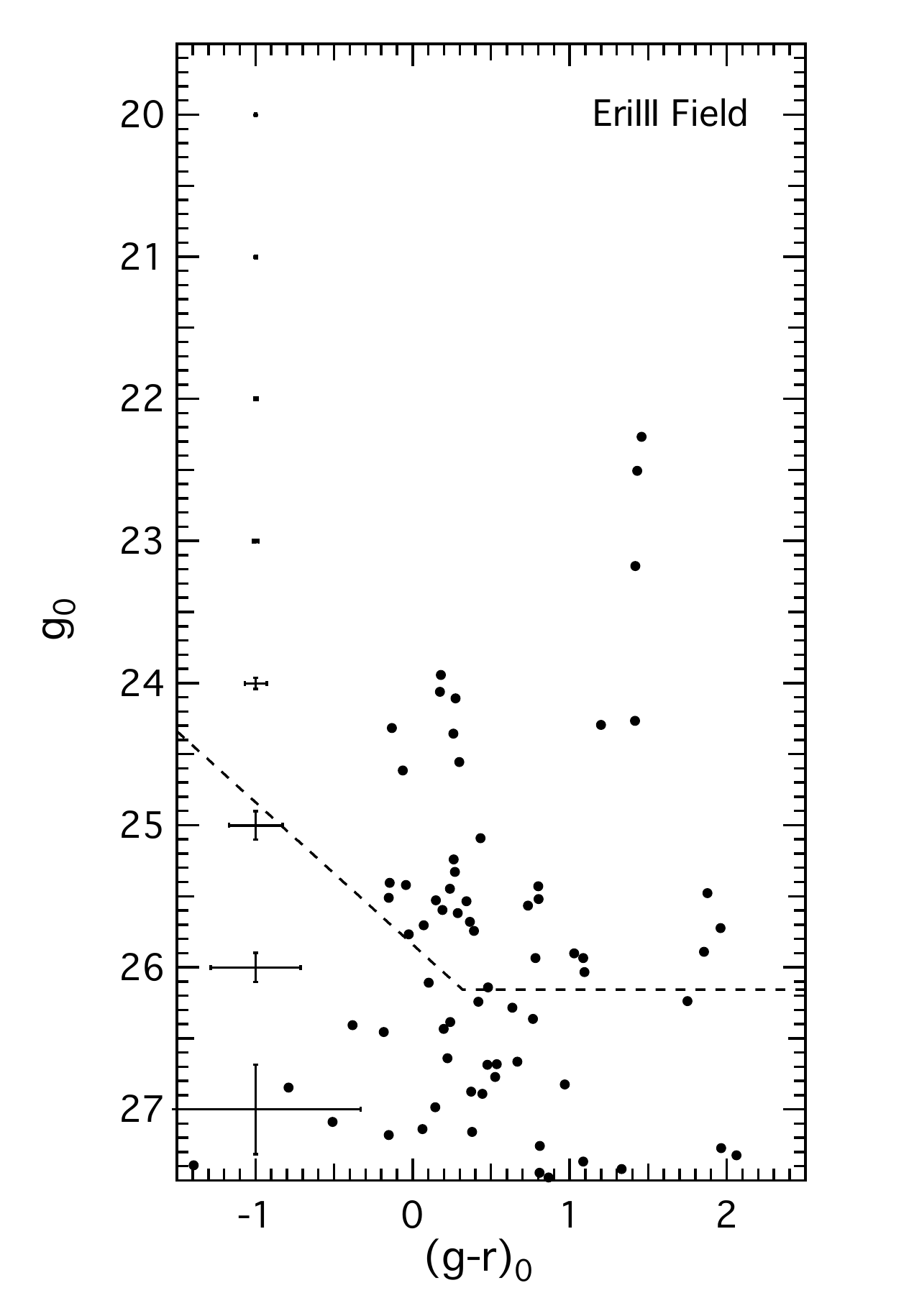}\hspace{-0.2cm}
\includegraphics[width=0.298\hsize]{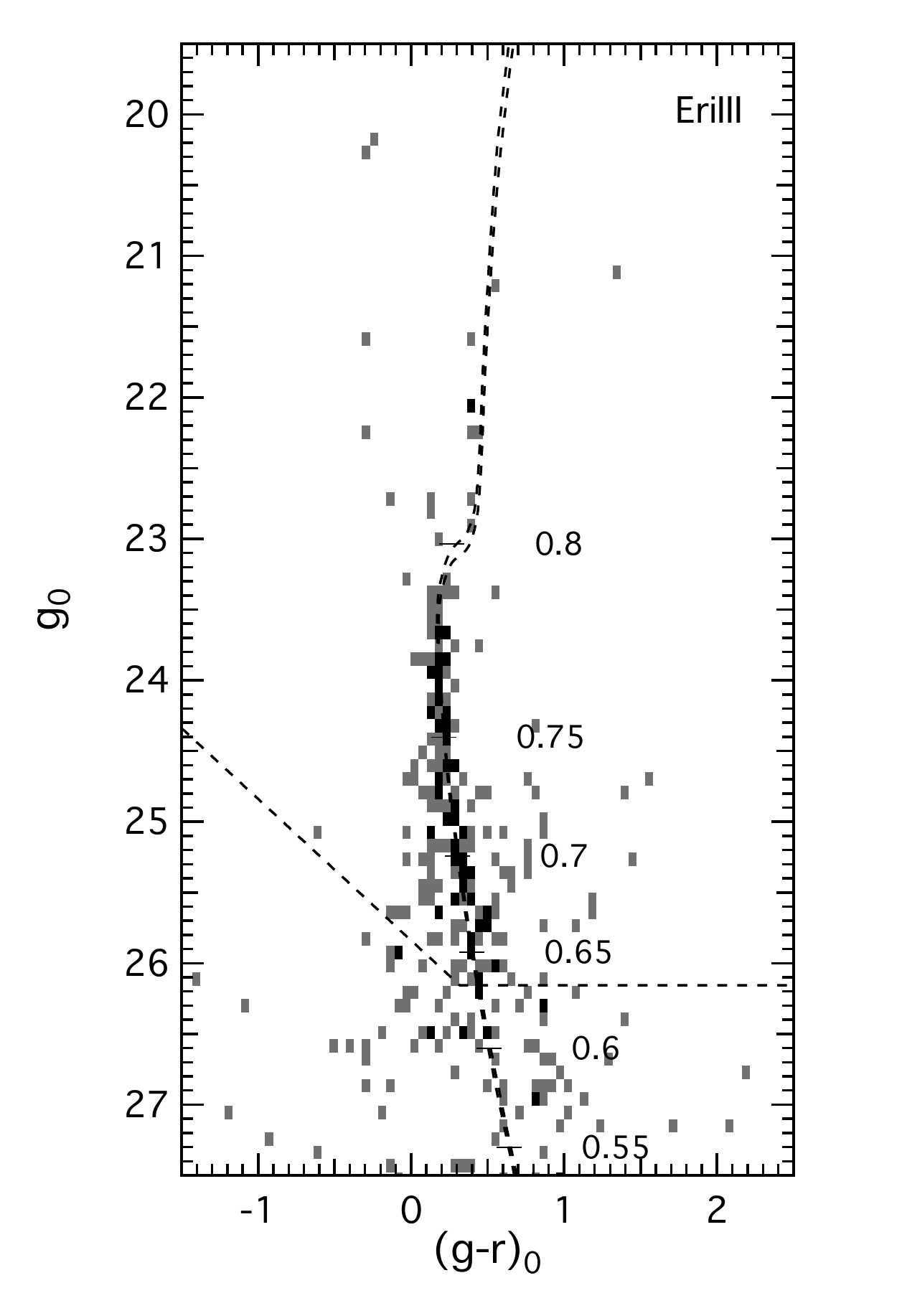}
\caption{{\it Left panel:} The colour-magnitude diagram of all stars within the ellipse centred on the nominal celestial coordinates of Eri\,III (Figure\,\ref{fig:EriIIIstellar_distribution}). 
{\it Middle panel:} Comparison CMD of stars between the inner and outer ellipses, showing the distribution of Galactic foreground stars in colour-magnitude space.
{\it Right panel:} Hess diagram of the foreground-subtracted CMD superimposed with the best-fitting Dartmouth isochrones as dashed lines bracketing the 1-$\sigma$ confidence level of the metallicity estimate. Both isochrones are 12.5\,Gyr, [$\alpha$/Fe]$=+0.2$, $m-M=19.80$\,mag with the left isochrone having an [Fe/H]$=-2.50$, while the right has an [Fe/H]$=-2.21$. Note: [Fe/H]=$-2.50$ is the lowest metallicity available for the Dartmouth model isochrones. The masses of main-sequence stars in solar mass units are marked to show the covered mass range. In all three panels the dashed line represents the 50-percent completeness limit as determined with artificial star tests and the MCMC method. 
}\label{fig:cmdiso_EriIII}
\end{center}
\end{figure*}

\begin{figure}
\begin{center} 
\includegraphics[width=0.99\hsize]{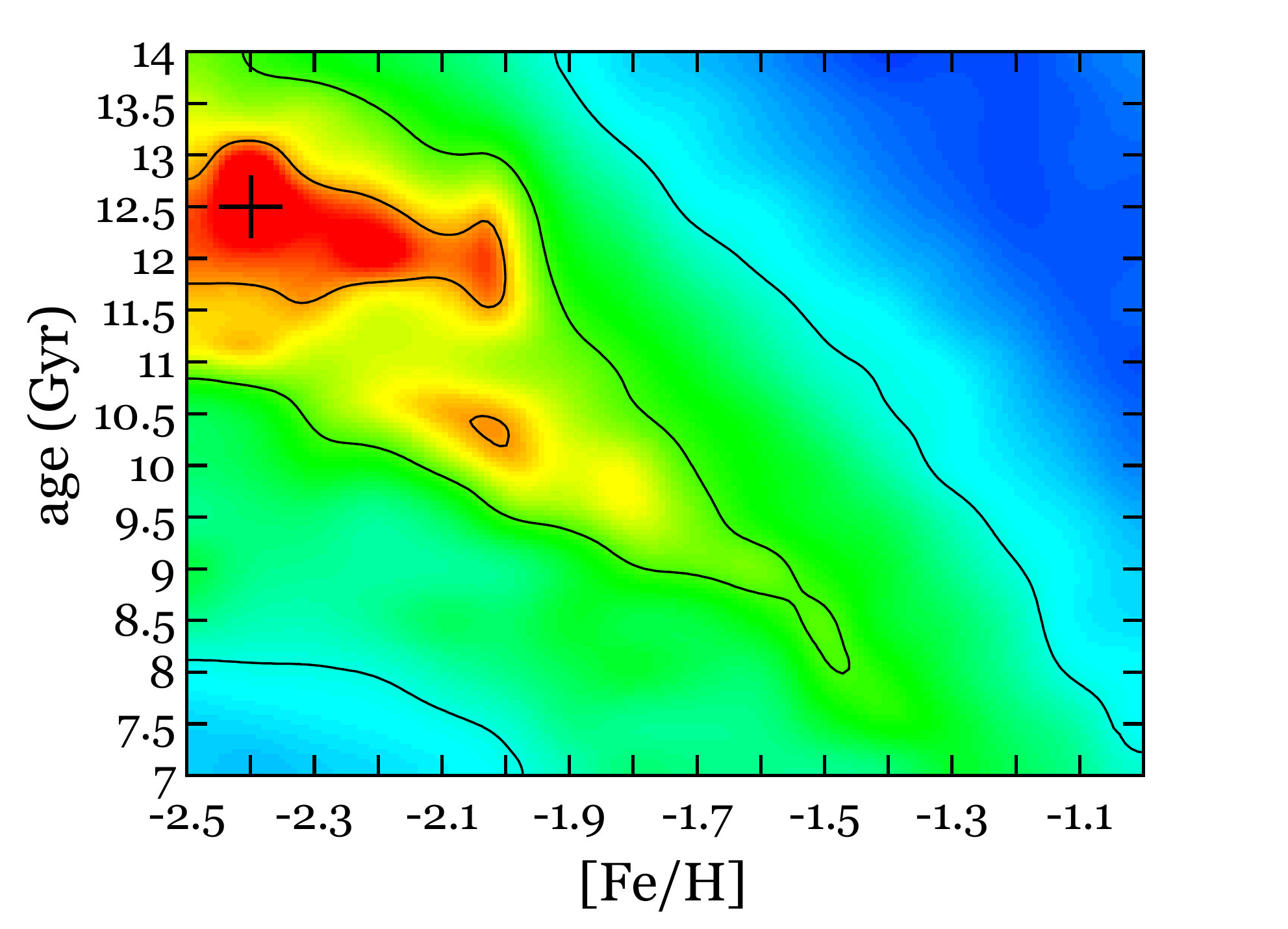}
\caption{Smoothed maximum likelihood density map in age-metallicity space for all stars within the 90\% ellipse around Eri\,III. Contour lines show the 68\%, 95\%, and 99\% confidence levels. The diagonal flow of the contour lines reflects the
age-metallicity degeneracy inherent to such an isochrone fitting procedure. The 1D marginalized parameters 
around the best fit with uncertainties are listed in Table~\ref{tab:EriIII_parameters}.\label{fig:EriIII_age_metal}}
\end{center}
\end{figure}

The properties of Eridanus\,III (Eri\,III) have been determined using the same procedure as outlined in $\S$\ref{sec:param_analysis}. The on-sky distribution of the Eri\,III stars (Figure~\ref{fig:EriIIIstellar_distribution}) and their corresponding radial profile (Figure~\ref{fig:EriIIIrad_profile}). The stellar population of those stars inside the 90\% ellipse (Figure~\ref{fig:cmdiso_EriIII}) and their most likely age and metallicity properties (Figure~\ref{fig:EriIII_age_metal}). Finally, we generate the luminosity function of the system, Figure~\ref{fig:LF_EriIII} and estimate its absolute magnitude. The parameters for EriIII are listed in Table~\ref{tab:EriIII_parameters}.

\subsection{Structural Parameters}\label{sec:EriIIIstruct}
Figure \ref{fig:EriIIIstellar_distribution} highlights the on-sky distribution of the Eri\,III ultra-faint dwarf galaxy candidate with stars selected based on their proximity to the best-fitting isochrone. The inner ellipse with a semi-major axis length of 3.9$r_h$ encompasses 90\% of the Eri\,III stellar population and the outer ellipse has a semi-major axis length of 5.5$r_h$. The size of the outer ellipse is chosen such that the area difference between the two ellipses is equivalent to the area of the inner ellipse. Stars located between the inner and outer ellipse are then used to populate the comparison field CMD. The two HB and nine RGB candidates are overplotted as cyan and red circles respectively while the four Blue Straggler (BS) candidates ($g-r\sim0$, $21.5<g_0<23.5$) are shown in green. The location of bright objects in the field from the {\sc AllWISE} survey are shown as open red circles, the size of which reflects their magnitude. These objects include both bright foreground stars and bright background galaxies.

Figure~\ref{fig:EriIIIrad_profile}, as per Figure~\ref{fig:DES1rad_profile}, shows the star number density in elliptical annuli around Eri\,III, where $r_e$ is the elliptical radius. Overplotted are the best-fitting Exponential (black dotted) and King (blue dotted) profiles using the modal values from the ML analysis. The error bars were derived from Poisson statistics. We derived a position angle $\theta=109^\circ\pm5^\circ$, an ellipticity $\epsilon=0.44^{+0.02}_{-0.03}$ and half-light radius of $r_h=8.3^{+0.9}_{-0.8}$\,pc. 
The values for the structure parameters are listed in Table\,\ref{tab:EriIII_parameters}.

\subsection{Stellar Population}\label{sec:EriIIIpop}
Figure~\ref{fig:cmdiso_EriIII} (left panel) shows the colour-magnitude diagram of all stars within $3.9r_h$ of the centre of Eri\,III. The middle panel, as per Figure~\ref{fig:cmdiso_DES1}, is the CMD of field stars outside the 90\,percent ellipse, covering the same area. The right panel shows the Hess diagram for the foreground-corrected Eri\,III CMD with the best-fitting Dartmouth isochrone superimposed. We note that the statistical decontamination was performed the same way as for DES1.

Figure~\ref{fig:EriIII_age_metal} shows the smoothed maximum likelihood density map of the age-metallicity space and the location of the best-fit is highlighted with a cross. The stars used to generate this map were selected from inside the 90\% ellipse as seen in Figure~\ref{fig:EriIIIstellar_distribution}. Similar to DES1, Eri\,III consists of an old (12.5\,Gyr), metal-poor ([Fe/H]$=-2.40$) stellar population with [$\alpha$/Fe$]_{\rm avg}=0.2$. Eri\,III is 20 percent further away at a heliocentric distance of 91\,kpc ($m-M=19.80$\,mag). 

Based on its CMD, Eri\,III appears to be a slightly more luminous system than DES1 with a clear hint of a RGB. Nine RGB star candidates are noticeable above the MSTO between $21.2<g_\circ<23.0$. Similar to DES1 there are two possible HB stars in Eri\,III ($g_\circ\approx 20.1$, $(g-r)_\circ\approx -0.22$)), as well as four Blue Stragglers. The sky positions of the RGB, HB, and BS stars are highlighted in Figure~\ref{fig:EriIIIstellar_distribution}.

\subsection{Luminosity Function and Total Luminosity}\label{sec:EriIIILum}

\begin{figure}
\begin{center}
\includegraphics[width=0.9\hsize]{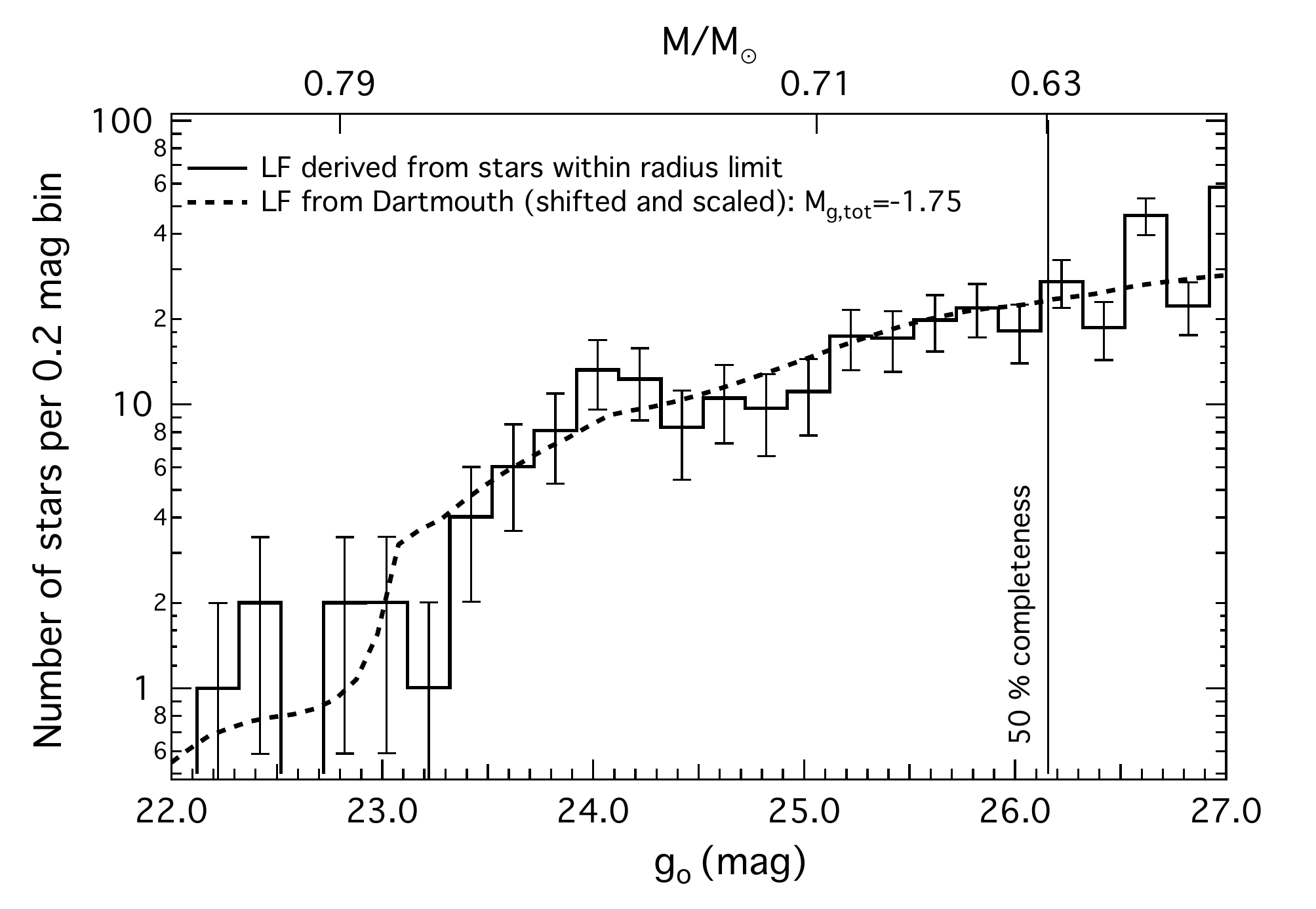}\hspace{-0.2cm}
\caption{Completeness-corrected Eri\,III luminosity function of all stars that are within the isochrone mask and within the 90\% ellipse (histogram). The best-fitting Dartmouth model luminosity function shifted by the distance modulus $19.80$\,mag and scaled to a total luminosity of $M_g=-1.75$\,mag is overplotted (dashed line).}.
 \label{fig:LF_EriIII} 
\end{center}
\end{figure}

The total luminosity of Eridanus\,III has been derived in the same manner as DES1 ($\S$\ref{sec:DES1Lum}) and presented in Figure~\ref{fig:LF_EriIII}. We calculated the integrated light by comparing the completeness-corrected observed LF with the Dartmouth model LF that corresponds to the best-fitting isochrone of 12.5\,Gyr, [Fe/H]$=-2.40$, and [$\alpha$/Fe]$=+0.2$. We measured a total $g$-band luminosity of $M_{g}=-1.75\pm0.2$. The integrated Dartmouth model LFs in $g$ and $V$ have a colour of $g-V=0.32$, which convert the $M_{g}$ magnitude into $M_{V}=-2.07$. 

For the same reasons as outlined in $\S$\ref{sec:DES1prop} a more realistic estimate for the uncertainty of the total luminosity of Eri\,III is $\sigma_{M_{V}}=0.50$. We also note that adding the fluxes of the HB and BS candidates would increase the total absolute magnitude to $M_V=-2.33$\,mag, well within the quoted uncertainty. All derived parameters presented in this section are summarized in Table\,\ref{tab:EriIII_parameters}.

\begin{table}
\caption{Derived properties and structural parameters of Eridanus\,III, , see $\S$\ref{sec:param_analysis} for details on the listed parameters. 
\label{tab:EriIII_parameters} }
{
\begin{center}
\begin{tabular}{l|ccccc}
\hline
 & \bf{Eridanus\,III} \\ \hline\hline
$\alpha_0$\,(J2000) & 
$02^\mathrm{h}22^\mathrm{m}45\fs3\pm0\fs5$ & 
 \\
$\delta_0$\,(J2000) & 
$-52^\circ 17'\, 05'' \pm 6''$ & 
 \\
$\theta$ (deg) & $109^{\circ}\pm 5^{\circ}$ \\
$\epsilon$ & $0.44^{+0.02}_{-0.03}$  \\
$r_c$ (arcmin) & $0.190^{+0.053}_{-0.049}$ \\
$r_h$ (arcmin) & $0.315^{+0.036}_{-0.027}$ \\
$r_t$ (arcmin) & $2.08^{+0.97}_{-0.91}$ \\
$N_*$  & $81\pm 14$ \\
$E(B-V)$ (mag) & 0.0223 \\     
$A_g$ & 0.084 \\
$A_r$ & 0.058\\
$(m-M)$ & $19.80\pm 0.04$  \\
$D_{\odot}$ (kpc) & $91\pm 4$ \\
$r_h$ (pc) & $8.6^{+0.9}_{-0.8}$   \\
age (Gyr) &   $12.5^{+0.5}_{-0.7}$  \\
$\langle [$Fe/H$] \rangle$ (dex) &  $-2.40^{+0.19}_{-0.12}$  \\
$[\alpha$/Fe$]_{\rm avg}$ (dex)&  $+0.2^{+0.1}_{-0.1}$ \\
$M_V$ (mag) & $-2.07\pm0.50$   \\
\hline\hline
\end{tabular}
\end{center}
}
\end{table}

\section{Properties of Tucana\,V}\label{sec:TucVProp}
\begin{figure}
\begin{center} 
\includegraphics[width=1.0\hsize]{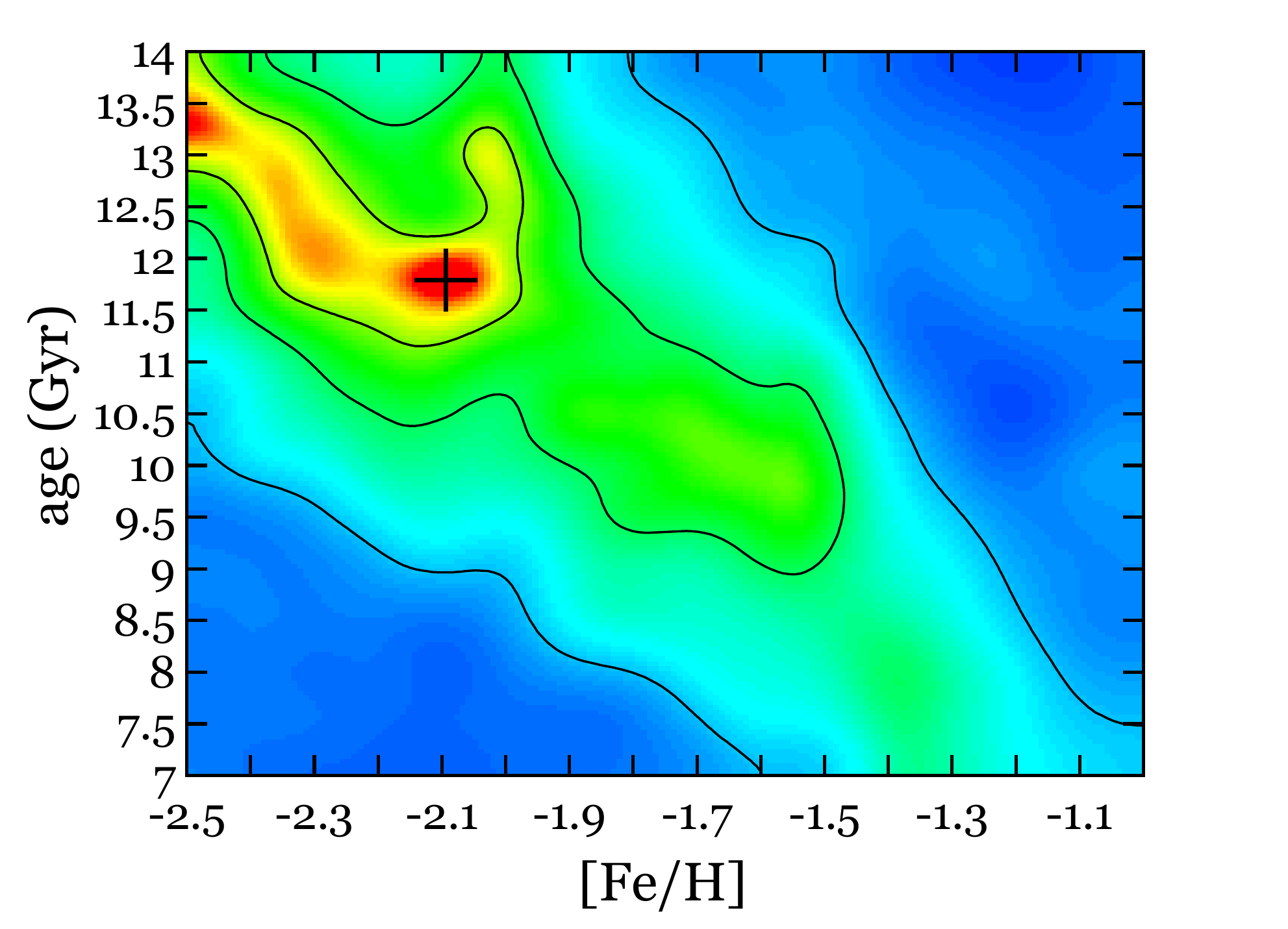}
\caption{ Smoothed maximum likelihood density map in age-metallicity space for all stars in the Tuc\,V field. The best-fitting Dartmouth isochrone has an age of 11.8\,Gyr, [Fe/H]=$-2.09$\,dex, [$\alpha$/Fe]=+0.4\,dex. Contour lines show the 68\%, 95\%, and 99\% confidence levels. 
\label{fig:TucVage_metal}}
\end{center}
\end{figure}

\begin{figure}
\begin{center} 
\includegraphics[width=0.5\hsize]{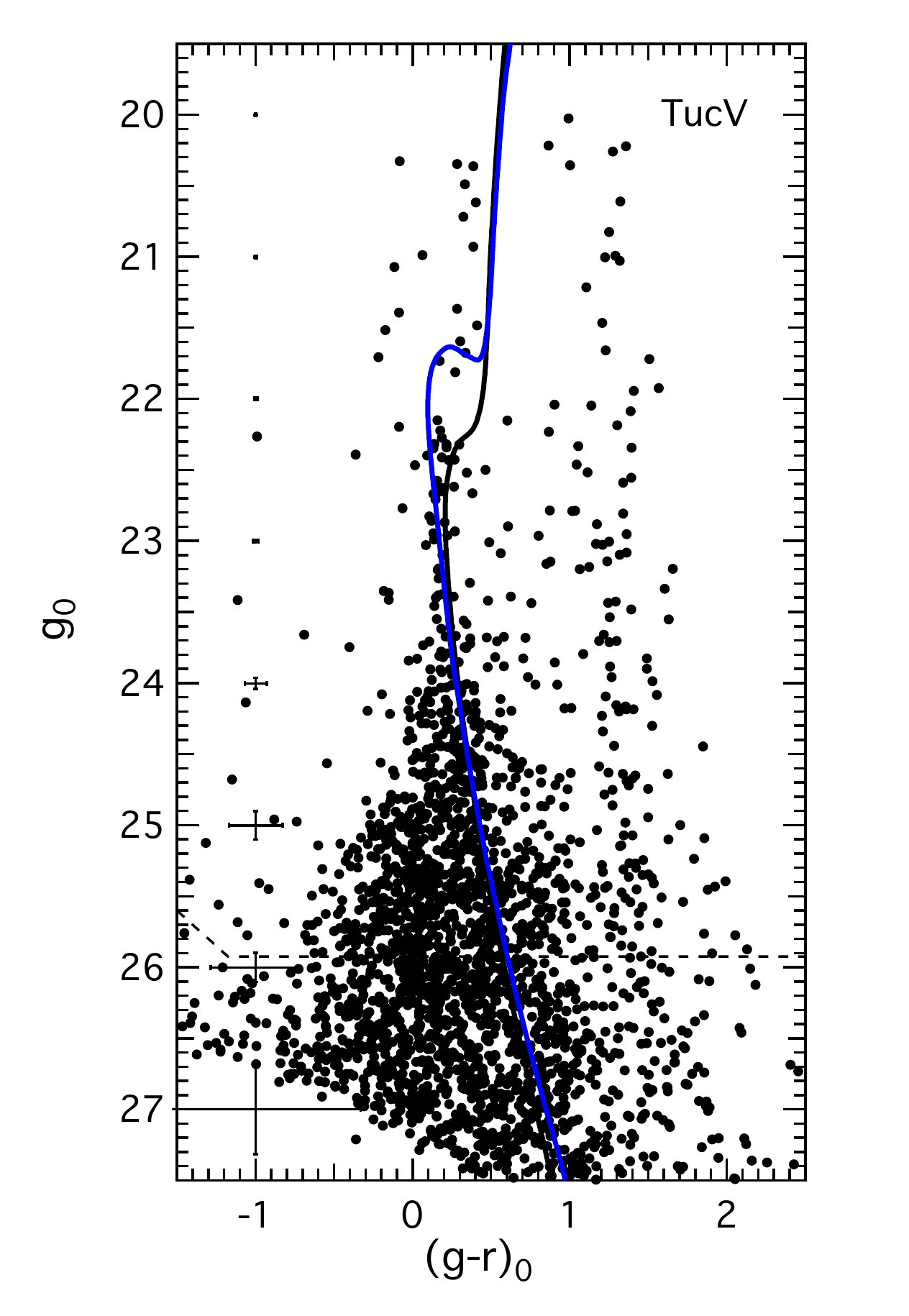}\hspace{-0.2cm}
\includegraphics[width=0.5\hsize]{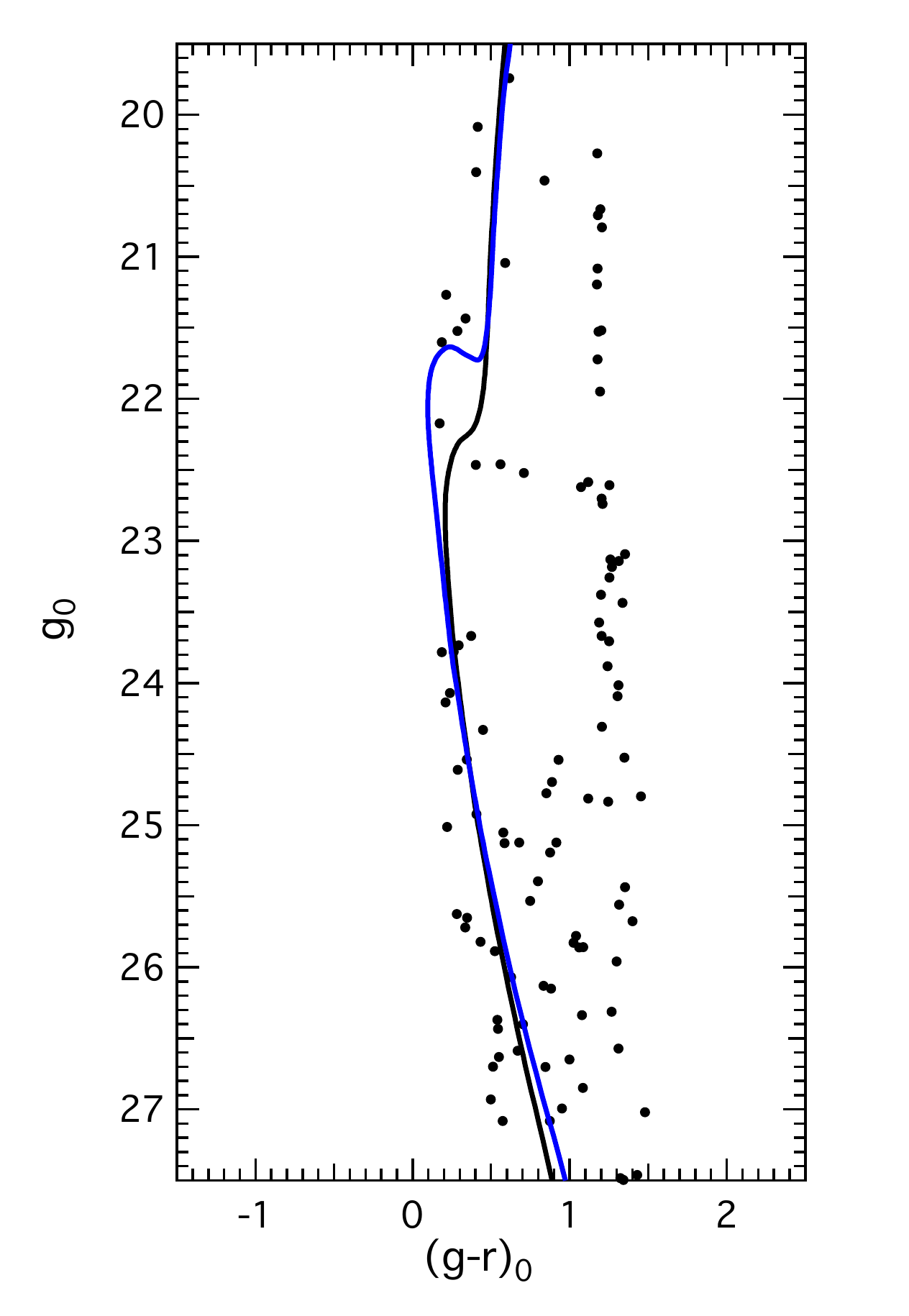}
\caption{{\it Left}: The colour-magnitude diagram of all stars in the Tucana\,V field with two isochrones superimposed. The best-fitting Dartmouth isochrone (black line) has an age of 11.8\,Gyr, [Fe/H]=$-2.09$\,dex, [$\alpha$/Fe]=0.4\,dex. The blue isochrone with an age of 6.0\,Gyr, [Fe/H]=$-1.30$\,dex ([$\alpha$/Fe]=0.0\,dex) corresponds to the Small Magellanic Cloud Northern Overdensity \citep{Pieres2017}. Both isochrones are shifted to a distance of 59.7\,kpc $(m-M=18.88)$. While the locations of the RGB and MS are effectively the same for the two isochrones, the main difference occurs around the MS turn-off, where the number of observed stars are sparse. {\it Right}: {\sc Galaxia} model prediction for the stellar population expected in the Tucana\,V GMOS-S field overplotted with the same isochrones as seen in the left panel.
}\label{fig:cmdinner_TucV}
\end{center}
\end{figure}

\begin{figure}
\begin{center} 
\includegraphics[width=1.0\hsize]{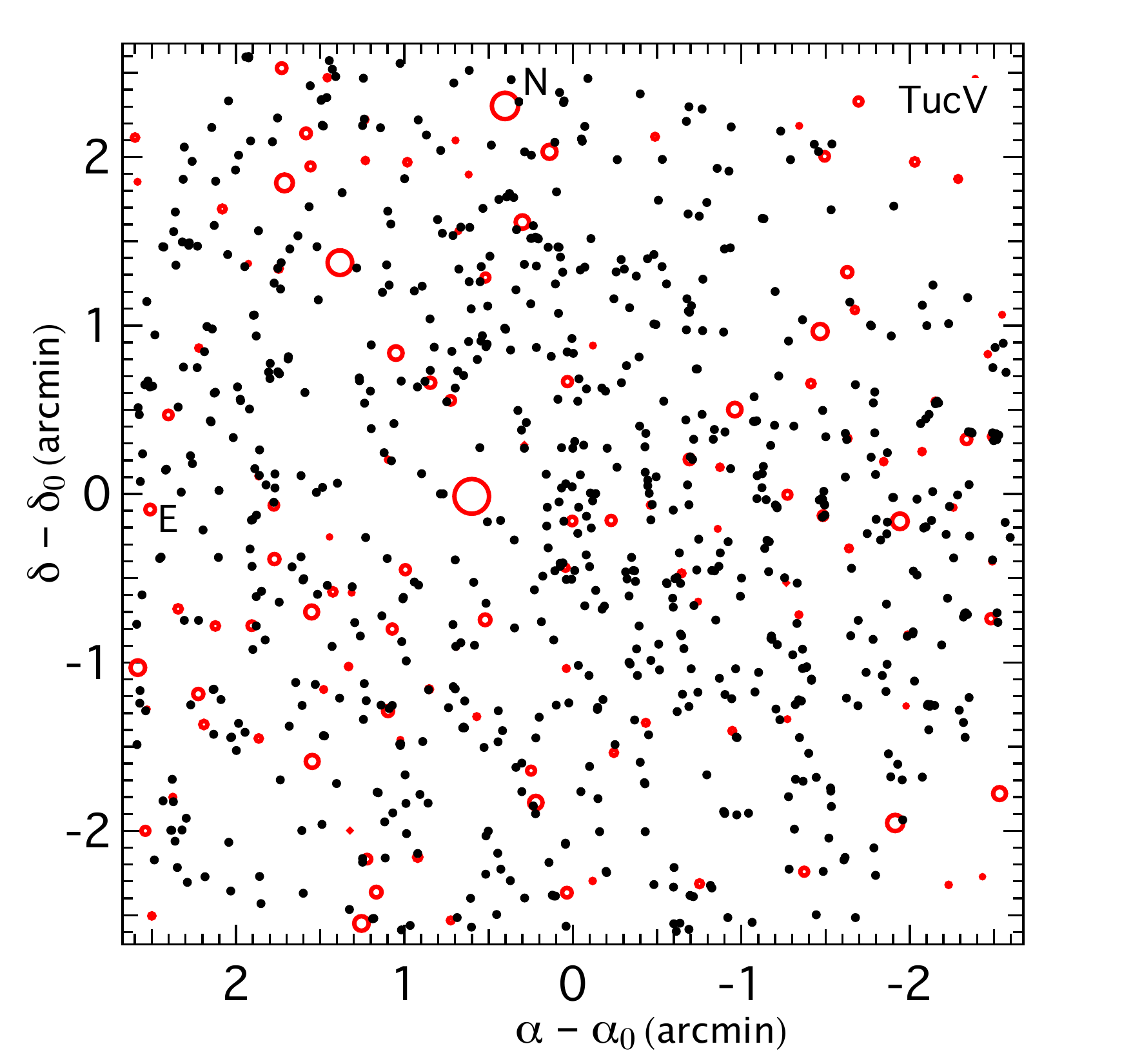}
\includegraphics[width=1.0\hsize]{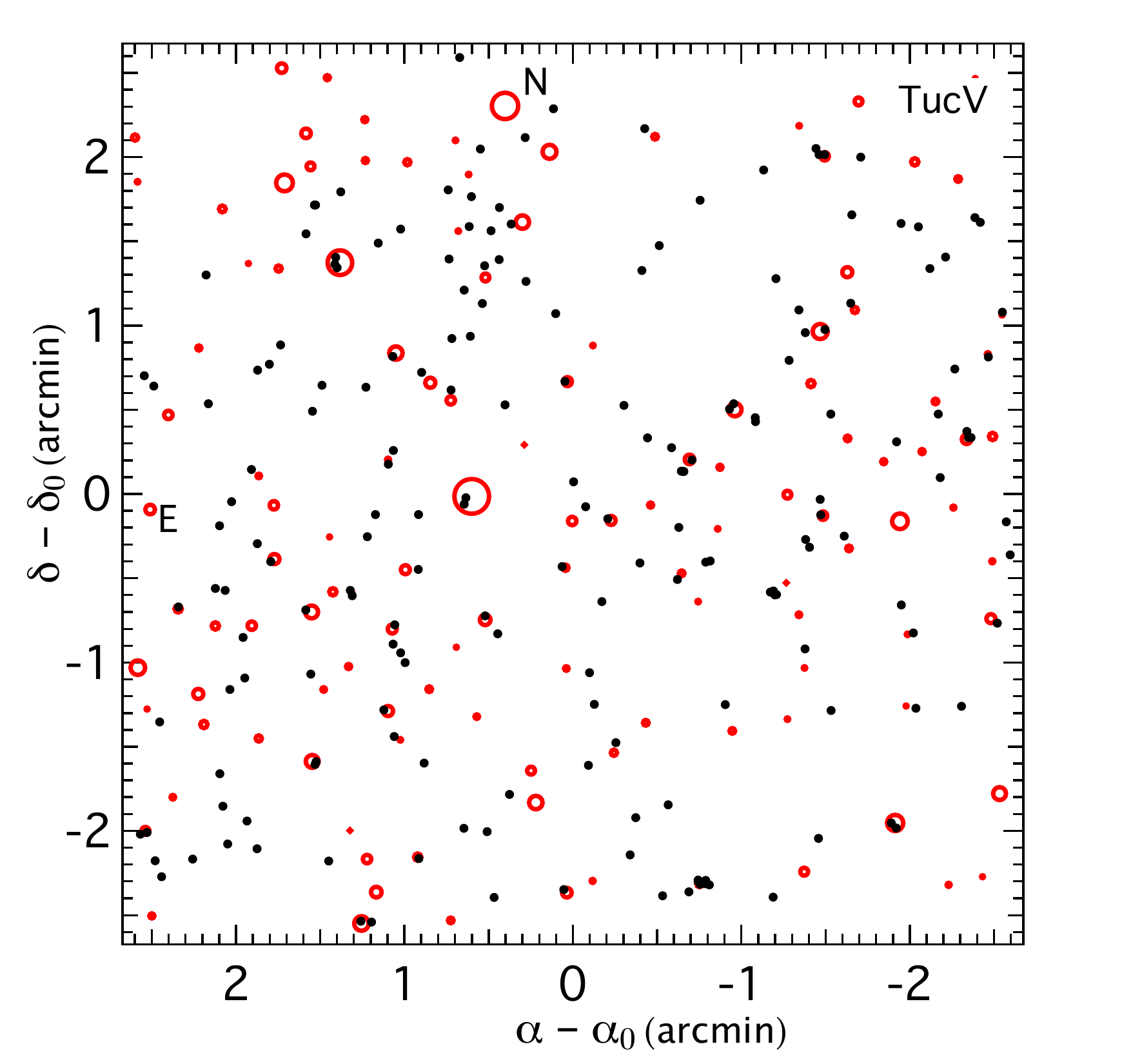}
\caption{{\it Top}: On-sky stellar distribution of the full field after selecting only stars that are sufficiently close to the best-fitting isochrone from Figure~\ref{fig:TucVage_metal}. This panel shows the distribution of stars in the field with the bright {\sc AllWISE} objects overplotted as open red circles scaled in size to reflect their magnitude. {\it Bottom}: The distribution of non-stellar objects selected in the same manner as the stars in the above panel overplotted with the {\sc AllWISE} objects. \label{fig:TucVstellar_distributionGMOS}}
\end{center}
\end{figure}

\begin{figure}
\begin{center}
\includegraphics[width=1\hsize]{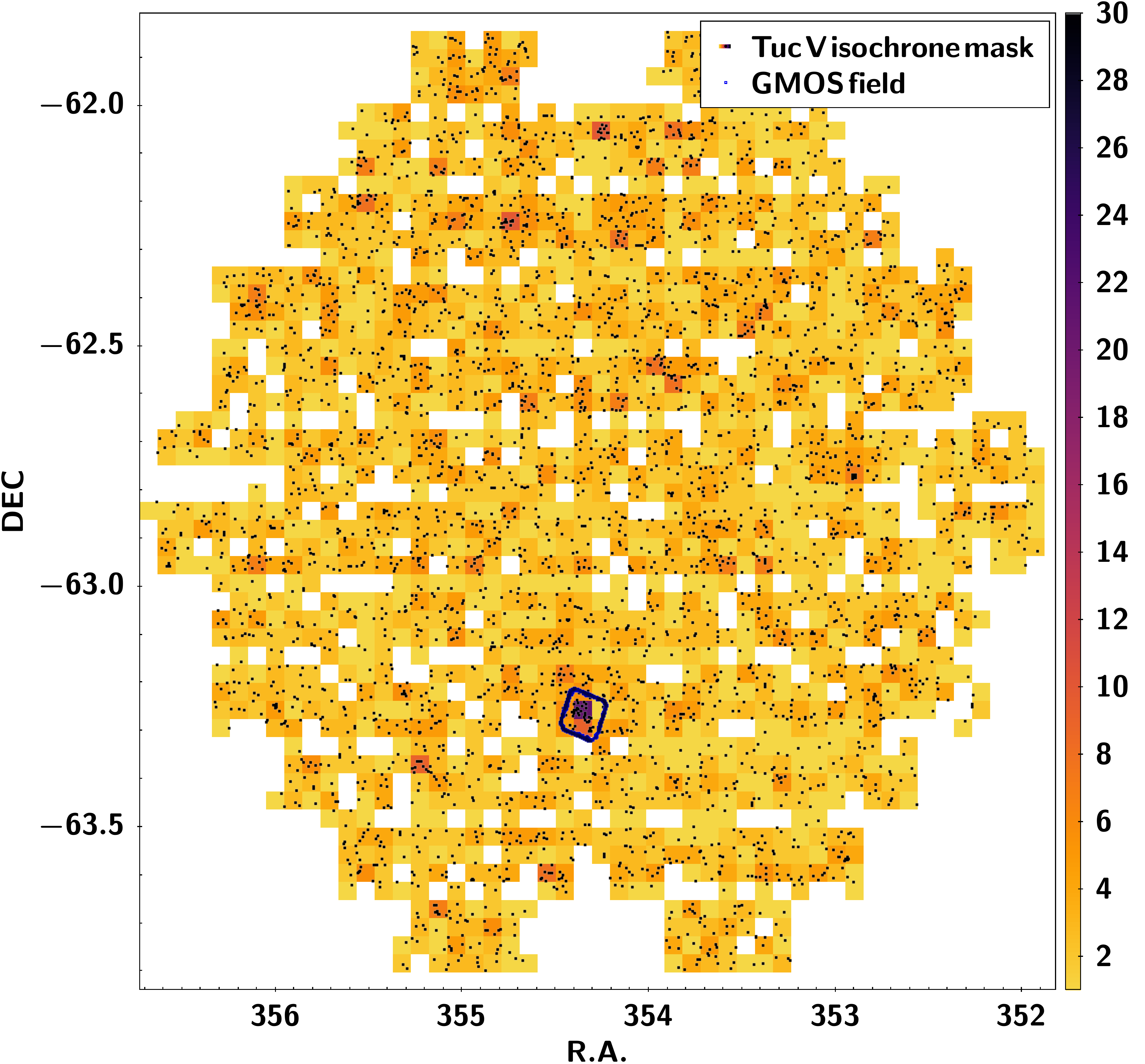}\hspace{-0.2cm}
\caption{This DECam field from the Dark Energy Survey contains the Tucana\,V ultra-faint dwarf galaxy candidate. Tuc\,V-like stars, selected with an isochrone mask similar to that shown in Figure~\ref{fig:cmdinner_TucV}, are plotted as a 2D density histogram. The small black points are stars in the field and the small square at ($\alpha,\delta$)$^\circ$ = ($354.35,-63.27$)$^\circ$ is the position of our Tuc\,V GMOS-S field. The significant overdensity at the Tuc\,V location coincides precisely with our GMOS-S field. There is no obvious stellar gradient or stream-like feature that extends beyond the GMOS-S field. The scale used here is an arbitrary stars per pixel where the pixel size has been chosen to merely highlight the overdensity in the entire DES field. North is up and East is to the left.}
 \label{fig:TucV_DESfield} 
\end{center}
\end{figure}

\begin{figure}
\begin{center} 
\includegraphics[width=1.0\hsize]{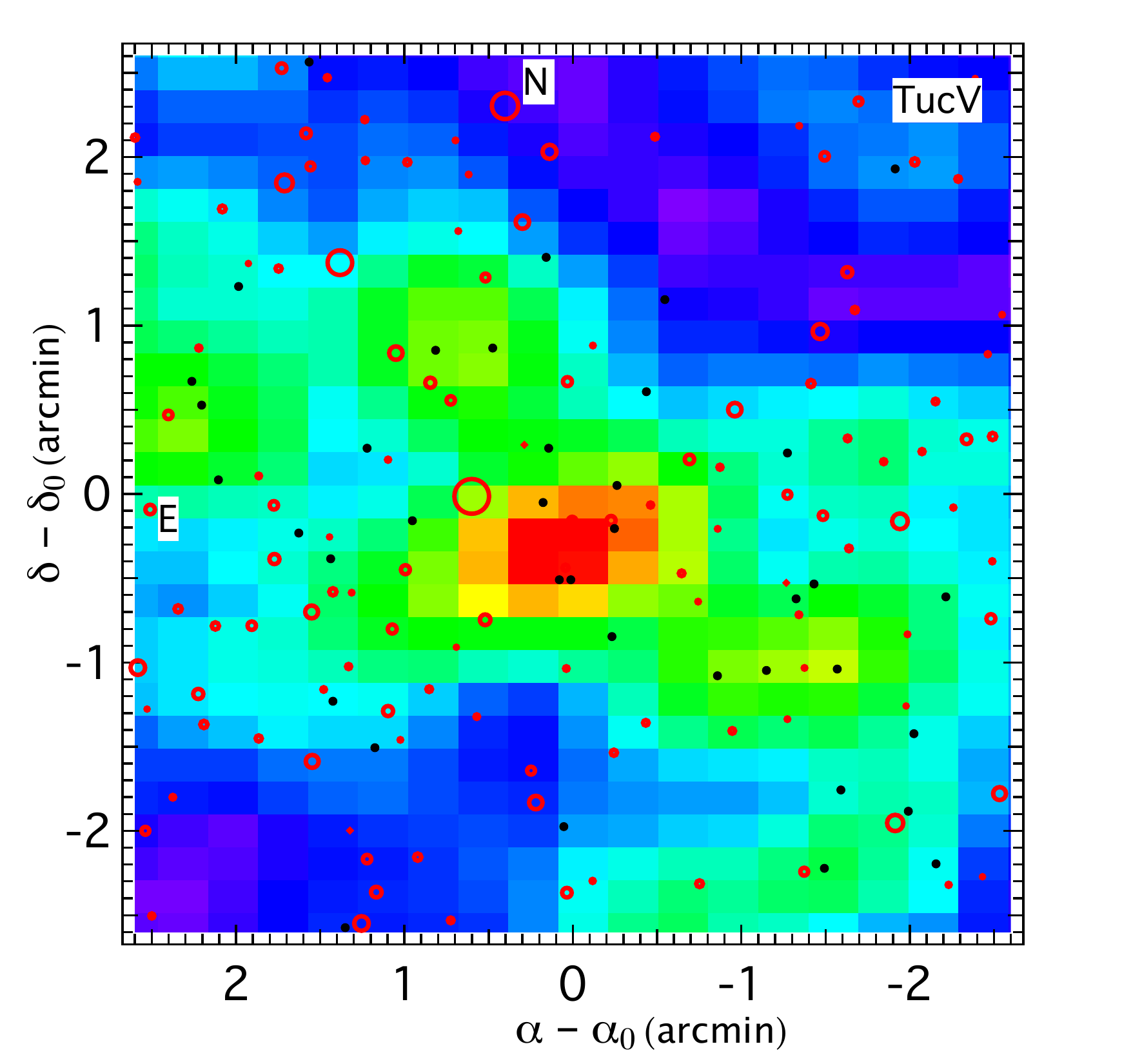}
\includegraphics[width=1.0\hsize]{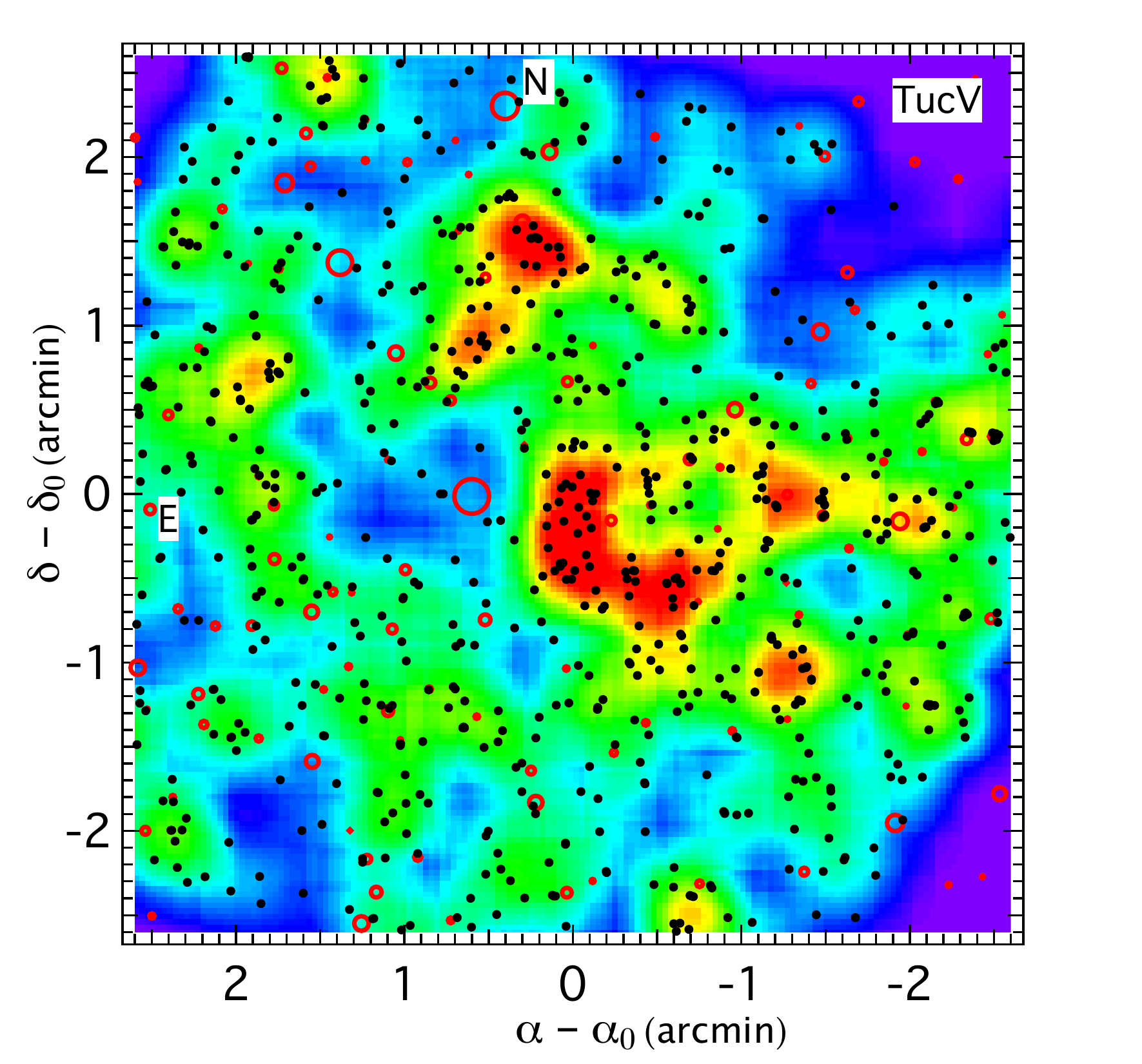}
\caption{ On-sky stellar distribution of the full field after selecting only stars that are sufficiently close to the best-fitting isochrone from Figure~\ref{fig:TucVage_metal}. The top panel shows the distribution of DES stars in the GMOS-S field and the hottest pixel has a value of $\sim 6$ stars per square arcminute. The lower panel is the new distribution revealed by the GMOS-S data. The regions of highest star density in red ($\sim 80$ stars per square arcmin) do not define a centre but are aligned along the same NE-SW direction as found in the DECam field from the top panel and the DES Survey (see Figure\,\ref{fig:TucV_DESfield}). The open red circles are the bright {\sc AllWISE} objects scaled to reflect their magnitude. Brighter objects have larger symbols. \label{fig:TucVstellar_distribution}}
\end{center}
\end{figure}

Tucana\,V (Tuc\,V), also known as DES\,J2337-6316, was reported as discovery in the second year of optical imaging data from the Dark Energy Survey
\citep{Drlica-Wagner2015}. Interestingly, it is not only the closest object known to the SMC in projection (see Figure 1) but also its heliocentric distance of $55\pm9$\,kpc is comparable.
The best-fit half-light radius as \citep{Drlica-Wagner2015} derived from an iterative MCMC analysis was found to be $r_h =1.0^{+0.3}_{-0.3}$\,arcmin and well matched by the $5\farcm5\times 5\farcm5$ GMOS-S field-of-view. However, looking at the false-colour image in Figure~\ref{fig:TucV}, there is no obvious stellar overdensity visible as opposed to both DES1 (Figure~\ref{fig:DES1}) and Eri\,III (Figure~\ref{fig:EriIII}). 

\subsection{Stellar Population}
Despite the false-colour image not revealing a clear overdensity in the field, the colour-magnitude diagram of the full GMOS-S field, right panel of Figure~\ref{fig:cmd_field}, nonetheless shows RGB and MS-like features. Given the absence of a well-defined overdensity and thus a centre (see Figure 7) in the base catalogue, we determine the age and metallicity of that population by fitting the entire field. Figure~\ref{fig:TucVage_metal} shows the maximum likelihood density map with the location of the best-fit model isochrone with an age of 11.8\,Gyr, [Fe/H]=$-2.09$ and [$\alpha$/Fe] = $+0.4$. In this field, we use E(B-V)$_{S\&F11}=0.0190$, $A_g = 0.072$ and $A_r = 0.050$ to extinction correct the data. This isochrone is a good match of the observed features in the CMD as can be seen in Figure~\ref{fig:cmdinner_TucV}. The associated distance of the Tuc\,V population is measured at $59.5$\,kpc, confirming the agreement to the SMC distance. 

The on-sky stellar distribution of Tuc\,V stars that are close to the best-fit isochrone can be seen in the top panel of Figure~\ref{fig:TucVstellar_distributionGMOS}  and in the bottom panel, the distribution of non-stellar objects in the same region of the CMD. In both panels, the bright {\sc AllWISE} objects are shown to highlight the location of the bright foreground stars and the bright background galaxies with larger circles representing brighter magnitudes. As with DES1 and Eri\,III, the bright objects in the Tuc\,V field generally do not correspond with any apparent low density regions in the stellar distribution. As seen in Figure~\ref{fig:stellar_distribution}, there is no concentrated overdensity in this field even after selecting likely Tuc\,V stars. The slight excesses visible by-eye are not sufficiently concentrated to be considered the core of an object.

The initial discovery of Tuc\,V was made with the Dark Energy Camera (DECam). In Figure~\ref{fig:TucV_DESfield} we show the locations of all stars that lie in the Tuc\,V isochrone mask selected over the entire DECam field (3 sqr deg). They are overplotted on the 2D star density histogram with our GMOS-S field shown as a box outline at $354.35$\,deg, $-63.27$\,deg. The figure confirms that there is indeed a significant stellar overdensity as reported by \citet{Drlica-Wagner2015} and that our GMOS-S field is in the correct location. In Figure~\ref{fig:TucVstellar_distribution}, after selecting only stars located close to the Tuc\,V isochrone and plotting their on-sky distribution, we can see how going from the shallow DES data ($g_{lim}\approx 23.0$, top panel) to the much deeper  GMOS-S data ($g_{lim}\approx 26.0$, bottom panel) breaks up the peak of the Tuc\,V structure into several hot spots with similarly high star densities. 
The improved number statistics from the deeper photometry reveals that there is no single overdensity in this field consistent with a cluster-like or dwarf galaxy-like morphology. Additionally, the location of the bright objects in the field do not correspond to the low density regions between the peaks of the Tuc\,V stellar distribution and therefore are not influencing our ability to accurately map this object. Returning to the wider DECam field (Figure~\ref{fig:TucV_DESfield}), there is also no evidence of a more diffuse stellar cluster or stream that would possibly make the GMOS-S field too small for such an object. Given the lack of a centre  in the Tuc\,V stellar distribution it is not possible to conduct a structure analysis for Tuc\,V. Nevertheless, we will discuss its potential nature in $\S$\ref{sec:TucVfalse}.

\section{Discussion}\label{sec:discussion}
\begin{figure}
\begin{center}
\includegraphics[width=\hsize]{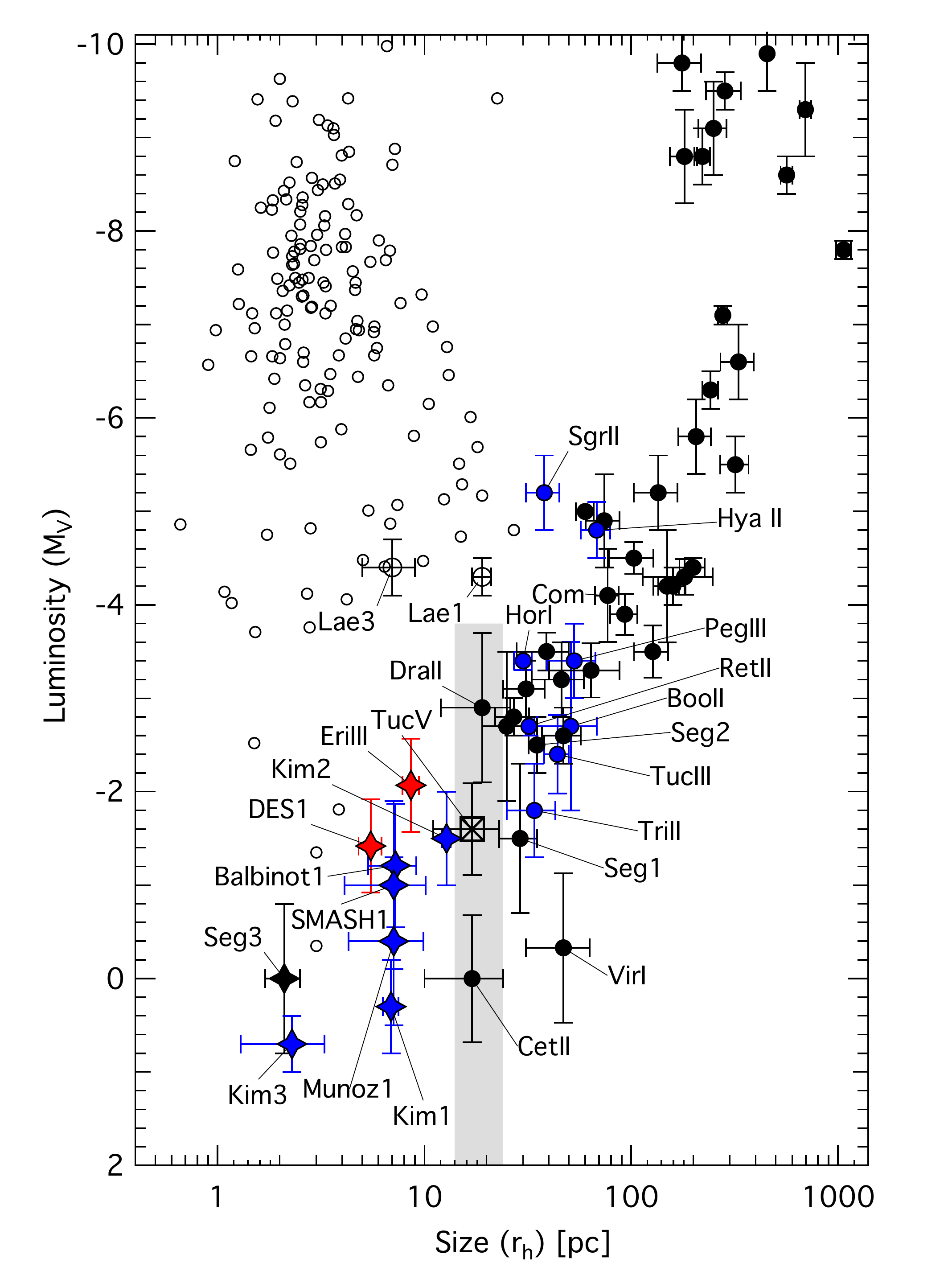}
\caption{ The size-luminosity (S-L) relation for classical Milky Way satellite galaxies (black symbols) complemented with data for Galactic globular clusters [\citet{Harris1996}, open circles] and recently discovered Milky Way satellites: Hya\,II, Kim\,1, 2 \& 3, Laevens\,1, Pisces\,II, Ret\,II, Hor\,I, Peg\,III, Segue\,1, Sgr\,II, Tri\,II, Balbinot\,1, Mu\~{n}oz\,1, SMASH\,1, Tuc\,III, and Bo\"{o}tes\,II (blue symbols). DES1 and Eri\,III (red diamonds) are found in a region occupied by star clusters. DES1 is close to Balbinot\,1, SMASH1, and AM4. The gray bar is part of the S-L plane in which Tuc\,V resides, which we discuss in $\S$\ref{sec:TUC} as ``The Trough of UnCertainty" (TUC). Objects to the left of the TUC are plotted as diamonds, objects in and to the right of the TUC are plotted as circles.
}
 \label{fig:SL-relation} 
\end{center}
\end{figure}

The fundamental properties of the stellar populations of DES1 and Eri\,III are remarkably similar. They have the same metallicity ([Fe/H]$ = -2.38^{+0.21}_{-0.19}$ vs [Fe/H]$ = -2.40^{+0.19}_{-0.12}$) and mean alpha abundance ([$\alpha$/Fe$]_{\rm avg} = +0.2^{+0.1}_{-0.1}$ for both), they have comparable ages (11.2$^{+1.0}_{-0.9}$ Gyr vs 12.5$^{+0.5}_{-0.7}$ Gyr). Structurally they also share similar properties: ellipticity (0.41$^{+0.03}_{-0.06}$ vs 0.44$^{+0.02}_{-0.03}$) and position angle (112$^\circ \pm3^\circ$ vs 109$^\circ \pm5^\circ$). Eri\,III ($r_h = 8.6^{+0.9}_{-0.8}$ pc) is about 1.5 times larger than DES1 ($r_h = 5.5^{+0.8}_{-0.7}$ pc) and consequently slightly more luminous ($M_V=-2.07\pm0.50$ vs $M_V = -1.42\pm0.50$). When it comes to their location in the Milky Way halo they are projected onto the trailing filaments of neutral hydrogen gas from the Magellanic Stream (see Figure\,1). However, both systems are more distant than the Magellanic Clouds.
DES1 ($D_{GC}=74\pm4$\,kpc) is 37\% and Eri\,III ($D_{GC}=91\pm$4\,kpc) is 69\% further away. They have similar angular separations (23.9$^\circ$ vs 22.3$^\circ$) and 3D distances (31.7 kpc vs 41.0 kpc) to the Small Magellanic Cloud.

DES1 is the less massive of the two and has demonstrably fewer stars than Eri\,III as is noticeable in the CMDs (Figures\,\ref{fig:cmdiso_DES1} and \ref{fig:cmdiso_EriIII}) and quantified by the parameter $N_*$ in Tables~\ref{tab:DES1parameters} and \ref{tab:EriIII_parameters}.
DES1 lacks an obvious red giant branch and the main sequence is less populated when compared to Eri\,III. Despite these differences, both objects have observed luminosity functions that are well matched with a Salpeter IMF and power law slope of $\alpha = -2.35$. This suggests that they always have been small stellar systems and have not lost significant amounts of mass.

DES1 and Eri\,III are found in the size-luminosity diagram (Figure~\ref{fig:SL-relation}, red diamonds) in a region dominated by ultra-faint star clusters. DES1's half-light radius and stellar content puts it close to the recently discovered objects Balbinot\,1 \citep{Balbinot2013} and SMASH1 \citep{Martin2016b}, while Eri\,III is the most luminous amongst objects with half-light radii less than 10\,pc. They are all significantly fainter than the bulk of the Milky Way globular clusters, plotted as open circles. The closest star clusters to DES1 are AM4 \citep{Carraro2009}, Koposov 1 \& 2 \citep{Koposov2007} in order of decreasing luminosity. Interestingly, \citet{Paust2014} have determined that Koposov 1 \& 2 are intermediate-age, open star clusters possibly related  to the Sagittarius dwarf galaxy, and \citet{Carraro2009} speculated that AM4 
might be associated with Sagittarius too. Since the bulk of objects in this part of the size-luminosity diagram around DES1 and Eri\,III are known to be star clusters and they appear distinct from the Milky Way population, the conclusion that Koposov 1, 2 and perhaps AM4 are related to the Sagittarius dwarf, raises the possibility that all of them are star clusters of non-Galactic origin.

\subsection{Metallicity [Fe/H] and Alpha Abundance [$\alpha$/Fe]}
\begin{figure}
\begin{center}
\includegraphics[width=0.98\hsize]{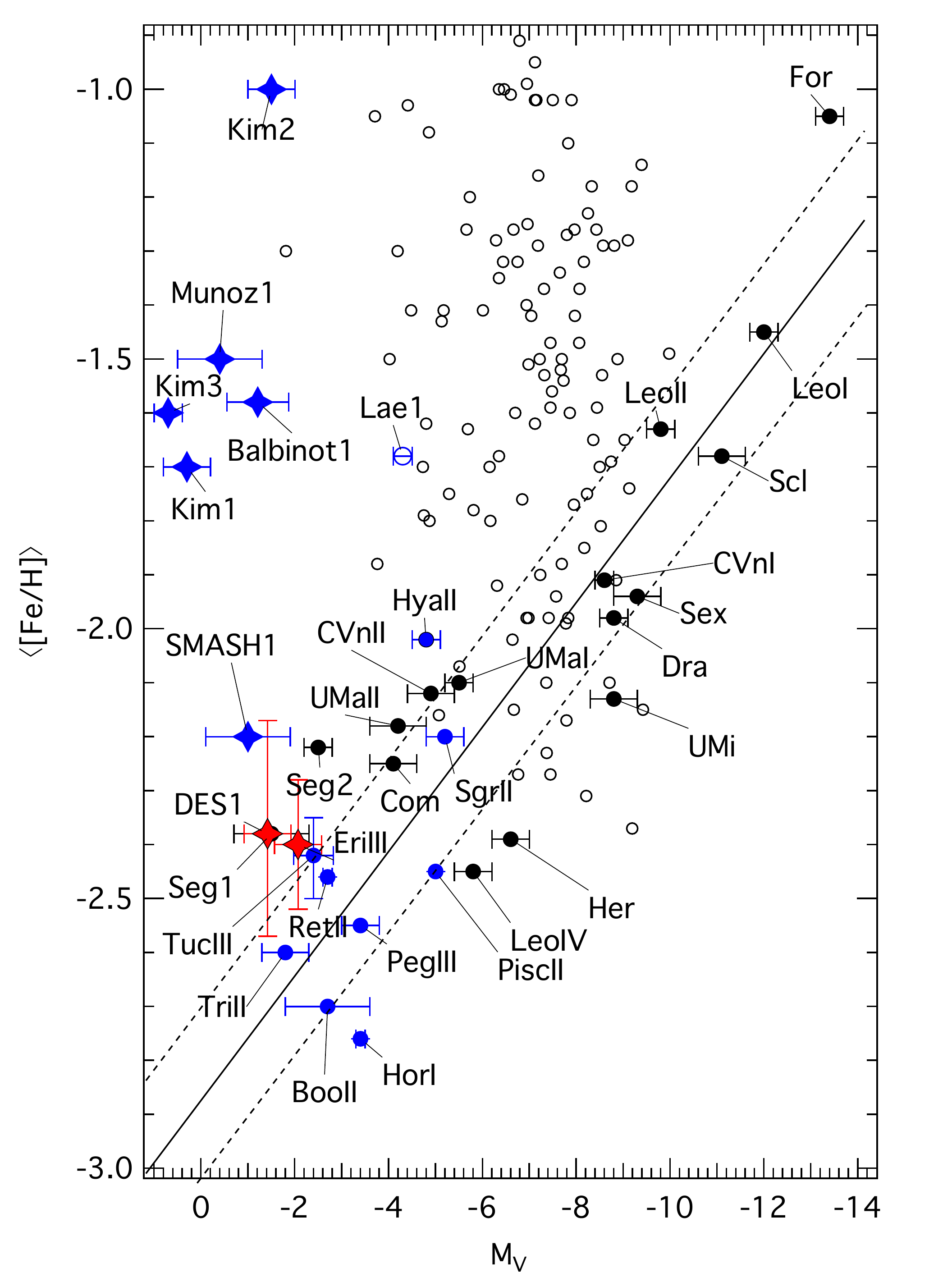}
\caption{ The luminosity-stellar metallicity relation for classical Milky Way satellite galaxies (black dots) complemented with data from the literature for Hya\,II, Kim\,1,2\&3, Laevens\,1, Pisces\,II, Ret\,II, Hor\,I, Peg\,III, Segue\,1, Sgr\,II, Tri\,II, Balbinot\,1, Mu\~{n}oz\,1, SMASH\,1 Tuc\,III, and Bo\"{o}tes\,II (blue dots). The black and dotted lines represent the least-squares fit and $1\sigma$\,rms from ~\citet{Kirby2013}, based on spectroscopically studied stars in 14 galaxies (Segue\,2 was excluded). DES1 and Eri\,III (red diamonds) do fall just outside of the 1$\sigma$ confidence band. DES1 is next to Segue 1, whereas the higher luminosity of Eri\,III moves this system closer to the LZ-relation. The open circles are the data from the Milky Way globular clusters, ~\citet{Harris1996}.}
 \label{fig:LZ-relation} 
\end{center}
\end{figure}

Milky Way satellite galaxies are known to follow a well-defined relationship between their total luminosity $M_V$ and average metallicity $\langle$[Fe/H]$\rangle$, e.g.~\citet{Kirby2013}.
Given the small intrinsic scatter the relation can be used, together with or as an alternative to the S-L relation, as diagnostic tool to discriminate between a dwarf galaxy and star cluster. Figure~\ref{fig:LZ-relation} shows this luminosity-metallicity parameter space. The black and dotted lines are the least-squares fit with the $1\sigma$ confidence band about the relation based on 13 galaxies taken from ~\citet{Kirby2013}. It corresponds to the fit of the data for the chemically best studied dwarf galaxies. We complemented that plot with data for new stellar systems: Hya\,II ~\citep{Martin2015}, Kim\,1 \citep{KimJerjen2015a}, Kim\,2 \citep{Kim2}, Kim\,3 \citep{Kim2016}, Laevens\,1, Pisces\,II ~\citep{Kirby2015}, Ret\,II \citep{Walker2015,Koposov2015b}, Hor\,I ~\citep{Koposov2015b}, Tri\,II ~\citep{Laevens2015b,Martin2016a}, Balbinot\,1 ~\citep{Balbinot2013}, Bo\"{o}tes\,II ~\citep{Koch2014}, Mu\~{n}oz1 ~\citep{Munoz2012}, and Peg\,III \citep{Kim2015b, Kim2016b}. 
DES1 and Eri\,III are just outside of the $1\sigma$ confidence band. They are close to Segue\,1 ~\citep{Belokurov2007} and Segue\,2 \citep{Belokurov2009}. 
The error bar for <[Fe/H]> brings DES1 also close to SMASH1. Segue 1, DES1, and Eri\,III share a similar total luminosity and mean stellar metallicity and also the same ellipticity ($0.48^{+0.10}_{-0.13}$  vs $0.41^{+0.03}_{-0.06}$ vs $0.44^{+0.02}_{-0.03}$ \citep{Geha2009}. The two fundamental differences are the half-light radius ($29^{+8}_{-5}$\,pc)  vs $5.5^{+0.8}_{-0.7}$\,pc vs. $8.6^{+0.9}_{-0.8}$\,pc) 
 
and the distance from the Milky Way. Segue\,1 is at a Galactocentric distance of 28\,kpc, while DES1 is 2.6 and Eri\,III 3.2 times further away. 

Segue\,1 \& 2 are both classified as ultra-faint dwarf galaxies based on the high mass-to-light (M/L) ratio 
estimates and a large intrinsic metallicity spread of more than 2\,dex (Segue\,1: \cite{Geha2009,Frebel2014}; Segue\,2: \citet{Kirby2013}. The metallicity spread is a signature observed in ultra-faint dwarf galaxies, while star clusters do not show that characteristic. DES1 and Eri\,III have 
[$\alpha$/Fe$]_{\rm avg}\sim0.2$\,dex consistent with the mean value observed for ultra-faint dwarf galaxies (Vargas et al.~2013, their Fig.~6)

There are two pieces of evidence that suggest DES1 and Eri\,III have a non-Galactic origin. First, they are at large Galactocentric distances which is unusual for star clusters, Galactic or non-Galactic. Only eight known clusters have $D_{GC}>70$\,kpc \citep{Kim2}. Secondly, they are close to the two dwarf galaxies Segue 1 \& 2 in LZ-space and only $\approx 0.2-0.3$\,dex more metal-rich than the LZ-relation predicts for systems of their luminosities. \citet{Belokurov2009} and \citet{Kirby2013} speculated that Segue 1 \& 2 may be the remnants of tidally stripped dwarf galaxies. DES1 and Eri\,III could be other examples of Milky Way satellite galaxies that came to be ultra-faint through tidal stripping but their small size implies that they were either intrinsically smaller to begin with or that they underwent significantly more tidal stripping than Segue 1 \& 2. Keeping in mind that the luminosity function of these systems, as discussed in the previous section, implies that they have not undergone much disruption during their lifetime. In which case, these would be the smallest known galaxies to fit this scenario. A key piece of evidence for this hypothesis would be a metallicity spread in the stellar population, which can be tested with spectroscopic follow up.

\subsection{Mass Segregation}\label{sec:mass_segregation}
\begin{figure}
\begin{center}
\includegraphics[width=0.98\hsize]{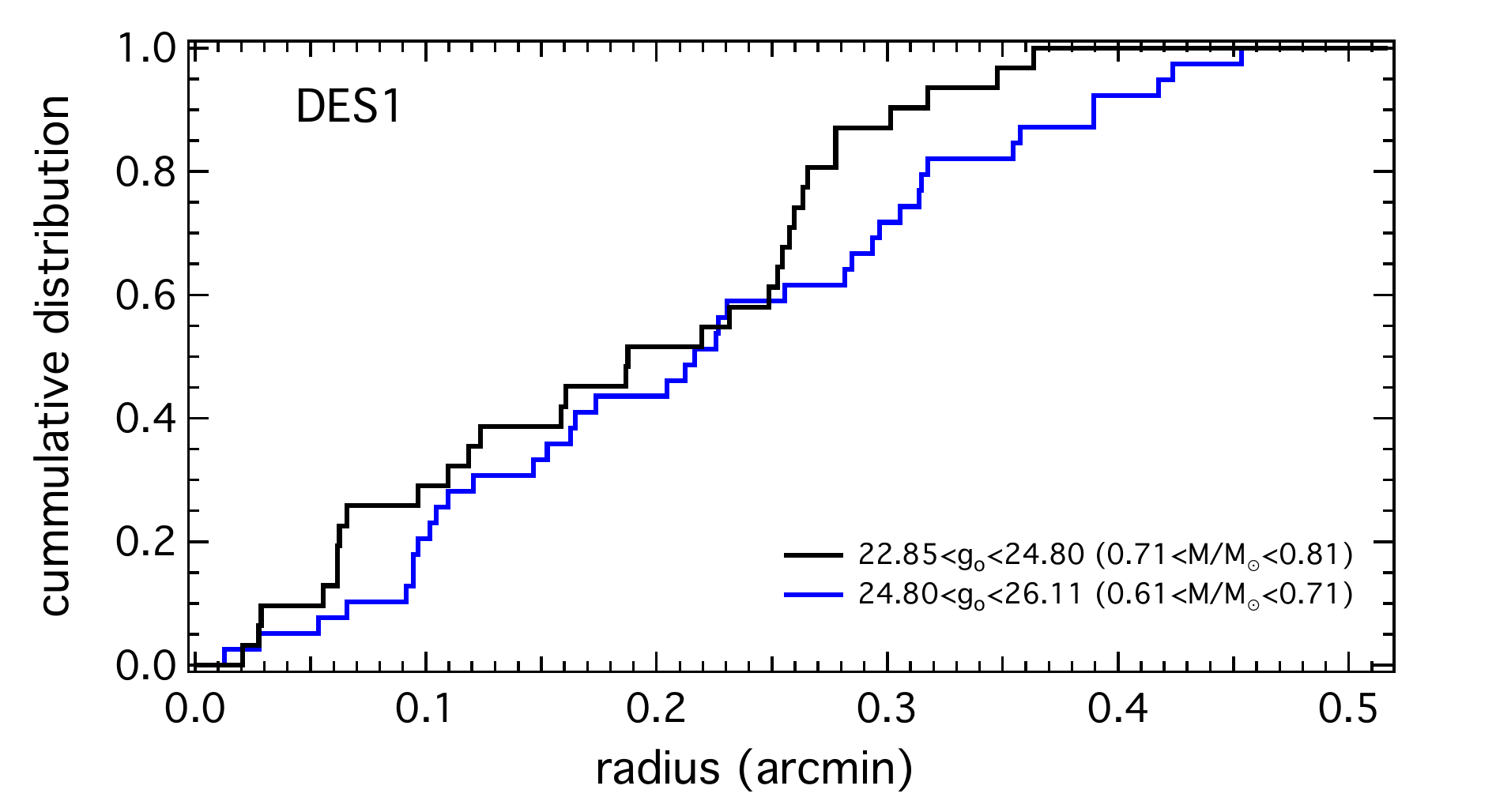}\hspace{-0.2cm}
\includegraphics[width=0.98\hsize]{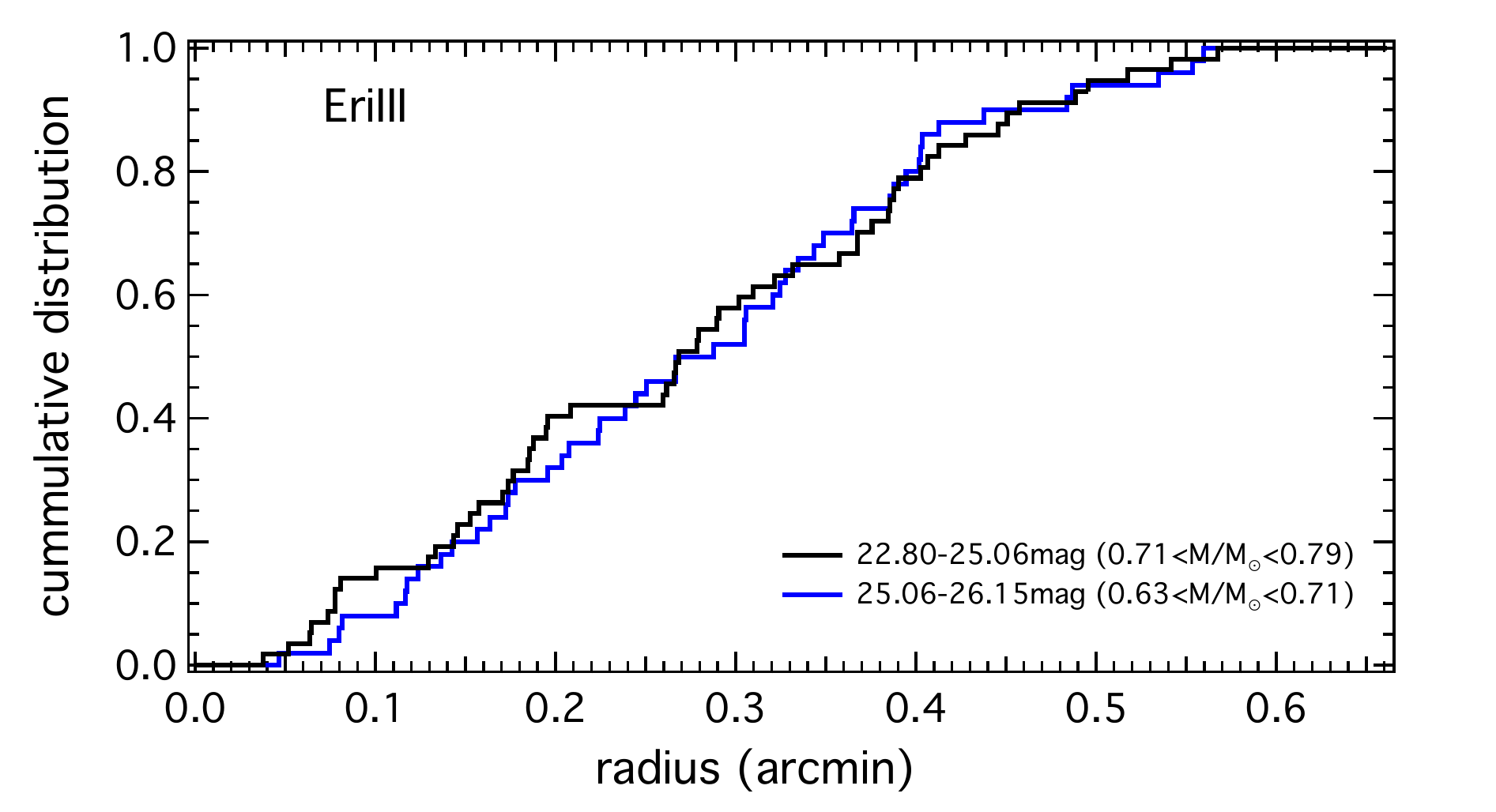}\hspace{-0.2cm}
\caption{Cumulative distribution functions for DES1 (top) and Eri\,III (bottom) main sequence stars out to 2r$_h$ from the nominal centre. The $\approx 3$ magnitudes from the MSTO down to the 50\% completeness limit were subdivided into two magnitude intervals that correspond to two stellar mass bins of equal size: $0.61<M/M_\odot<0.71$ and $0.71<M/M_\odot<0.81$ solar masses for DES1, and $0.63<M/M_\odot<0.71$ and $0.71<M/M_\odot<0.79$ solar masses for Eri\,III, respectively.
No evidence of mass segregation is found in the two stellar systems as concluded from the large $p$-values of a two-sided KS-test: $p=0.18$ and $p=0.92$, respectively.}
 \label{fig:mass_segr_DES1EriIII} 
\end{center}
\end{figure}

Similar to the analysis conducted for Kim\,2 \citep{Kim2} we performed a two-sample Kolmogorov-Smirnov test ~\citep{Massey1951}  to investigate whether MS stars with different masses in DES1 or Eri\,III follow the same spatial distribution. Figure~\ref{fig:mass_segr_DES1EriIII} shows the cumulative distribution functions for DES1 (top) and Eri\,III (bottom) main sequence stars out to 2r$_h$ from the nominal cluster centre. The $\approx 3$ magnitudes from the MSTO down to the 50\% completeness limit were subdivided into two magnitude intervals that correspond to two equal-size mass bins: $\Delta M/M_\odot=0.1$ for DES1, and $\Delta M/M_\odot=0.08$ for Eri\,III, respectively. For comparison, the confirmation of mass segregation in Kim\,2 utilised three mass bins of $\Delta M/M_\odot=0.1$. The KS test yields relative large $p-$values of $p=0.18$ and $p=0.92$, respectively, implying no evidence of mass segregation in the main-sequence population of the two stellar systems.

These results suggest that neither DES1 nor Eri\,III has experienced substantial mass loss from two-body relaxation and tidal stripping. This picture is also consistent with our finding that the observed LFs of DES1 and Eri\,III are well described with a model luminosity function using a Salpeter IMF.

We note that the stellar mass covered by main sequence stars in old ($\sim 10-12$\,Gyr), metal-poor ($-2.5<$[Fe/H]$<-1.5$) stellar populations are inherently small. Accessing a mass range of the order of $M/M_\odot=0.3$ would require photometry 4-5 magnitudes below the main sequence turn-off.

\subsection{DES1 \& Eri\,III: Star Cluster or Dwarf Galaxy?}\label{sec:SCorDG}
Using the derived properties of DES1 and Eridanus\,III, we attempt to determine what is the most likely nature of these objects. One of the main challenges in interpreting these data is that given the location of these objects in the multi-dimensional parameter space, we can expect the following possible explanations for their origins: Milky Way globular cluster, Milky Way dwarf galaxy, LMC/SMC star cluster or LMC/SMC dwarf galaxy.

{\it Size ($r_h$):} both DES1 and Eri\,III are small stellar systems with half-light radii less that 10\,pc. and as such are consistent with star clusters. In comparison to the few known globular clusters at comparable Galactocentric distances (Eridanus, AM1, Pal\,4, Pal\,3, NGC~2419, Pal\,14 -- in increasing size order), these clusters are $2.2-4.9$ times larger than DES1 and $1.4-3.2$ times larger than Eri\,III,  see \citet{Harris1996}. 

{\it Galactocentric Distance:} there are only a few known MW globular clusters at the distances of DES1 and Eri\,III (D$_{GC}=74$\,kpc and 91\,kpc) and so, it appears unlikely that they are Milky Way star clusters. Their distances are more compatible with Milky Way dwarf galaxies however given their proximity to the Magellanic Cloud system and their location in the trailing component of the Magellanic Stream provides two alternative views. They might be star clusters or dwarf galaxies that are in-falling with the LMC and SMC galaxies.

{\it Ellipticity ($\epsilon$):} MW globular clusters are spherical systems with a mean ellipticity $\langle\epsilon\rangle=0.08$ and a standard deviation of $\sigma_\epsilon=0.06$. The maximum ellipticities were measured for the clusters NGC\,6144 ($\epsilon=0.25$) and M19 ($\epsilon=0.27$) from \citet{Harris1996}. Hence, DES1 and Eri\,III are significantly more elliptical than all of the known Milky Way globular clusters. Their compact size effectively rules out them being ultra-faint dwarf galaxies and suggests that they are probably unusual star clusters, most likely dark matter free star clusters in the process of dissolution. Their elongated stellar distributions could be due to the tidal fields of the LMC/SMC or Milky Way. In this context, the elongated star cluster SMASH1 close to the LMC comes to mind.

{\it Luminosity Function and Mass Segregation:} The completeness-corrected luminosity functions for DES1 and Eri\,III are consistent with a Salpeter IMF. There is no evidence for a flatter LF through the lack of low mass stars. This suggests that DES1 and Eri\,III have always been small clusters and have not lost significant amounts of mass at the current stage. This picture is also consistent with the absence of mass segregation in the MS populations covering the approximate mass range $0.60<M/M_\odot <0.80$. 

{\it Size-Luminosity relation:} DES1 and Eri\,III are located in the ultra-faint regime of the size-luminosity relation as exhibited by the Milky Way satellite galaxies. However, that part of the parameter space ($M_V\gtrsim-2$, $r_h\lesssim 10$\,pc) is now populated by a number of ultra-faint star clusters. In terms of size, DES1 shares properties with other confirmed star clusters like AM4 and Balbinot\,1 but DES1 has twice the Galactocentric distance of those two. Eri\,III despite its brighter luminosity (M$_V$=$-2.07$) is closer in properties to star clusters than to dwarf galaxies. It is about five times smaller than the dwarf galaxy candidate Tri\,II, which has a similar luminosity, and Eri\,III is fainter and roughly half the size of the ultra-faint dwarf galaxy candidate Draco\,II. Eri\,III is probably a star cluster too.

{\it Luminosity-Metallicity relation:} DES1 has almost identical properties to the dwarf galaxy Segue1 despite being approximately six times smaller.  Although the error in [Fe/H] allows the possibility that DES1 is a dwarf galaxy, its luminosity suggests it is more likely to be a star cluster. Eri\,III is also found close to the $1\sigma$ confidence line of the dwarf galaxy LZ-relation ~\citep{Kirby2013} and has properties intermediate to Tucana\,III and DES1, although it is about five times smaller than Tuc\,III. With regard to the outer halo globular cluster population, only NGC~2419 ($D_\odot = 82.6$ kpc) has a similarly low metallicity ([Fe/H]$=-2.19$, \citet{Harris1996}), but this object is much more luminous ($M_V=-9.42$) and as such is considerably different.

DES1 and Eri\,III have equivalent luminosities to ultra-faint dwarf galaxies more than five times larger and this suggests that they reside in a gravitational potential that is, locally, more concentrated than those objects of similar luminosities (e.g. Segue\,1, Segue\,2, Triangulum\,II). Their metallicities however are considerably lower than Milky Way globular clusters and so we propose that DES1 and Eri\,III are most likely dissolving star clusters associated with the Magellanic Clouds. They are simply too distant and different to be considered part of the MW globular cluster system and both lie in close proximity to the Magellanic Clouds.

\subsection{Tucana\,V: a false positive ultra-faint candidate?}\label{sec:TucVfalse}
\begin{figure}
\begin{center}
\includegraphics[width=1\hsize]{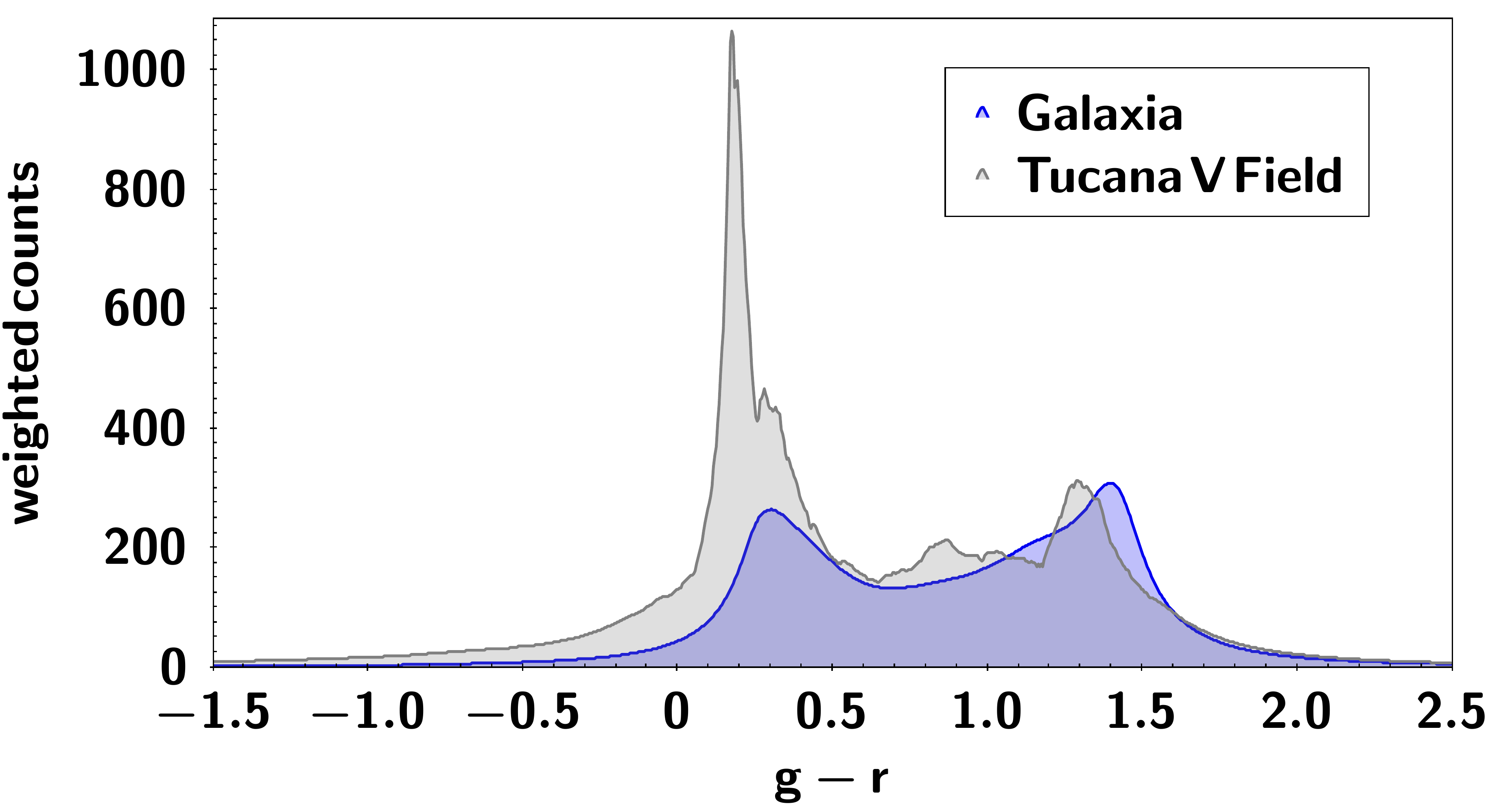}\hspace{-0.2cm}
\caption{Tuc\,V histogram of all stars within $20\leq g \leq24$ and $-1.5 \leq (g-r) \leq +2.5$ with the GMOS-S data presented in grayscale and the {\sc Galaxia} galactic model data for the Tuc\,V field ~\citep{Sharma2011} presented in blue. The counts have been scaled such that the expected number of stars from {\sc Galaxia} approximately matches the data in the region around $(g-r)\sim +1.5$. There is a clear excess of blue stars in the region of Tuc\,V "main-sequence" as compared to the model.}
 \label{fig:TucV_hist} 
\end{center}
\end{figure}

As demonstrated in $\S$\ref{sec:TucVProp}, our deep photometry of Tucana\,V did not reveal a stellar overdensity within the GMOS-S field and in particular, none that could correspond with the expected properties of Tuc\,V as outlined in ~\citet{Drlica-Wagner2015} (see Figure~\ref{fig:stellar_distribution}, right panel). Determining the best-fit stellar population to all of the stars in the field\footnote{Tuc\,V best-fit isochrone: $m-M$ = 18.8; $D_\odot$ = 59.7 kpc; Age $\sim$ 11.8 Gyr; [Fe/H] = $-2.09$\,dex; [$\alpha$/Fe] = $+0.4$\,dex.} and selecting only stars belonging to that isochrone also failed to recover any stellar overdensity that resembles a star cluster or dwarf galaxy radial profile (Figure~\ref{fig:TucVstellar_distribution}, bottom panel). This rules out a bound star cluster or dwarf galaxy as the explanation for the Tuc\,V phenomenon. However, Tuc\,V is not a false positive. These stars clearly belong to a coherent stellar population and the next obvious origins for this stellar excess is the Milky Way stellar halo, the Small Magellanic Cloud (SMC) stellar halo or a disrupted star cluster.

To test the possibility that the Tuc\,V stars are related to the Milky Way halo, we employed the {\sc Galaxia} code ~\citep{Sharma2011} to generate the expected stellar populations in the direction of Tuc\,V out to 100\,kpc. In Figure~\ref{fig:TucV_hist}, the histogram of ($g - r$) colours between the $g-$band magnitudes of 20 and 24\,mag are plotted for both our data (grey) and Galaxia model (blue). To scale the two data sets we use the observed number of red stars around $(g-r) \sim +1.5$ in the data as a calibration reference for the model. Aside from the small colour differences between the samples, it is clear that our Tuc\,V CMD contains significantly more stars in the color range $0.1<g-r<0.4$\,mag and that the main sequence in the Tuc\,V data cannot be attributed to Galactic halo stars. 

Another possible origin for the Tuc\,V stars is an extended structure related to the Small Magellanic Cloud. If we take the centre of the SMC from~\citet{deGrijs2015} as ($\alpha$,$\delta$) = ($15.129^\circ$, $-72.720^\circ$) at a distance of $D_\odot = 61.94$\,kpc and the Tuc\,V field position as ($l,b$) = ($316.31^\circ$,$-51.89^\circ$) at a distance of $D_\odot = 59.7$\,kpc, we can compute the three-dimensional distance using the Astropy python package\footnote{http://www.astropy.org ~\citep{Astropy2013}} finding $D_{SMC}\sim$13 kpc ($\sim 12\fdg1$). Investigating the entire DECam field in which the Tuc\,V stellar excess resides, we selected all stars that occupy the same region of the colour-magnitude diagram as the Tuc\,V isochrone and plot their distribution on the sky (see Figure~\ref{fig:TucV_DESfield}). There are no obvious stellar streams or gradients in the field. The significant excess of Tuc\,V stars is very localised within the GMOS-S field from this dataset and represents the only significant peak in the DECam field. 

The closest known SMC-halo substructure is the SMC Northern Overdensity (SMCNOD) from ~\citet{Pieres2017} which is 8$^\circ$ or 8\,kpc from the SMC.  If we overlay the SMCNOD isochrone with a distance of $\sim$59.7\,kpc on our Tuc\,V CMD (Figure~\ref{fig:cmdinner_TucV}, blue line), we find it is roughly consistent with the photometry. This isochrone requires the MSTO to be around $g\sim 22$\,mag but given the small size of the GMOS-S field it is possible that the turn-off from the best-fit isochrone is being biased by the small grouping of stars around $g\sim 22.5$. Those MSTO stars are spread across the entire field with only about half of them in locations that could be considered {\it inside} Tuc\,V. From Figure~\ref{fig:TucVage_metal}, it is clear that a 6\,Gyr, [Fe/H]$=-1.3$\,dex SMCNOD isochrone is broadly consistent with the age-metallicity degeneracy of this stellar population.  At this distance, although it is outside the nominal 7$\fdg$5 break radius seen in the SMC halo density profile (Fig.~4 from \citet{Pieres2017}), could these Tuc\,V stars be part of the SMC-halo? The SMC halo does show an increasing excess in star counts along some position angles (e.g. P.A.$\sim$200$^\circ$, Fig.~4 of \citet{Pieres2017}) and it seems that even the typical SMC halo density profile has not yet reached the background at the angular distance of Tuc\,V. Taking all these factors into account, it appears Tuc\,V  could be either a chance grouping of SMC halo stars or represents an SMC-halo substructure like the SMCNOD. It is {\it not an ultra-faint stellar system} as originally thought.

Finally, Tuc\,V could be a tidally disrupted star cluster in its final stages of dissolution. It is uncertain how the final stage of a disrupting star cluster would appear on the sky and unfortunately such analysis is beyond the scope of this paper. Spectroscopic analysis of the Tuc\,V stars would be desirable to confirm that the sub-groups seen in Figure~\ref{fig:TucVstellar_distribution} are chemically similar and likely to have a common origin.

Tucana\,V is not an ultra-faint dwarf galaxy candidate in the classical sense leaving us to speculate about other systems reported in the literature that have strong similarities to Tuc\,V. In particular, the colour-magnitude diagram of the Draco\,II dwarf galaxy candidate \citep{Laevens2015b}, from their figure 1 (middle panel), has the same truncated main sequence like Tuc\,V and the isochrone fit struggles to fit this feature because it is constrained by the apparent Red Giant Branch. If we consider the Field CMD and the Object CMD above the main sequence turn-off ($i_{P1.0} < 20$\,mag) we see that they are almost identical and the on-sky distribution is small, sparse ($r_h = 2\farcm7^{+1.0}_{-0.8}$) and not centrally concentrated. As with Tuc\,V, there is obviously a coherent stellar population within the field, as seen by the manifest main sequence in the CMD, but it lacks a conspicuous progenitor. Draco\,II is somewhat isolated on the sky from other large dwarf galaxies but even though it is relatively high above the Galactic Plane at $(l,b)$ = ($98.3^\circ$, $+42.9^\circ)$, it is well within the Milky Way stellar halo ($D_{GC}= 22\pm3$\,kpc). And as such may represent an equivalent chance overdensity of Milky Way halo stars in much the same manner that Tucana\,V may be a chance overdensity of SMC Halo stars. Similarly, Cetus\,II ~\citep{Drlica-Wagner2015}, $(l,b)$ = ($156.48^\circ$, $-78.53^\circ$) at D$_{GC} \sim$ 32\,kpc, with a half-light radius of only $r_h \sim 1\farcm9$, is another interesting case. As candidate systems become fainter and less populated, deep follow-up photometry of all the ultra-faint dwarf galaxy candidates is crucial to better undestand the nature of these systems and use their properties to refine and improve the current search algorithms. 

\subsection{The Trough of Uncertainty}\label{sec:TUC}

In the previous section, we identify the Tucana\,V stellar excess as a likely chance overdensity in the Small Magellanic Cloud Halo or a dissolving star cluster, and after examining the colour-magnitude diagrams of other ultra-faint dwarf galaxy candidates concluded Draco\,II and Cetus\,II may also represent systems of that nature. Serendipitously, all of these objects appear in the same part of the size-luminosity diagram. In Figure~\ref{fig:SL-relation}, we highlight the region (gray box) encompassed by these objects showing that either side of the region there is an apparent clear distinction between those objects tentatively identified as star clusters (to the left) and dwarf galaxies (to the right). To honour Tucana\,V as the potential prototype false-positive identification, we label this region the Trough of UnCertainty (TUC). The TUC illuminates the region of the S-L plane where potential false-positive identifications might be occurring and might also serve to divide the S-L plane between star clusters and dwarf galaxies.

If we compare objects on either side of the TUC in the luminosity-metallicity relation (Figure~\ref{fig:LZ-relation}), we find that those to the left of the TUC (plotted as diamonds) do not fall on the LZ-relation in Figure~\ref{fig:LZ-relation} while those to right of the TUC (plotted as circles) are consistent with the relation. It is important to note that many of the objects below $M_V=-4$ only have preliminary metallicity estimates at this stage and so the true scatter of the relation at these metallicities is uncertain. While we have raised possible issues with Draco\,II and Cetus\,II, it is obvious from Figure~\ref{fig:SL-relation} that simply removing Tuc\,V, a clear gap has opened between the star clusters and dwarf galaxies in this relation. Deep photometry of the objects in the TUC is crucial to resolving their status and it seems that the TUC is delimiting the border between star clusters and dwarf galaxies.  

Inside the Trough of UnCertainty, the objects there are either of the specific dimension where dissolving star clusters are still coherent enough to be mistaken for dwarf galaxies or the point at which random fluctuations within the predominantly smooth stellar halos can be detected. While Tuc\,V is a cautionary tale of interpreting data dominated by small number statistics, it has highlighted the fascinating intersection in the size-luminosity plane where star clusters meet dwarf galaxies.

\section{Conclusion}
This deep photometry of DES1 and Eridanus\,III have revealed them to be old, small and highly elliptical stellar populations with very low metallicity at the outer reaches of the Milky Way. Our analysis has shown that their observed luminosity functions and lack of any apparent mass segregation both point to systems which have undergone very little tidal stripping. Coupling these results together, we conclude that they are most likely star clusters infalling with the Magellanic Clouds. In this regard, they join SMASH1 and Tucana\,III as probable MC star clusters. 

Tucana\,V represents a different challenge. There {\it is} an excess of stars in this field compatible with a single stellar population however, this overdensity does not have a well-defined centre and does not follow a radial density profile consistent with any other ultra-faint system. Ever since \citet{Klypin1999} highlighted the "Missing Satellites" problem, the race has been on to scrutinize the low mass end of the Milky Way's galaxy luminosity function. The expectation has been that there should be hundreds of smaller satellites of the Milky Way if the hierarchical formation scenario of a $\Lambda$Cold Dark Matter universe was to be validated. While the latest all sky surveys have delivered a few dozen more candidates, as these systems become smaller and fainter it becomes even more important to confirm their status as dwarf galaxies. With so few known systems in the ultra-faint regime, any incorrectly classified objects could skew the distribution away from the true solution. In this paper, we have shown that Tucana\,V is almost certainly one of these false-positive candidates and we propose that it is either a disrupted SMC star cluster or an anomaly in the SMC stellar halo. 

We speculate that Draco\,II and Cetus\,II may yet be other false-positive detections and along with Tuc\,V occupy a region of the size-luminosity plane (TUC) that may be peculiar to this sort of object. Our results demonstrate that the shallow discovery data are potentially insufficient in unambiguously identifying objects like Tuc\,V and that deeper follow-up observations are crucial to avoid further false-positives from entering the sample of Milky Way satellites.

\section{Acknowledgements}
The authors would like the thank the anonymous referee for their work in improving this manuscript.

BCC and HJ acknowledge the support of the Australian Research Council through Discovery project DP150100862.

BCC would like to thank Andrew Dolphin for his assistance with {\sc dolphot} for the photometry. 

This paper is based on observations obtained at the Gemini Observatory (GS-2016B-Q-7), which is operated by the Association of Universities for Research in Astronomy, Inc., under a cooperative agreement with the NSF on behalf of the Gemini partnership: the National Science Foundation (United States), the National Research Council (Canada), CONICYT (Chile), Ministerio de Ciencia, Tecnolog\'{i}a e Innovaci\'{o}n Productiva (Argentina), and Minist\'{e}rio da Ci\^{e}ncia, Tecnologia e Inova\c{c}\~{a}o (Brazil). 

This research has made use of: "Aladin sky atlas" developed at CDS, Strasbourg Observatory, France. \citep{2000A&AS..143...33B, 2014ASPC..485..277B}; the AAVSO Photometric All-Sky Survey (APASS), funded by the Robert Martin Ayers Sciences Fund; {\sc Topcat} in exploring and understanding this dataset \citep{2005ASPC..347...29T}; Astropy, a community-developed core Python package for Astronomy (Astropy Collaboration, 2013); SIMBAD database, operated at CDS, Strasbourg, France.

This project used public archival data from the Dark Energy Survey (DES). Funding for the DES Projects has been provided by the U.S. Department of Energy, the U.S. National Science Foundation, the Ministry of Science and Education of Spain, the Science and Technology Facilities Council of the United Kingdom, the Higher Education Funding Council for England, the National Center for Supercomputing Applications at the University of Illinois at Urbana-Champaign, the Kavli Institute of Cosmological Physics at the University of Chicago, the Center for Cosmology and Astro-Particle Physics at the Ohio State University, the Mitchell Institute for Fundamental Physics and Astronomy at Texas A\&M University, Financiadora de Estudos e Projetos, Funda\c{c}\~{a}o Carlos Chagas Filho de Amparo \`{a} Pesquisa do Estado do Rio de Janeiro, Conselho Nacional de Desenvolvimento Cient\'{i}fico e Tecnol\'{o}gico and the Minist\'{e}rio da Ci\^{e}ncia, Tecnologia e Inova\c{c}\~{a}o, the Deutsche Forschungsgemeinschaft and the Collaborating Institutions in the Dark Energy Survey. The Collaborating Institutions are Argonne National Laboratory, the University of California at Santa Cruz, the University of Cambridge, Centro de Investigaciones En\'{e}rgeticas, Medioambientales y Tecnol\'{o}gicas-Madrid, the University of Chicago, University College London, the DES-Brazil Consortium, the University of Edinburgh, the Eidgen\"{o}ssische Technische Hochschule (ETH) Z\"{u}rich, Fermi National Accelerator Laboratory, the University of Illinois at Urbana-Champaign, the Institut de Ci\`{e}ncies de l'Espai (IEEC/CSIC), the Institut de F\'{i}sica d'Altes Energies, Lawrence Berkeley National Laboratory, the Ludwig-Maximilians Universit\"{a}t M\"{u}nchen and the associated Excellence Cluster Universe, the University of Michigan, the National Optical Astronomy Observatory, the University of Nottingham, the Ohio State University, the University of Pennsylvania, the University of Portsmouth, SLAC National Accelerator Laboratory, Stanford University, the University of Sussex, and Texas A\&M University. 

This publication makes use of data products from the Wide-field Infrared Survey Explorer, which is a joint project of the University of California, Los Angeles, and the Jet Propulsion Laboratory/California Institute of Technology, funded by the National Aeronautics and Space Administration.





\end{document}